\documentclass[pdflatex,sn-basic,referee]{sn-jnl}% Default

\usepackage{graphicx}%
\usepackage{multirow}%
\usepackage{amsmath,amssymb,amsfonts}%
\usepackage{amsthm}%
\usepackage{mathrsfs}%
\usepackage[title]{appendix}%
\usepackage{xcolor}%
\usepackage{textcomp}%
\usepackage{manyfoot}%
\usepackage{booktabs}%
\usepackage{algorithm}%
\usepackage{algorithmicx}%
\usepackage{algpseudocode}%
\usepackage{listings}%
\begin{document}

\title{Electron Weibel instability and~quasi-magnetostatic structures in~an~expanding collisionless plasma}

\author*[1]{\fnm{Vladimir V.} \sur{Kocharovsky}}\email{kochar@ipfran.ru}

\author[1]{\fnm{Anton A.} \sur{Nechaev}}\email{a.nechaev@ipfran.ru}

\author[1]{\fnm{Mikhail A.} \sur{Garasev}}\email{garasev@ipfran.ru}

\affil[1]{\orgname{Institute of Applied Physics, Russian Academy of Sciences}, \orgaddress{\city{Nizhny Novgorod}, \postcode{603950}, \country{Russia}}}

\abstract{
We outline transient quasi-magnetostatic phenomena associated with Weibel-type instabilities, mainly, the formation and decay of the current sheets and filaments. We consider a collisionless anisotropic plasma cloud with hot electrons expanding into an inhomogeneous background plasma taking into account an external magnetic field in different geometries entailing 
(i) a hot-electron spot of a circular or cylindrical form within an initial-value problem or a finite-time injection of electrons from a target surface, 
(ii) inhomogeneous layers of cold background plasma of different densities and spatial scales, 
(iii) an external magnetic field with three orientations: perpendicular to the target or along it, directed either across or parallel to a long axis of the hot-electron spot. 
We heed typical laser-plasma experiments. 
We outline development of the principle current structures linked with distinct forms of the anisotropic electron velocity distribution using particle-in-cell modeling of the instability process for diverse sets of the attributes (i)--(iii). Applications to the analysis of laboratory and space plasma problems involving an explosive development of the small-scale filamentation of the electric current and self-consistent magnetic turbulence are discussed.
}


\maketitle

\section{Introduction}
\label{sec:intro}

\subsection{The scope of problems under consideration}
\label{sec:Scope}

The review aims at a classical problem of the decay of a strong discontinuity in plasma and a similar problem of the injection of a plasma with hot electrons into a rarefied cold plasma or into a vacuum with a magnetic field in the absence of significant particle collisions. The main goal is the transient phenomena of the formation of an anisotropic electron velocity distribution and generation of small-scale current structures, primarily current sheets and filaments. These kinetic phenomena, in contrast to the various beam instabilities known for a generation of high-frequency electric fields of plasma waves (see, e.g.,~\citep{Akhiezer1975, Gary1993}), are largely due to quasi-magnetostatic aperiodic instabilities and not accompanied by development of significant high-frequency oscillations or waves (see, e.g.,~\citep{Huntington2017, Zhou2018, Stepanov20_LO20, Zhang2022_PoP, Zhang2022_PNAS}). Such phenomena exist in both laboratory and space plasmas and can't be described in a magnetohydrodynamic approach. They remained poorly studied until recent experiments, observations and numerical simulations made it possible to advance our understanding of their nature.

The review is limited to the analysis of the main stage of such processes, in which the dynamics of electron currents plays a leading role, and ion currents do not have time to develop due to a large inertia of heavy ions. This situation is typical for the plasma  with rather cold, weakly nonequilibrium ions. Their current (arising both due to an own dynamics of ions and an inductive action of the electron current) could reach a magnitude of the electron current only at the final, not so interesting stage, when the latter has already died out many times over. Strong magnetic fields generated at the main transition stage are turbulent in nature, reach, in typical laser experiments, the mega-Gauss and higher values, and cause effective interaction between charged particles in the course of the decay of a strong discontinuity in a fairly rarefied plasma in the absence of their Coulomb collisions. According to the analytical estimates, numerical simulations and laser ablation experiments, such magnetic fields of self-consistent currents of hot and cold electrons quickly lead to a spatial separation of counter flows of particles and suppress beam instability of plasma (Langmuir) waves. In this case, the high-frequency electric field plays a minor part in the decay of a strong discontinuity and its energy is much smaller than the energy of the quasi-static magnetic field. Such situation is qualitatively different from the well-studied decay of a weak discontinuity, where the electrons obey a Boltzmann distribution. The decay of a weak discontinuity in plasma does not lead to the generation of magnetic turbulence, so that the kinetics of nonequilibrium ions and the generation of ion-acoustic solitons, which include a self-consistent electric field, immediately come to the fore (see, e.g.,~\citep{Sagdeev1966, Medvedev2014, Grismayer2006}).

In an effort to make the qualitative picture of the occurring phenomena more clear, we will mainly rely on two-dimensional (2D, or more precisely 2D3V, see below) particle-in-cell (PIC) modeling, bearing in mind both the initial-value and boundary problems. It implies homogeneity along one of the Cartesian coordinates on the surface of the target, which is irradiated by a laser pulse and generates an expanding plasma with hot electrons. (For simplicity's sake, we consider all ions to be protons.) This formulation of the problem involves cylindrical focusing of laser radiation, which must also be femtosecond in order to provide the preferential heating of electrons, but not ions (see, e.g.,~\citep{Schou2007_LaserAblation}). However, in a number of cases, we present the results of a fully three-dimensional (3D) modeling to confirm the qualitative physical statements made within 2D modeling.

\subsection{Manifestations of current and magnetic microstructures}
\label{sec:Manifest}

In a general case of a collisionless nonequilibrium plasma possessing an anisotropic distribution of particle velocities it is typical the presence of multi-scale turbulent quasi-magnetostatic structures of various types, which prescribe, to a great extent, not just the kinetics of individual particles but also the dynamics of macroscopic quasi-neutral plasma formations and their large-scale inhomogeneous structures~\citep{Baumjohann2012, Treumann2009, Marcowith2016}.
The most important mechanism for the emergence of such a turbulent situation is the Weibel instability~\citep{Weibel1959, Davidson1989, Kocharovsky2016_UFN}, which is aperiodic and accompanies a variety of transient processes in space and laboratory (laser) plasma.
This instability is common for processes of an expansion of a nonequilibrium plasma into the background, such as jets and shock waves or flare phenomena in the active regions of the chromosphere and corona of the Sun and a number of other stars or planets~\citep{Baumjohann2012, Medvedev1999, Gruzinov2001, Lyubarsky2006, Spitkovsky2008, Balogh2013, Zaitsev2017}. Weibel-type aperiodic instabilities, with or without an external magnetic field, have been studied both at the linear stage and taking into account nonlinear processes of saturation of the growth of a small-scale magnetic field and its subsequent decay~--- although often in the abstract formulation of the simplest initial-value problem of a homogeneous plasma with anisotropic particle velocity distributions; see subsection~\ref{sec:Open problems}.

Lately, thanks to advances in the femtosecond high-power lasers, it became possible to study Weibel-type instabilities in laser plasma~\citep{Quinn2012, Huntington2015, Huntington2017, Scott2017, Li2019, King2019, Gode2017, Ruyer2020, Zhang2022_PNAS}. This opens up new ways for the observation of quasi-magnetostatic turbulence and modeling associated phenomena in weakly collisional astrophysical, magnetospheric and gas-discharge plasmas.
In the cited experimental works, an expanding cloud of plasma has been formed via an ablation of targets by means of laser pulses lasting from tens of femtoseconds to subnanoseconds with energies of the order of $1 - 10^3$~J, respectively.
Advanced experiments~\citep{Stepanov2018_LO2018, Zhou2018, Stepanov20_LO20, Ngirmang2020, Zhang2020, Zhang2022_PoP}, numerical simulations~\citep{Thaury2010, Schoeffler2016, Nechaev20_RadiophysEn, Nechaev20_FPen, Garasev22_GA_MagnEn, Garasev22_JPP, Garasev22_GA_InjEn, Stepanov22_TVTen, Kocharovsky23_DANen}, and our analytical estimates reveal creation of a strong magnetic field up to the mega-Gauss level (and higher) within a time period less than or of the order of~$1$~ns in a wide range of parameters, even when a nonequilibrium plasma is produced by means of less energetic femtosecond pulses with a typical energy $\sim 10$--$100$~mJ. In this case, the ions remain quite cold and the heating of a significant part of the target electrons by powerful radiation leads to the emergence of a long-term flow of their high-energy fraction from the central part of heated dense plasma into the surrounding space, filled, as a rule, with background plasma of a much lower density. In such processes, an important role could be played also by initially cold electrons, which originate from the background plasma and acquire a directionally predominant velocity being heated up anisotropically. Hence, these electrons become capable of changing the nature of the Weibel instability caused by the hot electrons.

Kinetic processes initiated by the expansion of plasma with hot electrons into cold plasma can play a significant part in the aforesaid explosive phenomena taking place in the magnetospheres of planets and stars. There are other similarly interesting problems related to the generation of a turbulent magnetic field, for example, the ejection of plasma with hot electrons from a collapsing coronal loop into a background magnetoactive plasma in the course of a solar flare or the collision between so-called magnetic clouds of hot and cold plasma in a highly inhomogeneous stellar wind \citep{Yoon2017, Zaitsev2017, Nakamura2018, Albertazzi2014, Marsch2006, Dudik2017, Lazar2022}.

\subsection{Anisotropy of electron velocity distribution and the initial stage of a strong discontinuity decay}
\label{sec:Anisotropy}

The expansion of a highly nonequilibrium laser plasma into a background equilibrium plasma and the development of the resulting collisionless shock wave are discussed in detail in~\citep{Garasev2017_JETPLen, Nechaev20_FPen}.
As has been shown, the flattening of a sharp initial plasma profile occurs predominantly in its steeper sections, so that the scales of an inhomogeneity gradually level out and, on average, the plasma profile becomes exponential with a single, ever-increasing scale. The density drops monotonically all the way to the compacted layer in the shock.
This layer can contain an appreciable fraction (up to~$\sim 10$\%) of the energy of a hot plasma cloud and is electrostatic in origin. Its structure, together with the structure of a narrow region preceding the shock front, is created and maintained by means of the electric field formed by separated charges of cold ions and hot electrons. Behind and in front of the shock front, a plasma beam instability and ion-acoustic waves as well as the growth of small-scale quasi-electrostatic fields~\citep{Nechaev20_FPen, Zhang2018, Moreno2020, Malkov2016} develop due to the counter flows of charged particles.

Soon after the beginning of plasma expansion the distribution of electron velocities becomes highly anisotropic, particularly under the shock front, so that the quasi-stationary magnetic fields appear. 
Their energies exceed the energy of the electric fields by a few orders of magnitude. Their spatial size is also significantly larger. In this case, the other situation, which is considered below, is usually realized. In fact, the magnetic field energy does not become greater than a few percent of the hot plasma energy and the cloud expansion is not affected by these fields, at least, until they change the anisotropic electron velocity distribution and, hence, the structure of the shock front. Yet, the energy of the resulting magnetic fields is enough for microstructuring the counter flows of particles and a radical separation of these flows in space, which greatly suppresses the beam instability and stops the growth of electric fields.

\subsection{Mechanisms of the magnetic field generation}
\label{sec:Mechanisms}

 There are three groups of mechanisms of the magnetic field generation in the aforementioned situation. They differ in the localization, rise time, spatial scales and orientations of the wave vectors of the spatial spectrum features of the generated fields. Weibel temperature and filamentation instabilities, taking place due to the anisotropic velocity distribution of ions, do not have time to reveal themselves, just as they cannot support the observed small spatial scales of magnetic field disturbances.

Currents of hot electrons, rapidly arising and flying out of the molten target material mostly across its boundary, and the return electrical currents of cold electrons, which partially compensate for the hot-electron currents (since they flow in the counter direction) and also fly up along the surface of the target to its heated part, create the large-scale azimuthal magnetic field component oriented along the target surface. (This effect is called a fountain effect~\citep{Sakagami1979, Kolodner1979, Albertazzi2015}).
This component is localized predominantly outside the cloud of expanding plasma, far enough from the heated region. A time scale of rising the fountain currents is of the order of the dimension of heated region ($\sim 30$--$300$~µm) divided by the speed of hot electrons and amounts to $0.1$--$10$~ps (see~\citep{Sarri2012, Albertazzi2015, Shaikh2017, King2019, Palmer2019}).

Inside a denser plasma, which could be collisional, the large-scale electron currents as well as related azimuthal magnetic field are developed mainly due to non-collinearity of gradients of hot electron density and their temperature. This mechanism, called the thermoelectric effect or Biermann battery, provides a rapid increase in the magnetic field over a time. It is dictated by the thermal speed of electrons and characteristic scale of temperature change, which usually is of the order of the dimension of heating region~\citep{Schoeffler2016}. It has been studied in detail experimentally and theoretically, in particular, within the kinetic approach~\citep{Schoeffler2016, Albertazzi2015, Schoeffler2018, Fox2018}.

The third group consists of the aforementioned kinetic instabilities of the Weibel type~\citep{Weibel1959}, for example, thermal and filamentation intabilities, caused by the anisotropic distribution of velocities of both the cooling hot electrons originating from the expanding plasma and warming up cold electrons of the background, which also create counter flows. Such instabilities ensure the growth of a small-scale (typically, with a size less than or of the order of $10$~µm) magnetic field in the collisionless expanding plasma on the time scale greater than or of the order of $10$~ps, originally, in the denser plasma region under the shock front, but then in a fairly extended volume preceding the front. In this case, the nature of the anisotropy of an expanding nonuniform plasma and features of Weibel-type instabilities have not been studied yet.

\subsection{Various Weibel-type instabilities}
\label{sec:Types}

Works~\citep{Schoeffler2016, Schoeffler2018, Fox2018, King2019, Gode2017} examine the generation of magnetic fields due to the filamentation instability developing at the front of an expanding plasma cloud, which is actually 1D in nature. In this region the electron velocity distribution becomes anisotropic in virtue of local counter currents of electrons from a cold background, compensating for the charge of the quickly escaping hot electrons and together with them forming two-stream distributions. Such ''beam'' electron velocity distributions are unstable due to the Weibel mechanism and produce small-scale electric currents. These currents flow in the direction of plasma expansion and acquire a spatial modulation along expansion front. At the same time, the generated magnetic fields are perpendicular to the expansion direction (see~section~\ref{ch:chWeibInj}).
A similar structure of magnetic fields was observed in~\citep{Ruyer2020} far from the laser-heated region, where the mentioned two-stream distributions are formed by flows of hot electrons, on the contrary, returning to the target surface through the background plasma near it.
The paper~\citep{Zhou2018} explored a magnetic field of a similar structure which was experimentally detected already at short times $\sim 1$~ps in an expanding plasma, produced by a femtosecond laser pulse, and lived for at least a few picoseconds. Such a rapid field generation has been attributed to a filamentation Weibel instability of the two-stream electron velocity distribution with a flow of relativistic electrons appearing due to a ponderomotive action of laser pulse.

A dissimilar mechanism of the magnetic field self-generation in an expanding plasma due to Weibel instability is suggested by the authors of~\citep{Thaury2010}. Their 1D and 2D simulations show that if a flat layer of plasma with hot electrons, initially having an isotropic distribution of velocities, expands into a vacuum, then the thermal energy of hot electrons is lost and, predominantly, the dispersion of their velocities in the expansion direction decreases. In the plane oriented perpendicular to the expansion direction, the effective temperature stays almost intact and retains its original value. Such a disk anisotropy of the velocity distribution function gives rise to Weibel instability which generates the magnetic field component in the specified plane. Its magnitude is modulated in the direction of expansion.
A Weibel-type mechanism is also analyzed in the paper~\citep{Stockem2014} describing two colliding electrostatic shocks in counter flows of laser plasma. The conclusion is that hot electrons trapped in electrostatic potential wells between the fronts increase their effective temperature in the direction of motion of the shock waves. This ''needle-shaped'' bi-Maxwellian anisotropy results in the longitudinal current filaments with wave vectors orthogonal to the indicated direction. The magnetic field is orthogonal to both the filaments and the wave vectors.
In both papers, only the simplest, plane-layered scenario of a 1D plasma expansion was studied.

Finally, the work~\citep{Ruyer2020} demonstrates the instability of a cloud of hot electrons, subjected to scattering within the simulation plane into a cold dense plasma and anisotropic cooling. The authors employ a hybrid numerical simulation of a 2D3V PIC-MHD type using the Ohm's law. According to their results, the instability leads to the generation of a radial magnetic field, modulated in azimuth, and currents consistent with this field, flowing orthogonal to the specified plane and forming flat sheets.
This mechanism of field generation is similar to that considered below by the method of completely kinetic modeling. However, the authors aimed to explain the experiment on laser ablation and placed the simulation plane on the target's surface, not orthogonal to it. This did not allow them to trace the dynamics of the expansion of electrons into a rarefied plasma in space above the target and, therefore, describe the picture of fields and currents there.

\subsection{Origin of the magnetic turbulence in laser-plasma settings}
\label{sec:Laser}

An interplay of all of the above mechanisms of magnetic field formation in actual experiments deserves special study since it can lead to a complex evolution and superposition of quasi-magnetostatic structures with different scales. Below, only part of this problem is touched upon, relating to two characteristic instabilities of the Weibel type, namely, the thermal instability of anisotropically cooling electrons of an expanding plasma and the thermal or filamentation instabilities of warming electrons in the background. Only the main electron stage of the magnetic field evolution is discussed~--- the growth, saturation of growth and slow decay. In this case, often, just as for the Weibel instability in a homogeneous plasma, the current dynamics in the presented calculations is actually quasi-static. The point is that its characteristic time is longer than the scale of current's inhomogeneity divided by the speed of typical electrons, and the effect of the inductive electric field is negligible. The study of a longer-term stage is not carried out because, generally speaking, it requires a difficult and necessarily 3D analysis of the coordinated dynamics of ions, including analysis of the energy redistribution between electrons and ions and their trajectories' mixing under the influence of the emerging magnetic field (especially significant in the region under the  shock fronts, if they form). Moreover, a well-studied field of the high-energy-density plasma created by means of the nanosecond laser ablation \citep{} is also not discussed in the review because it is related mainly to magnetohydrodynamic, not kinetic phenomena, and deals with plasma of hot electrons and also hot, not cold, ions.  

Note that if ions contain much more energy than electrons and have an anisotropic velocity distribution, then due to a similar ion instability of the Weibel type in a plasma, either homogeneous or expanding, the filamentation of currents will also occur and quasi-magnetostatic turbulence develop. Similar problems have been widely studied for various ion velocity distribution functions, e.g., bi-Maxwellian, Kappa and two-stream types, particularly in relation to the problem of the emergence of collisionless shock waves, the structure of which is determined by a relatively small-scale self-consistent magnetic field (see, e.g.,~\citep{Kato2008, Sironi2011, Fox2018, Nishigai2021, Kropotina2023}). 
Then, the generated magnetic field is affected significantly by electrons only if their energy is of the order of the energy of ions (cf.~\cite{Moiseev1963, Medvedev1999, Lyubarsky2006, Achterberg2007}; for the analysis of a special case of an electron-positron plasma, see \citep{Chen2023}.)

The latter is demonstrated on the basis of a long-term modeling of the initial-value problem for a plasma with the same initial energy content and bi-Maxwellian velocity distributions of electrons and ions~\citep{Borodachev2017_RadiophysEn}. In this case, the development of the ion instability is prevented by the magnetic field arose due to the electron one. This field scatters ions, decreases the degree of anisotropy of their distribution, and magnetizes the electrons, thus, limiting the increase of the magnetic field by ion currents. However, the long-term maintenance, evolution of the structure and decrease in the energy of the quasi-stationary magnetic turbulence, which becomes increasingly large-scale, are due to the ion currents induced by the gradually decaying magnetic field and, over time, beginning to dominate over the electron ones.

The saturation criteria for the aperiodic Weibel-type instability, depending on the energy, type and degree of anisotropy of the particle velocity distribution function, were studied only qualitatively for a number of characteristic cases (see, e.g.,~\citep{Yang1994, Achterberg2007, Kocharovsky2016_UFN, Kuznetsov22_FPen} and references therein). If one does not talk about the extreme case of relativistic plasma, then the quasi-magnetostatic turbulence that arises after instability saturation is weakly nonlinear and its energy content does not exceed a value $\sim 10\%$ of the initial energy content of nonequilibrium plasma particles. Yet, the small-scale turbulence could considerably change the particle kinetics and the evolution of plasma structures, especially those related to larger-scale magnetic fields, including reconnection of the field lines. The assessment of saturation growth as well as the nonlinear self-consistent development of the small-scale and larger-scale magnetic fields in differently shaped clouds of expanding plasma with hot electrons are demonstrated in the following sections. Relevant open problems are also discussed in the last, concluding section.

Observation of the kinetic phenomena under consideration is possible primarily in experiments with laser plasma created by ablation of a flat target during the incidence of a powerful beam of femtosecond laser radiation. Determination of the structure of generated magnetic fields with a resolution of the order of microns at subpicosecond time scale is feasible by means of the laser pulse diagnostics with a controlled time delay. Interferometric measurements of the phase shift of radiation make it possible to obtain a large-scale distribution of plasma density~\citep{Stepanov2018_LO2018, Stepanov20_LO20}, and its small-scale inhomogeneities can be revealed by the method of shadowgraphy (see, for example,~\citep{Plechaty2009}). By measuring the rotation of the polarization plane of radiation and changes in its ellipticity, which occur in a magnetic field when choosing different propagation paths of the diagnostic beam due to the Faraday and Cotton--Mouton effects, respectively, it is possible to qualitatively determine the structure of the arising magnetic field and its magnitude~\citep{ Borghesi1998, Shaikh2017, Chatterjee2017, Zhou2018, ForestierColleoni2019, Stepanov20_LO20}.
An idea of the parameters of hot electrons escaping from the target can be obtained using measurements of X-ray radiation or directly the electron spectrum~\citep{Langdon1980, Shaikh2017, Stepanov20_LO20}.

\subsection{The contents of the review}
\label{sec:Contents}

We have in mind an analysis of fundamental quasi-magnetostatic phenomena in collisionless nonequilibrium plasma using the emerging opportunities of experimental research and numerical modeling of laser plasma. We describe the occurring physical processes within the framework of four simplest combinations of the three main geometric attributes of the problem posed:
(i) a circular or cylindrical form of a hot-electron spot in the case of either the initial-value problem or finite-time injection of electrons from a target surface,
(ii) inhomogeneous layers of cold background plasma of different densities and spatial scales,
(iii) an external magnetic field with three orientations: perpendicular to the target or along it, directed either across or parallel to a long axis of the hot-electron spot.

In sections \ref{ch:raspad} and \ref{ch:chWeibMagn} we consider the initial-value problem on the decay of a localized strong discontinuity within two scenarios. First, the plasma region containing hot electrons borders a more rarefied cold plasma without a magnetic field. Second, it borders a vacuum equipped with an external magnetic field parallel to the target surface, i.e., parallel to the discontinuity plane. Sections \ref{ch:chWeibInj} and \ref{ch:DAN23} deal with the injection of a plasma with hot electrons through the boundary of a localized region into a half-space with colder plasma, the density of which monotonically decreases either in the absence of magnetic field (section \ref{ch:chWeibInj}) or in the presence of magnetic field directed along the target plane (section \ref{ch:DAN23}). Section \ref{sec:concl} contains conclusions and elaboration on some features of the evolution of quasi-magnetostatic Weibel turbulence, as well as relevant open questions, including in relation to cosmic plasma.

\section{The decay of a strong discontinuity in plasma and generation of small-scale current structures} 
\label{ch:raspad}

\subsection{Kinetic phenomena in the plasma discontinuity decay}
\label{sec:raspad:intro}

Consider the case of very different electron temperatures on opposite sides of the initial sharp boundary between two regions of plasma with different particle densities and the same particle composition. 
This situation naturally arises in the femtosecond laser ablation of various targets \citep{Albertazzi2015,Shaikh2017, Zhou2018, Shukla2020, Ngirmang2020} and is relevant to flares in the stellar coronas and coronal mass ejections \citep{Zaitsev2017, Dudik2017, Srivastava2019}, for instance, in the course of reconnection of magnetic field lines and injection of hot electrons at the boundary of the flare region.

If the differences in the temperatures and densities of ions (protons) on different sides of the original boundary are not large (two times or less), i.e., the discontinuity is weak, then the process largely comes down to the excitation of ion-acoustic waves and solitons responsible for the structure and evolution of the shock front according to the standard hydrodynamic description \citep{Medvedev2014, Srivastava2019, Kakad2016, Pusztai2018, Patel2021}. If the differences in ion densities are many orders of magnitude, i.e., the plasma with hot electrons is factually expanding into a vacuum, then the shock front does not arise and the decay of the expanding plasma boundary occurs in the kinetic regime. It is described by the collisionless Vlasov equation for nonequilibrium velocity distribution functions of electrons and ions.

In the case of decay of a moderately strong discontinuity in a plasma with hot electrons, when the number densities of ions on opposite sides of the boundary differ by many times, but not by many orders of magnitude, kinetic effects and multi-flow of particles are also significant and lead to a number of phenomena that have been clarified only recently~\citep{Malkov2016, Garasev2017_JETPLen, Nechaev20_FPen, Nechaev20_RadiophysEn}. Thus, according to \citep{Nechaev20_FPen, Nechaev20_RadiophysEn}, at the front of the resulting quasi-electrostatic shock, due to the reflection of rarefied plasma ions and the emergence of multi-flow, a compacted plasma layer with ions of both fractions is formed. It is accompanied by a highly anisotropic distribution of electron velocities under the shock front and in front of it. As a result, the Weibel instability and the generation of a small-scale magnetic field take place. Similar kinetic phenomena associated with the anisotropic cooling of hot electrons in an expanding plasma occur in the course of the ''plasma--vacuum'' discontinuity decay in the presence of an external magnetic field, which plays the ''braking'' role of the rarefied plasma and also leads to the stratification of hot electron currents \citep{Stepanov22_TVTen}. The latter scenario is described in section \ref{ch:chWeibMagn}.

The method and parameters for numerical simulations of the decay of a strong discontinuity formed by a dense plasma with hot electrons and a rarefied cold plasma are described in subsection \ref{sec:chWeibExp:InitCond}.
Subsections~\ref{sec:chWeibExp:AnisType} and~\ref{sec:chWeibExp:FieldStruct} tell of the anisotropy of the electron velocity distribution of two fractions and the general structure of emerging currents and magnetic field in different regions of the plasma.
Subsection~\ref{sec:chWeibExp:correl} outlines estimates and properties of the Weibel instability associated with the dominant fraction of the initial hot electrons at linear and nonlinear stages of formation of a small-scale magnetic field, including a pronounced correlation of its structure with the spatial structure of the anisotropy of electron velocity distribution.

\subsection{An initial-value problem for numerical modeling}
\label{sec:chWeibExp:InitCond}

\begin{figure}[!b]
	\includegraphics[width=1.00\linewidth]{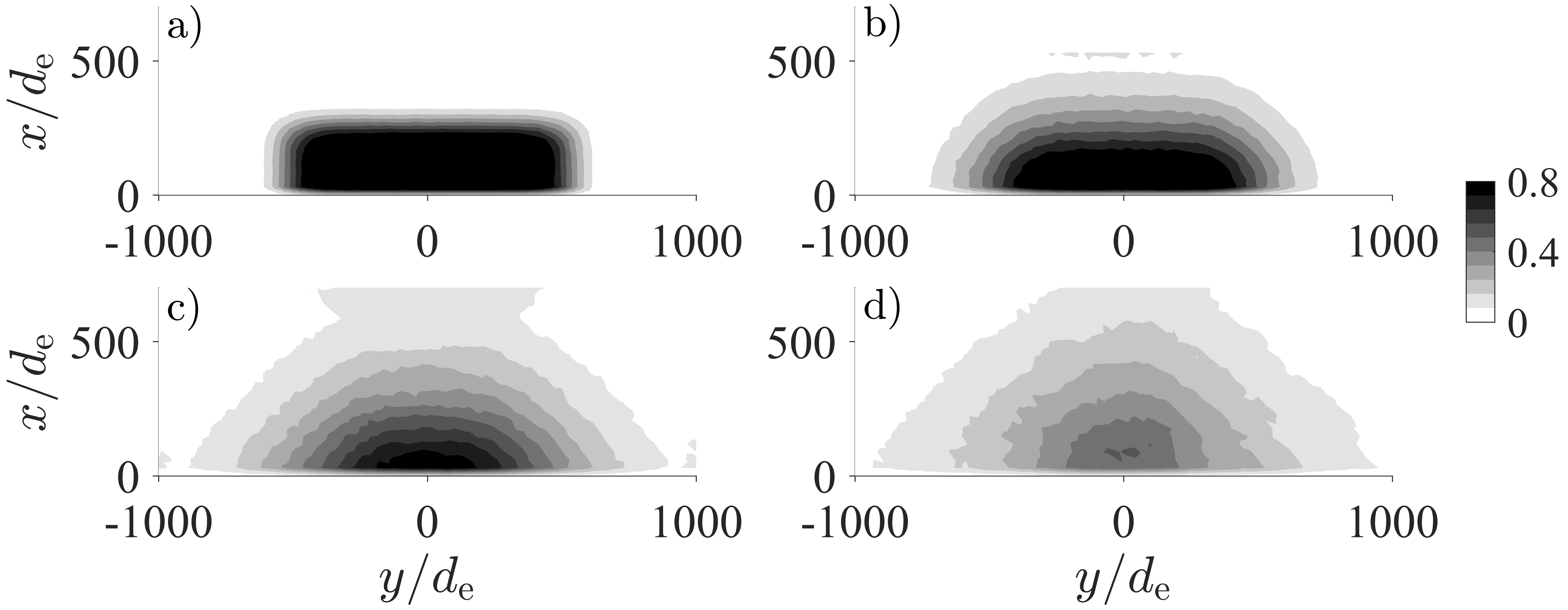}
	\centering
	\caption{
		Profiles of the normalized plasma number density at consecutive times: a)~$t = 300 \,\omega_\mathrm{p}^{-1}$, b)~$1500 \,\omega_\mathrm{p}^{-1}$, c)~$2800 \,\omega_\mathrm{p}^{-1}$, d)~$4300 \,\omega_\mathrm{p}^{-1}$. 
        Parameters of the 2D simulation: $n_{\mathrm{bg} 0} = 0.01 \, n_0$, $L \approx 270 \, d_\mathrm{e}$, $m_\mathrm{i} /m_\mathrm{e} = 100$, the dimension of the simulation domain in the $xy$~plane is $850 \,d_\mathrm{e} \times 3400 \,d_\mathrm{e}$.
	}
	\label{fig1}
\end{figure}

A generic scenario we consider is shown in Fig.~\ref{fig1}. It features an appearance of the anisotropy due to a decrease in the dispersion of velocities of hot electrons within the simulation $xy$-plane caused by the expansion of plasma at an almost constant temperature in the direction of $z$~axis along which the system is homogeneous.
The~$x$ axis of Cartesian coordinates is chosen to be orthogonal to the target surface.

Here and in following sections of the review, we present calculations carried out by means of the PIC code EPOCH~\citep{Arber2015_EPOCH}, both in the three-dimensional (3D3V) and two-dimensional (2D3V) geometries, taking into account all of the three vector components of the fields and particle velocities. In the latter case dependence of physical quantities on the coordinate~$z$ is absent and the simulation domain lies in the~$xy$ plane. Note that for some other distributions of nonequilibrium plasma there are a number of limited 3D3V calculations with due account for the Weibel instability, e.g.,~\citep{Silva2006, Dieckmann2009, Ruyer2015}.

Let us consider a plasma cloud in the form of a half-cylinder. Suppose a dense (main) plasma cloud consists of cold ions and hot electrons with a maximum number density of~$n_0 = 10^{20}$~cm$^{-3}$, has a semicircular or a quasi-rectangular boundary form (see~below) and is immersed in a rarefied background plasma of the two-orders-of-magnitude less number density, $n_{\mathrm{bg} 0} = 10^{-2}\, n_0$. (When the value of $n_0$ is increased to $10^{21}$~cm$^{-3}$ or the background density is decreased by one or two orders of magnitude, the phenomenon of magnetic field generation does not change qualitatively.)
At the initial moment, the particles obey the isotropic Maxwellian velocity distributions. Namely, the temperature of the main dense plasma electrons is $T_0 = 2.5$~keV, the background electrons -- $T_{\mathrm{bg} 0} = 50$~eV, all of the ions -- $T_{\mathrm{i} 0} = 3$~eV.
For these parameters, according to estimates in subsection~\ref{sec:chWeibExp:correl}, the expected growth rates as well as optimal wave numbers associated with the Weibel thermal or filamentation instabilities of background electrons are much smaller, proportional to their density, than that associated with the Weibel thermal instability of the cloud electrons. This makes the emergent structures qualitatively different from those obtained in a number of works~\citep{Schoeffler2016, Schoeffler2018} in which the decay of a plasma discontinuity has been simulated for much higher temperatures of hot cloud electrons and background plasma densities. 
Note that a 3D3V counterpart of the 2D3V modeling should deal with a long half-cylinder of dense plasma, and not a semisphere.

To speed up calculations, in all simulations presented a model ratio of the masses of ions and electrons $m_\mathrm{i}/m_\mathrm{e} = 100$ is adopted . Numerous calculations in 1D geometry with a mass ratio within $m_\mathrm{i}/m_\mathrm{e} = 100$\,--\,$50\,000$ have shown that its value does not qualitatively affect the formation of an electrostatic shock wave and dynamics of the electron Weibel instability under its front, only delaying the start time of each of them by $(m_\mathrm{i}/m_\mathrm{e})^{1/2}$ times, as per the fact that the expansion rate of the nonequilibrium plasma with hot electrons and cold ions is defined by the ion-acoustic speed.

Below, in subsections~\ref{sec:chWeibExp:AnisType}--\ref{sec:chWeibExp:correl}, we elaborate on the typical cases of expansion of a dense plasma cloud with the boundary of a quasi-rectangular type defined by the initial density profile $n_\mathrm{L} = n_0 \exp \left( -x^8/L^8 - y^8/(2L)^8 \right)$, where $L \approx 270 d_\mathrm{e}$, for definiteness (Fig.~\ref{fig1}).
A grid of $850 \times 3400$ cells is used; the cell side length is~$1.4 \, d_\mathrm{e}$, the number of particles per cell is $40$. Hereinafter $d_\mathrm{e} = ( T_0 / m_\mathrm{e} )^{1/2} \omega_\mathrm{p}^{-1}$ is the initial Debye radius of hot electrons; $\omega_\mathrm{p} = ( 4\pi e^2 n_0 / m_\mathrm{e} )^ {1/2}$ is their initial plasma frequency, calculated for the number density~$n_0$, $e$ is the elementary charge. At the lower border of the computational domain, i.e., at the surface of cold target $x=0$, the condition of reflection of particles and fields (as at the border with an ideal conductor) is used; the remaining boundaries are set open (absorbing particles and waves).

\subsection{The character of the electron-velocity anisotropy}
\label{sec:chWeibExp:AnisType}

Simulations show~\citep{Nechaev20_FPen}) that the aforementioned plasma discontinuity decay leads to the formation of a quasi-electrostatic shock wave propagating with an almost unchanged speed, no more than twice the local ion-acoustic velocity in the plasma under the shock wave front.
Everywhere along the resulting smooth profile of the plasma density (excluding the shock front itself, characterized by a jump in the electrostatic potential), the quasineutrality takes place, so that the electric currents of relatively slowly moving heavy ions are almost completely compensated by the currents of fast electrons shielding them.
Since the ions are cold and heavy and the currents' two-stream character develops only in a rarefied plasma in front of the shock wave front and in a narrow compacted layer directly below the front, then exclusively nonequilibrium electrons are responsible both for the fast creation of the magnetic field within the main part of the expanding plasma and its further saturation.

 Various fractions of the nonequilibrium electrons (both from the background and main plasma), forming directed flows, including counter flows, and anisotropically cooling or heating due to movement in the quasi-electrostatic large-scale fields and turbulent small-scale fields of plasma waves created by them, can lead to kinetic instabilities responsible for the growth of some components of magnetic field, differing in direction, structure and scale of inhomogeneity.

In the parameter range under consideration, the calculations and estimates show a low efficiency of a type of Weibel instability called filamentation and caused by counter flows of electrons. These flows of various electrons can be pulled under the front of the shock wave by a large drop in the electrostatic potential of the double layer, which arises in the very beginning of the discontinuity decay and exists there afterwords. The electrons can then return back being accompanied by the flow of hot electrons, which are partial reflected from the reverse potential drop near the target and come from there.
In the works~\citep{Schoeffler2016, Schoeffler2018}, such a filamentation instability, which arises in virtue of the relative motion of cold and hot electrons, was modeled under conditions of expansion of a sufficiently large cloud through a fairly dense background plasma with a density only several times less than the initial laser plasma density. The electron temperatures in these two fractions ($\sim$ 20~keV and 50 eV, respectively) were chosen to differ by 400 times.
On the contrary, for the parameters adopted in the calculations presented here and characteristic of most experiments on the femtosecond laser ablation, the background plasma has a density about by two orders of magnitude less than the density of the main plasma, and the electron temperatures of two plasmas (of order of 3 keV and~50~eV) differ by a factor less than a hundred. Hence, the required electron anisotropy is not achieved, so that the corresponding instability has a growth rate much smaller than that of the Weibel instability associated with the hot electron anisotropy of the main plasma.

In virtue of the intrinsic limitation of 2D simulations, the partial electron temperature~$T_z$ in the direction of $z$ axis of homogeneity of the system does not change much at all the considered times of formation and saturation of the magnetic field growth. For the cold background electrons, the temperature~$T_z$ increases by no more than 3 or 4 times. For the initial hot electrons, it decreases by no more than 2 or 3 times.
Both changes occur, apparently, due to weak inductive electric fields generated by varying magnetic field in the simulation $xy$-plane. At the same time, as will be shown below, after a certain time since the start of expansion, there occurs a pronounced increase (by 1 or 2 orders of magnitude) and decrease (also by 2 or 3 times) of effective transverse temperatures in the simulation plane, $T_{x,y}$, for the cold background electrons and initial hot-cloud electrons, respectively.
The background electrons has a strong ''disk'' anisotropy (in virtue of a huge ratio of their temperatures $T_{x,y} / T_{z}$) which is clearly seen in 2D calculations. However, the instability does not take place because there are no disturbances with wave vectors along the homogeneity axis~$z$ \citep{Weibel1959, Davidson1989}. For disturbances with wave vectors in the simulation plane, the instability is weak since it is caused by another anisotropy. It is determined by the ratio of the transverse temperatures, $T_{x}$ and $T_{y}$, which is not large. Therefore, it does not considerably affect the growth rate of Weibel instability, set by electrons of the main dense plasma.
In 3D modeling of the expansion of a hemispherical cloud, a large ''disk'' anisotropy of background electrons does not arise, since the temperature $T_z$ varies in the same way as the temperature $T_y$, growing by~one--two orders of magnitude and approaching the initial temperature $T_0$ of the electrons of the main plasma.

In subsections~\ref{sec:chWeibExp:AnisType}--\ref{sec:chWeibExp:correl} we mainly describe creation of magnetic fields via Weibel currents of hot electrons of the initial laser plasma cloud flowing parallel to the surface of the target. For a semicircular or quasi-rectangular shape of the initial discontinuity in the plasma, the Weibel currents arising during the expansion process are directed predominantly along the $z$~axis. That corresponds to the maximum partial temperature~$T_z$, which remains close to the initial value until the instability is saturated. The wave vectors of the growing magnetic fields lie in the $xy$ plane. In various regions, they are oriented mainly along the direction of the minimum partial temperature.

It is the anisotropy that causes formation of Weibel currents through the spatial differentiation of electrons with different velocities under the influence of a self-consistent magnetic field. There is no any noticeable redistribution of the electron density at the characteristic times of the instability development, and the plasma remains quasi-neutral.
In calculations, such a redistribution  of hot electron density does not exceed a few percent of the total ion density and is anticorrelated with the spatial redistribution of the background electron density, which varies several times within the expanding plasma, remaining less than the ion density by one-two orders of magnitude.

\begin{figure}[!b]
	\includegraphics[width=1.00\linewidth]{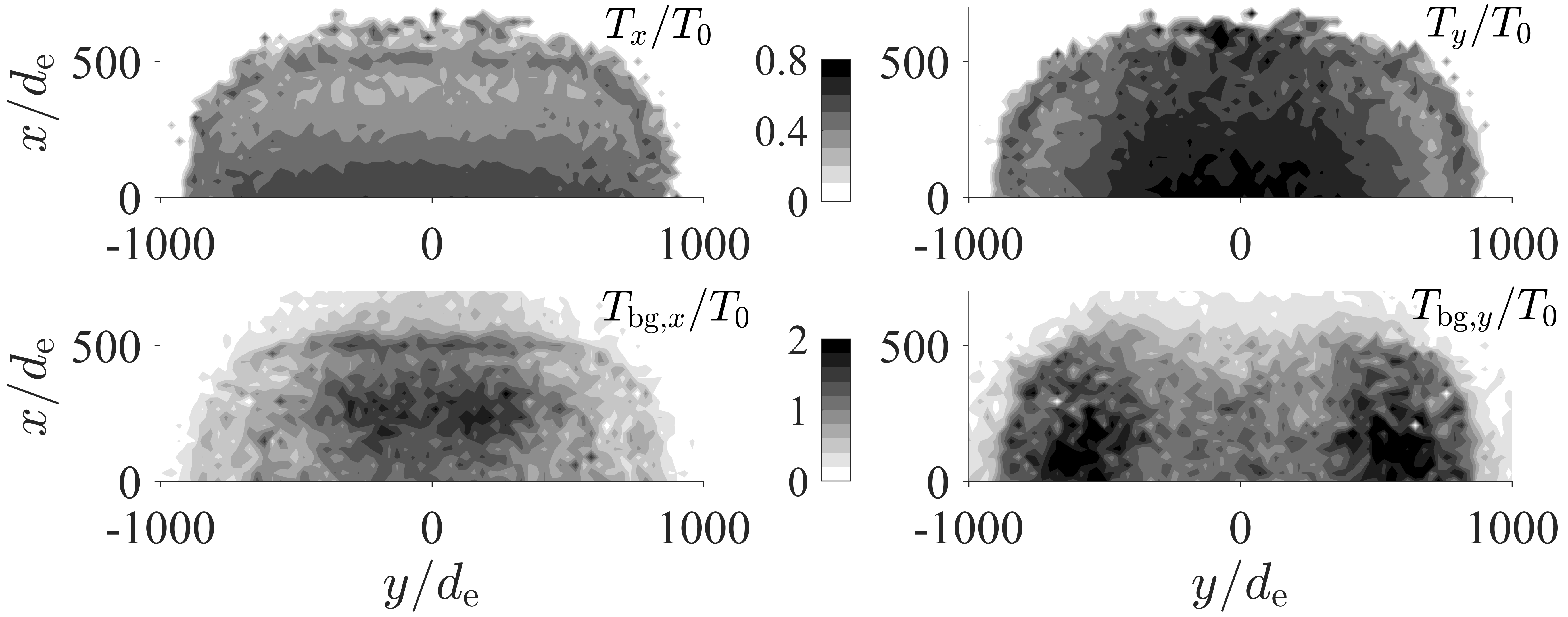}
	\centering
	\caption{
		Distributions of the normalized effective temperatures of hot electrons $T_{x} / T_0$ (top left) and~$T_{y} / T_0$ (top right) at the time moment $t = 1500\,\omega_\mathrm{p}^{-1}$; bottom panels show the same for the background electrons.	
	}
	\label{fig2}
\end{figure}

During the collisionless expansion of a plasma, the temperatures of electrons moving along the~$x$ and~$y$ axes, i.e., the dispersion of velocities in the indicated directions, decrease, while the temperature along the~$z$ axis remains almost constant~\citep{Thaury2010}.
(Numerical modeling in the 1D case, i.e., for the decay of a flat layer of plasma with hot electrons, demonstrates similar results.) Based on calculations with a quasi-rectangular initial shape of the plasma cloud, Fig.~\ref{fig2} shows how the partial electron temperatures $T_x$ and~$T_y$ are distributed in the $xy$ plane for the main plasma and background at the moment of time approximately in the middle of the linear stage of Weibel instability. In the course of the quasi-adiabatic expansion, mainly across the long axis, the temperature $T_x$ of the cloud electrons in this direction within a wide band of the central part drops more strongly, to about $0.3 \, T_0$, than the temperature $T_y \sim 0.5 \, T_0$ in the transverse direction. Its fall, also to about~$0.3\,T_0$, takes place mainly in the vicinity of the left and right edges of the expanding cloud, where the temperature $T_x$ decreases to values close to~$0.5\,T_0$, i.e., not as strong as in its central part.

It is worth noting that the background electrons are heated several times more strongly and their effective temperatures $T_{\mathrm{bg}, x}$ at the center of the cloud and $T_{\mathrm{bg}, y}$ at its left and right edges reach the initial temperature of hot electrons~$T_0$ or even $2T_0$. This heating is due to the quick formation of a flow of background electrons towards the shock wave, accompanied by their subsequent reflection from the region of excess negative charge near the target, and occurs at a much earlier stage, corresponding to the very formation of the shock wave, several tens of plasma periods after the onset of discontinuity decay.
However, as already noted, due to the low number density of the background plasma within the considered variants of discontinuity decay, its influence on the Weibel instability appears to be negligible, despite the strong anisotropy.

\begin{figure}[!t]
	\includegraphics[width=0.8\linewidth]{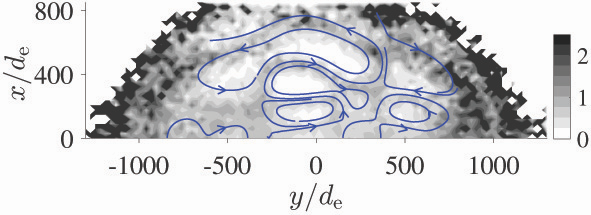}
	\centering
	\caption{
		Distribution of the anisotropy degree of hot electrons $A_\mathrm{e} = T_z / T_x - 1$ at the moment $t = 2800 \, \omega_\mathrm{p}^{-1}$. 
        Some magnetic field lines in projection on the $xy$ plane are shown in blue.			
	}
	\label{fig3}
\end{figure}

The hot cloud electrons contribute to Weibel instability in accord with the degree of their anisotropy $A_\mathrm{e} = T_z / T_x - 1$. Its distribution is shown in Fig.~\ref{fig3} at the moment of magnetic field saturation in the case of the initially quasi-rectangular cloud (see~subsection~\ref{sec:chWeibExp:InitCond}). Typical in-plane magnetic field lines are also given. The evolution and type of the electron anisotropy are related to spectral and spatial properties of this field in subsection~\ref{sec:chWeibExp:correl}.
Here we just note a pronounced stratification of anisotropy associated with the resulting current filaments, directed perpendicular to the expansion plane, along the axis $z$, in which the temperature $T_z$ is reduced, and the transverse temperatures $T_x$ and $T_y$ are increased. This occurs because the angle between the $z$ axis and electron velocities at the nonlinear stage of instability development increases on average under the influence of the magnetic field. This field also captures particles with small longitudinal (along the $z$ axis) velocities in bounce oscillations~\citep{Yang1994, Achterberg2007} in the region of their minima, i.e., current maxima. These slow particles are accumulated there and, hence, decrease the local effective temperature~$T_z$.
The degree of anisotropy is greater within the region of the compacted plasma layer near the shock front, where the transverse and longitudinal velocities of electrons have not yet been redistributed by a rather weak magnetic field.

\subsection{Main features of emerging quasi-static fields and currents}
\label{sec:chWeibExp:FieldStruct}

Even taking into account the nature of the anisotropy of hot electrons, which is inherent to the problem on the decay of strong discontinuity in a collisionless plasma and is revealed in the previous subsection, elucidating the Weibel mechanism for generating self-consistent currents and magnetic fields turns out to be a difficult task and requires numerical modeling. The point is that (i) the anisotropy is non-stationary and inhomogeneous in the expanding region of plasma and (ii) geometry of the region itself and the non-uniform plasma density in it are also non-stationary.

For a semicircular initial plasma cloud, the evolution of the components $\vec{B}_\perp$ and $B_z$ of the magnetic field in the simulation $xy$-plane and across it (along the $z$ axis of homogeneity of the system) has a different character. 
The component~$B_z$ appears very quickly, over a few tens of plasma periods, originally, chiefly before the shock front in a rarefied plasma, and then below the front in a dense plasma. Such generation cannot be Weibel (cf.~\citep{Schoeffler2016}) due to the absence of dense enough electron flows or substantial longitudinal anisotropy of the plasma (both in the dense cloud and rarefied background). Apparently, it is caused by the complementary currents of cold and hot electrons, formed due to the fountain effect and/or Biermann battery.

\begin{figure}[!t]
	\includegraphics[width=0.4\linewidth]{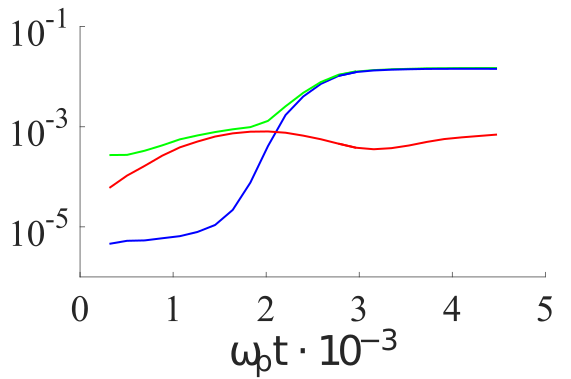}
	\centering
	\caption{
		Time dependence of the energy associated with the magnetic field components $B_\perp$ (blue) and $B_z$ (red), as well as the total energy of electric and magnetic fields (green), all normalized to the initial plasma energy.
	}
	\label{fig4}
\end{figure}

Currents, which create a much stronger Weibel field, flow predominantly along the $z$ axis in filaments occupying regions where the magnitude~$B_\perp$ in the $xy$ plane is small. The electrons, which make up these current filaments, experience bounce oscillations due to reflections from magnetic ''walls'' with a~large value of $B_\perp$. It happens when the instability gets to the nonlinear stage~\citep{Yang1994, Achterberg2007}.
As Fig.~\ref{fig4} demonstrates, the energy content of such configurations of magnetic fields and currents with the representative scales, progressively increasing as the shock moves, varies very little. This is primarily in virtue of a large reservoir of hot electrons appeared near the target and characterized by a fairly high, on average, and slowly declining degree of the electron anisotropy.
The latter provides a long-term pumping of increasingly large-scale current filaments and the magnetic field matched with them, while the small-scale current filaments, which have reached saturation, and the field associated with them gradually decay. 
For a semicircular cloud, electron temperatures along the~$x$ and~$y$ axes decline in a similar way (subsection~\ref{sec:chWeibExp:AnisType}). Hence, throughout the nonlinear stage, the wave vectors of developed field disturbances are oriented mainly radially (cf.~\citep{Ruyer2020}).

\begin{figure}[!t]
	\includegraphics[width=0.75\linewidth]{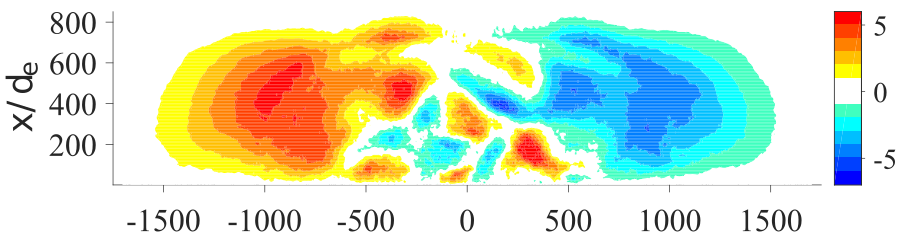}
	\includegraphics[width=0.75\linewidth]{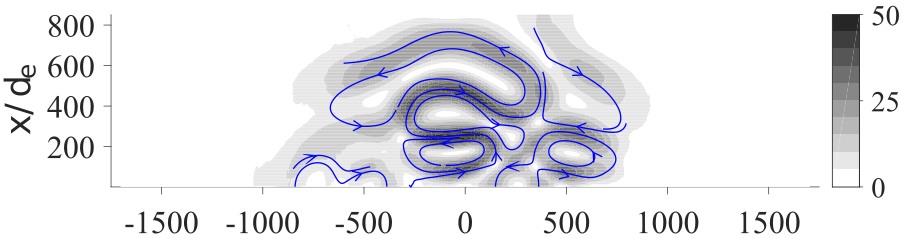}
    \includegraphics[width=0.757\linewidth]{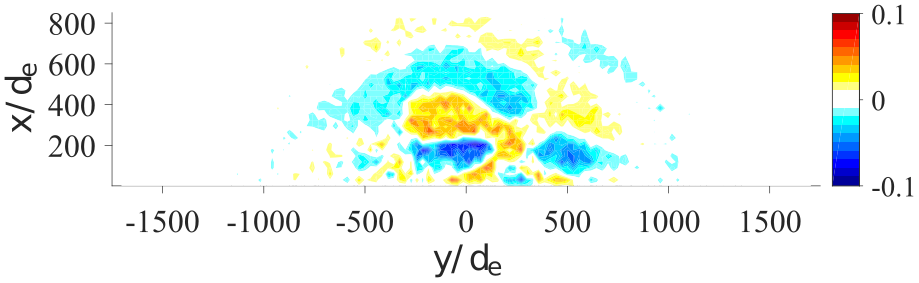}
	\centering
	\caption{
		Distributions of the magnetic field component $B_z$ (top panel), magnitude of the in-plane magnetic field $B_\perp = \left( B_x^2 + B_y^2 \right)^{1/2}$ (middle, typical field lines are shown in blue), both in Teslas, and normalized current density $j_z \, ( e n_0 )^{-1} \, (2 T_0 / m_\mathrm{e} )^{-{1/2}}\!$ (bottom) at the moment $t = 2800 \, \omega_\mathrm{p}^{-1}$.
	}
	\label{fig5}
\end{figure}

\begin{figure}[b]
	\includegraphics[width=0.75\linewidth]{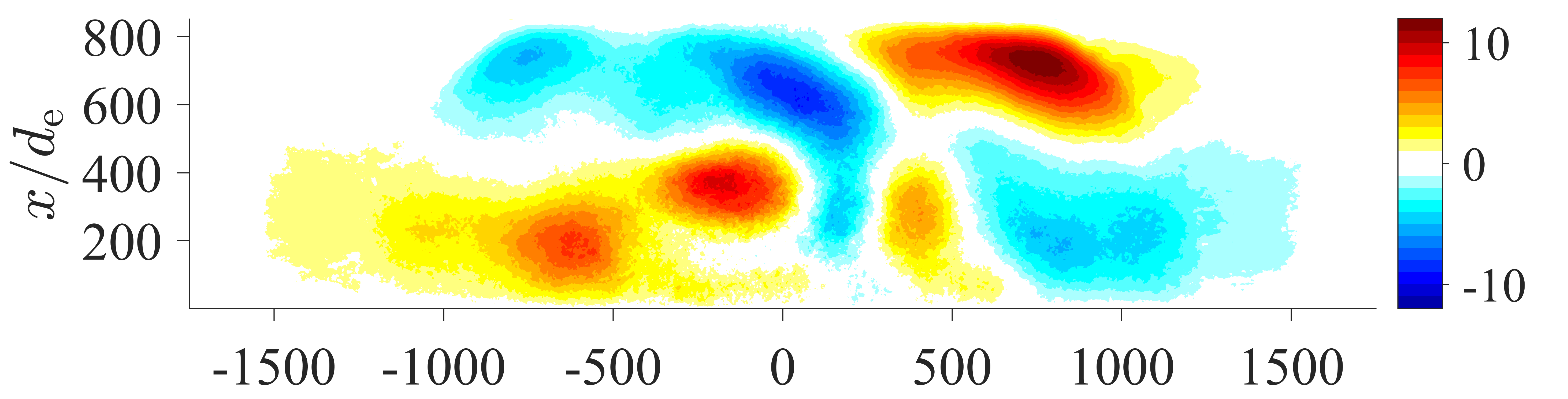}
	\includegraphics[width=0.75\linewidth]{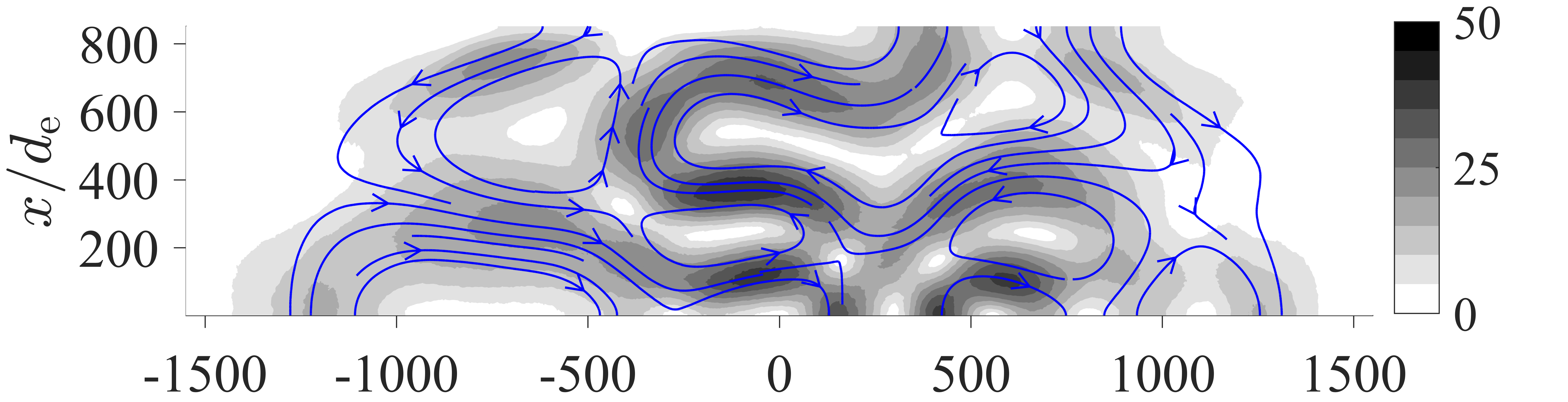}
    \includegraphics[width=0.75\linewidth]{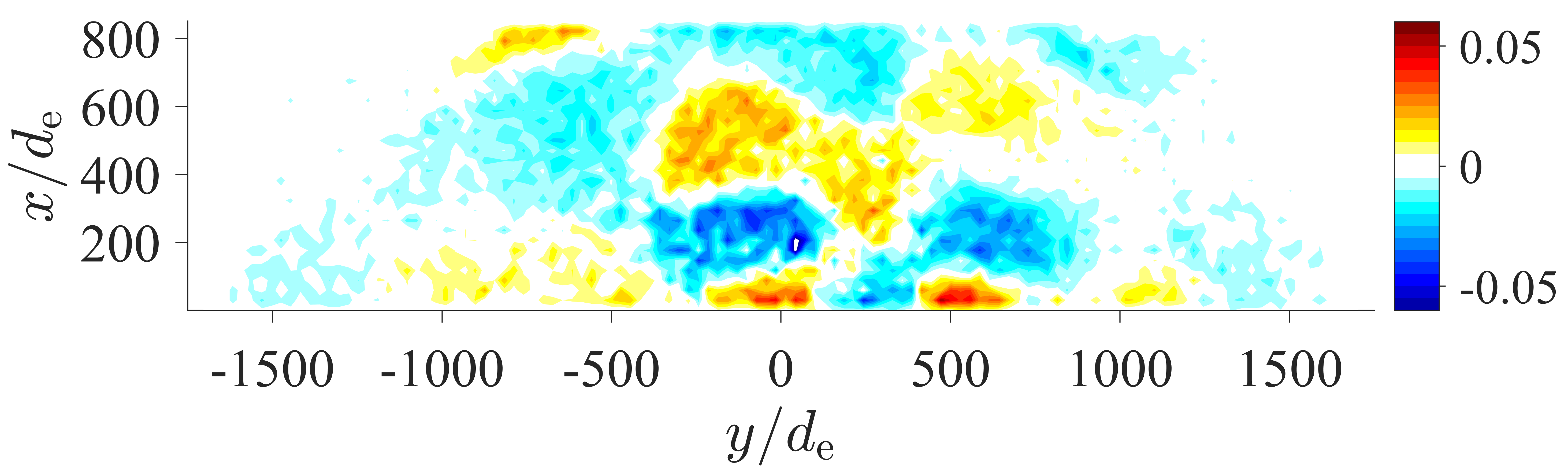}
	\centering
	\caption{
		The same as in Fig.~\ref{fig5}, but at the moment $t = 4300 \, \omega_\mathrm{p}^{-1}$.			
	}
	\label{fig6}
\end{figure}

For a quasi-rectangular plasma cloud containing hot electrons, defined in subsection~\ref{sec:chWeibExp:InitCond} and considered below,
a largely similar picture of joint growth (Fig.~\ref{fig4}, right panel), saturation, and nonlinear evolution (cf. Fig.~\ref{fig5} and Fig.~\ref{fig6}) of two quasi-stationary field components~-- a strong, Weibel-generated $B_\perp$ (in the simulation $xy$-plane) and a weak, larger-scale~$B_z$ (along the homogeneity axis~$z$)~-- takes place. 
In this case, the spatial structure of Weibel current filaments and the corresponding magnetic field, of course, turns out to be far from being as azimuthally isotropic as for a semicircular cloud, and is dictated by the plane-layered nature of the electron anisotropy (see subsection~\ref{sec:chWeibExp:AnisType} and Figs.~\ref{fig2},~\ref{fig3}), until the shock wave travels to a distance exceeding the dimension of the quasi-rectangle. This occurs at times at least $2$--$3$ times longer than the instability saturation time, when the spatial spectrum of the current density $j_z$ acquires a pronounced quasi-isotropic continuous part complementing discrete components. Starting from the saturation stage and up to the aforementioned times, i.e.,~up to $\omega_\mathrm{p} t \lesssim 6000$, as is clear from Figs.~\ref{fig5},~\ref{fig6} (bottom panels), the spectrum of the current~$j_z$ is dominated by one somewhat blurred component. This components has almost zero $y$-wavenumber and is related to almost homogeneous magnetic field with a scale of the order of the initial dimension of the plasma cloud.

Contrary to the case of a semicircular cloud, the magnetic field component $B_z$ begins to grow much faster and reaches much higher value than that of the component $\vec{B}_\perp$, prevailing over it midway through the linear stage of the instability (see~Fig.~\ref{fig4}). Then the energy amount stored in the $B_z$ component declines. When Weibel instability saturates, alike the case of a semicircular cloud, the representative value of $B_z$ remains a few times weaker than $B_\perp$ for a long time.
For both of these and  other similar cases, the modeling proves that the total magnetic energy at the saturation stage approaches several percent of the initial kinetic energy of plasma particles. It stays at such a high level for quite a long time, considerably surpassing the energy of electric field.

\subsection{Spatial correlations between the magnetic field and the degree of the electron-velocity anisotopy}
\label{sec:chWeibExp:correl}

Following the theory of Weibel instability known for a homogeneous bi-Maxwellian plasma with an order-of-unity degree of anisotropy \citep{Weibel1959, Davidson1989, Thaury2010, Vagin2014, Kocharovsky2016_UFN}, one can employ the standard estimates for the optimal scale of field perturbations, $\lambda_0$, their maximum growth rate, $\omega_0$, and magnitude of the saturating field, $B_\mathrm{s}$, implying an equality between $\lambda_0 / 2$ and the gyroradius of a typical electron, $r_\mathrm{L} \sim \left( T_\perp m_\mathrm{e} c ^2 \right)^{1/2} \! \left( e B_\mathrm{s} \right)^{-1}$, 
\begin{eqnarray}
	\lambda_0 \sim \frac{2 \pi c}{\omega_\mathrm{p}} \, \left( \frac{n_0}{n_\mathrm{e}} \right)^{1/2} 
	\left( \frac{3}{A_\mathrm{e}} \right)^{1/2} ,
\label{eq:chWeibExp:eqLambdaTheory}
	\\
	\omega_0 \sim \omega_\mathrm{p} \left( \frac{n_\mathrm{e}}{n_0} \right)^{1/2} 
	\left( \frac{2 T_\perp}{m_\mathrm{e} c^2} \right)^{1/2} 
	\left( \frac{A_\mathrm{e}}{3} \right)^{3/2} 
	\frac{2}{\sqrt{\pi}} \,
	\frac{1}{ 1 + A_\mathrm{e} } ,
\label{eq:chWeibExp:eqOmegaTheory}
	\\
	B_\mathrm{s} \sim \frac{2 c}{e}\, 
	\frac{\left( m_\mathrm{e} T_\perp \right)^{1/2}}{\lambda_0}
	\sim 6400 \, \frac{\left( T_\perp / T_0 \right)^{1/2}}{\lambda_0 / d_\mathrm{e}} \ [\text{T}] .
\label{eq:chWeibExp:eqSaturTheory}
\end{eqnarray}
Here $A_\mathrm{e} = T_z / T_\perp - 1$ is the degree of hot electron anisotropy, $n_\mathrm{e}$ their local number density, $T_{z,\perp}$ their effective temperature along the $z$ axis and orthogonal to it, respectively, and $c$ the speed of light in vacuum. 
As per the velocity distribution stated in subsection~\ref{sec:chWeibExp:AnisType}, the current filaments should be directed along the $z$ axis and have the representative scales greater than or of the order of~$\lambda_0$. The associated with them magnetic field $\vec{B}_\perp$ lies in the $xy$ plane.

Eq.~(\ref{eq:chWeibExp:eqLambdaTheory}) gives an order of magnitude estimate from below because the electron distribution is inhomogeneous, non-stationary and not exactly bi-Maxwellian. At the moment of the onset of instability development, the scale of plasma inhomogeneity is comparable to~$\lambda_0$. So, the expressions~(\ref{eq:chWeibExp:eqLambdaTheory})--(\ref{eq:chWeibExp:eqSaturTheory}), with the local values of temperatures and plasma frequency, are valid only approximately, by the order of magnitude.

The expressions above contain the parameters of the main plasma only. The background electron contribution is easy to estimate by means of the Weibel instability's dispersion relation for the wave vectors in the $xy$ plane in a plasma containing two fractions of the bi-Maxwellian electrons having orthogonal anisotropy axes (as well as heavy ions; see, for example, \citep{Borodachev2017_RadiophysEn}).
It appears that for the following representative ratios of number densities, $n_\mathrm{e} / n_\mathrm{bg} \approx 20$, and temperatures, $T_\perp / T_{\mathrm{bg}\perp} \approx 1/4$, of these fractions and values of the degree of anisotropy $ A_\mathrm{e} \approx 1.5$ and $T_{\mathrm{bg},z} / T_{\mathrm{bg}\perp} - 1 \approx -1$ in the region of space and at the moment corresponding to the beginning of field generation (in particular, $x / d_\mathrm{e} \approx 300$, $\omega_\mathrm{p} t \approx 1500$ for the case of a quasi-rectangular cloud), the background electrons contribute no more than 10\% to the optimal scale and growth rate of instability~(\ref{eq:chWeibExp:eqLambdaTheory})--(\ref{eq:chWeibExp:eqOmegaTheory}). However, they can contribute up to~40\% to the total anisotropy of electron velocity distribution.

The creation of magnetic field via Weibel instability begins as soon as the degree of anisotropy $A_\mathrm{e}$ becomes large enough in a sufficiently extended region which dimension exceeds the scale given in Eq.~(\ref{eq:chWeibExp:eqLambdaTheory}). According to modeling (see~Fig.~\ref{fig1}), this is achieved by a time of the order of $1000 \, \omega_\mathrm{p}^{-1}$. By that time, the degree of anisotropy becomes of the order of unity, $T_z / T_\perp \sim 2$--$3$, plasma with hot electrons occupies a region of radius $\sim 400 \, d_\mathrm{e}$, and Weibel scale approaches $ \lambda_0 \sim 300 \, d_\mathrm{e}$. In this situation, the linear stage of instability (exponential growth of magnetic field energy) has a duration of the order of $600 \, \omega_\mathrm{p}^{-1} \sim 3 \, \omega_0^{-1} $, as per Fig.~\ref{fig4}.
The value of the created magnetic field $B_\perp \sim 40$~T shown in Figs.~\ref{fig3},~\ref{fig4} is also consistent with the estimate in Eq.~(\ref{eq:chWeibExp:eqSaturTheory}). Thus, these estimates testify in favor of the Weibel mechanism as the origin of the observed magnetic field generation.

In the aforesaid scenario of the self-consistent generation of the field, the Weibel instability plays a decisive part it develops, for the selected set of parameters, faster than other instabilities. This fact is verified in a number of 2D and 1D calculations for various initial forms of the dense plasma cloud with hot electrons.
Note that 1D calculations, in which the inhomogeneity of physical quantities is allowed only along the direction of expansion, predetermine the plane-layered nature of field and exclude a number of scenarios for its generation, including the ones due to filamentation instability or fountain and thermoelectric effects, but have practically no effect on the buildout of Weibel instability due to temperature anisotropy.
However, only 2D simulations make it possible to account for the dependence of the generated field's structure on the type of the arising electron anisotropy that, in turn, depends on the initial form of the dense plasma cloud with hot electrons.

\begin{figure}[t]
	\includegraphics[width=0.6\linewidth]{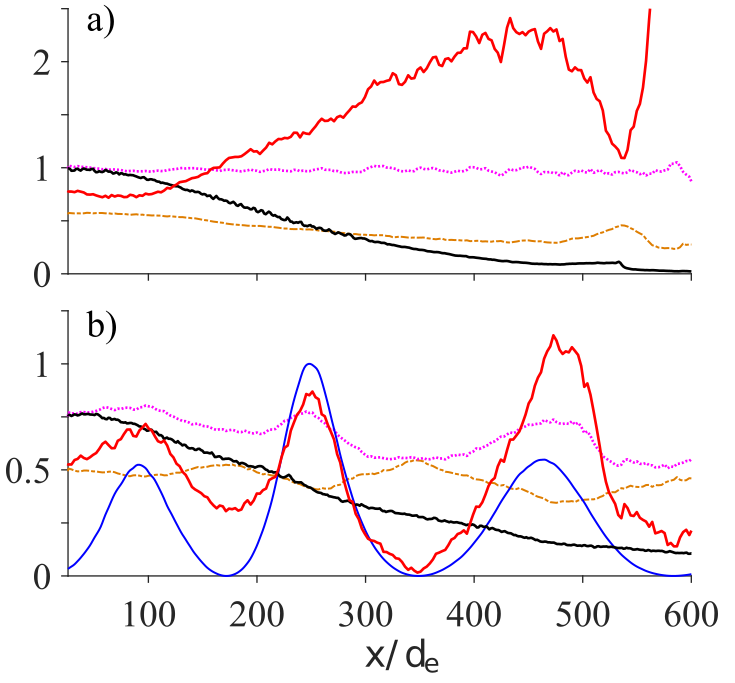}
	\centering
	\caption{
		a)~Anisotropy degree of hot electrons (red), $A_\mathrm{e} = T_z / T_x - 1$, their longitudinal (brown) and transverse (magenta) normalized temperatures, $T_x / T_0$ and~$T_z / T_0$, and normalized ion number density (black), $n_\mathrm{i} / n_0$, all in dependence on the longitudinal coordinate~$x$ at the moment $t = 1500 \, \omega_\mathrm{p}^{-1}$, when the front of the density bump is at $x / d_\mathrm{e} \approx 550$. 
        b)~Same, but at the moment $t = 2800 \, \omega_\mathrm{p}^{-1}$, when front of the density bump is at $x / d_\mathrm{e} \approx 700$. The blue line shows the normalized energy density of the in-plane magnetic field, $B_\perp^2 / (8 \pi n_0 T_0)$. 
		All profiles are obtained by averaging respective quantities over the coordinate~$y$ within the range $[-200,\,80] \, d_\mathrm{e}$ and, for clarity sake, are smoothed by the moving frame of a size~$10 \, d_\mathrm{e}$ along~$x$.
	}
	\label{fig7}
\end{figure}

Let us compare the spatial structures of the current density, temperature anisotropy, and magnetic field in the vicinity of the symmetry axis (vertical axis $y$) as the functions of $x$ at two moments -- at the start of the exponential field growth, Fig.~\ref{fig7}, and at the saturation of this growth shown in Figs.~\ref{fig2}, \ref{fig3}, \ref{fig5},~\ref{fig6}. For clarity's sake, we average the above quantities over a small range $[-200,\,80]\,d_\mathrm{e}$ of the~$y$ coordinate. Let us consider a quasi-rectangular plasma cloud with hot electrons, the initial dimension of which along the~$y$ axis was four times larger than the initial dimension along the~$x$ axis. Then the plasma expansion has a quasi-1D character in the $x$ direction that mainly reduces the related temperature $T_x$ and keeps the temperature $T_y$ intact. Hence, the wave vectors of emerged magnetic field disturbances are predominantly oriented along the axis $x$.
During the linear stage of instability in Fig.~\ref{fig7} the temperatures $T_x$ and $T_y$ inside the expanding plasma are nearly homogeneous and the change in the degree of anisotropy $A_\mathrm{e}$ is barely visible. However, when the development of Weibel field prevailing over other magnetic field components, gets saturated, say at $t \approx 2800\, \omega_\mathrm{p}^{-1}$, both the electron temperature $T_x$ and specifically the temperature $T_z$ in those regions of plasma, where current filaments appear, change significantly. Subsequently, the degree of anisotropy there becomes small, retaining a considerable value just in places where the field magnitude is about its maximum value.
The related spatial correlation between the degree of electron anisotropy and the magnetic field value at the saturation stage, depicted in Fig.~\ref{fig3} for the strong discontinuity decay, has been first identified in the work~\citep{Nechaev20_RadiophysEn} and is seen even in 1D3V modeling.

\begin{figure}[t]
	\includegraphics[width=0.55\linewidth]{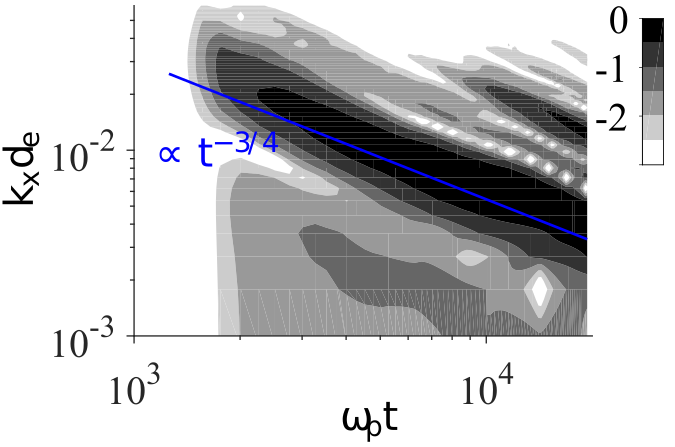}
	\centering
	\caption{
		Time evolution of the spatial spectrum of transverse magnetic field as per 1D3V simulations: The density plot shows the logarithm of the power spectral density. 
        The characteristic wavelength, and so is the scale of current density~$j_z (x)$ distortion, grows with time following the power law $\propto t^{3/4}$.
	}
	\label{fig8}
\end{figure}

Evolution of spatial spectra of Weibel currents in the described inhomogeneous problem demonstrates a self-similar, power law behavior at the nonlinear stage. It reminds the nonlinear behavior of Weibel instability in the homogeneous problem (see~\citep{Borodachev2017_RadiophysEn, Nechaev23_JPP, Kuznetsov23_JETPen}). Yet, for the strong discontinuity decay it is considerably altered by the inhomogeneous and non-stationary character of an expanding plasma. The consequent law of spectrum evolution depends to a large extent on the plasma parameters on opposite sides of the discontinuity, and on the boundary shape. In the case of decay of a quasi-plane layered discontinuity, this law is established by 1D3V modeling fairly good.
The evolution of the $x$-profile's spectrum of the $B_y$-component of the field is presented in Fig.~\ref{fig8}. 
According to modeling, the spatial scale (determined by the maximum of the spectrum) grows with time following a significantly altered power law, with the exponent different from that of the $1/2$ power law known for the initial problem describing Weibel instability of the homogeneous plasma in the case of a ''needle-shaped'' electron anisotropy~(see~\citep{Borodachev2017_RadiophysEn, Nechaev23_JPP, Kuznetsov23_JETPen}). The distinctions are due to the disproportion between the rate of Weibel instability, determined by electrons, and the rate of inhomogeneous plasma expansion, which is largely dictated by heavy ions.

Previously (cf. \citep{Sarri2011, Schoeffler2016, Fox2018, Zhou2018, Ruyer2020}) the Weibel scenario, associated with the inevitable temperature anisotropy of hot electrons arising due to plasma expansion, had not been considered as the leading scenario for creation of strong (mega-Gauss or higher) fields and studied by the particles-in-cell method in 2D geometry, cf.~\citep{Thaury2010, Stockem2014, Ruyer2020}. A primary attention had been paid to the filamentation instability of counter flows of ions or electrons. Here we spell out the possibility and main principles of the Weibel mechanism of a fast (picosecond) generation of strong magnetic field with small (micron) scales of inhomogeneity during expansion of dense laser plasma containing hot electrons into a cold rarefied plasma.

\section{Expansion of plasma with hot electrons into a vacuum subjected to an external magnetic field}
\label{ch:chWeibMagn}

\subsection{An external magnetic field versus a background plasma}
\label{sec:chWeibMagn:intr}

Let us now consider a different geometry of the initial-value problem. Suppose that near the flat boundary of homogeneous plasma and vacuum there is a long half-cylinder region with initially non-uniformly heated electrons. The axis of the half-cylinder is situated on the surface of heated plasma layer. Let an uniform external magnetic field is present everywhere and is directed along the discontinuity boundary in particle densities. This external field tends to prevent expansion of hot electrons. It changes the character of their anisotropic cooling, which is linked with slowly moving ions of a dense plasma and occurs even if a rarefied background is absent, that is, in the course of expansion into a vacuum, while an electrostatic shock, in fact, is not formed.
The Weibel-type instability developing in this case and creation of magnetic fields of different scales lead to a variety of quite universal effects that have not been properly studied yet (cf.~\citep{Thaury2010, Moreno2020, Dieckmann2018, Fox2018}). 
Within the magnetohydrodynamic (MHD) approach for a plasma with considerable particle collisions (cf.~\citep{Moritaka2016, Priest2014, Plechaty2013}), the currents would flow inside a thin surface layer setting apart a region of an almost non-deformed external field and a plasma with a weak field. 

On the contrary, in the absence of collisions, the currents appear and exist for a long time in an entire volume of expanding plasma.  
As numerical simulation of the stated nonstationary problem shows, the magnetic field is not weakened crucially everywhere in an expanding plasma cloud, but acquires a complex structure and even could be directed opposite to the external field. In a vacuum, outside the expanding cloud the value and direction of the magnetic field also could vary.
Since phenomena of this kind greatly depend on the actual geometry of the electron heating region and are very diverse, they are not discussed in detail here. The review is devoted to one but universal aspect of magnetic field structuring, namely, the one related to the small-scale pinching of the currents.
Weibel-type instabilities and especially the filamentation instability have been repeatedly studied numerically in the case of colliding magnetoactive plasma flows (see, for example,~\citep{Chang2008, Spitkovsky2008, Sironi2013, Sironi2009, Bret2009} and references in subsections~\ref{sec:Laser} and~\ref{sec:chWeibInj:Interpret}). Still, detailed calculations of the thermal-type Weibel instability for the aforementioned decay of the inhomogeneously heated discontinuity between magnetized plasma and vacuum, had been missing and will be presented below for the first time.

An initially heated region of plasma could experience strongly nonequilibrium expansion only if hot electrons with the initial isotropic temperature $T_0$ and number density $n_0$ have the density of kinetic energy, $n_0 T_0$, exceeding that of the external magnetic field, $B_0^2 / (8 \pi)$. The expansion is characterized by the ion-acoustic speed $(T_0 / m_\mathrm{i})^{1/2}$ defined by the hot electron temperature and the mass of ions. (In the absence of hot electrons, only slow expansion is possible at a speed less than the thermal speed of cold ions.)
If the inequality is strong, $n_0 T_0 \gg B_0^2 / (8 \pi)$, that is, the hot electrons' pressure is much greater than the magnetic field pressure, the external magnetic field has a weak effect on the global plasma density profile, only slightly deflecting the plasma flow along the field lines.
A weak external field is quickly displaced by currents in the plasma. As a result, in the average field remaining after displacement, $B_\mathrm{r}$, the inverse electron gyrofrequency $m_\mathrm{e} c/(e B_\mathrm{r})$ and gyroradius of energetic electrons become significantly (logarithmically) larger than a typical growth time and inverse wave number of Weibel instability, respectively. In this case, the influence of the external field on the structure of currents and magnetic fields formed inside this cloud is practically eliminated, although the appeared magnetic fields can be many times greater in magnitude than the external field.

As is shown below, kinetic and specifically Weibel mechanisms are capable of producing currents of various scales in the transient processes in a plasma containing hot electrons, experiencing non-uniform scattering in an external magnetic field, and do provide a rich set of spatial quasi-magnetostatic structures. They are determined to a great extent by the magnitude of external field which can be within a wide range, depending on plasma parameters. This makes such phenomena important for numerous laboratory and space situations (see sec.~\ref{sec:concl}). One can simulate these phenomena on a qualitative level in the experiments with a magnetized laser plasma created via ablation of a flat target by means of a femtosecond laser pulse. As mentioned above, it almost immediately heats just electrons (typically to a temperature of the order of some keV) in a restricted near-surface region and creates truly collisionless plasma there. The surrounding parts of the target and plasma in them remain cold, with a temperature ranging from a few to tens of~eV, and the existing vacuum or highly rarefied preplasma (with a number density typically less than $10^{17}$~cm$^{-3}$) stay almost non-contaminated~\citep{Romagnani2008, Quinn2012, Gode2017}.

Below in this section we outline numerical modeling of precisely this kind of the initial-value problem in a laser plasma. In subsection~\ref{sec:chWeibMagn:InitCond}, a detailed formulation of the problem in the most simplified geometry is given.
Subsection~\ref{sec:chWeibMagn:strong} presents the simplest formulas used to estimate the conditions for the development and spatiotemporal scales of Weibel instability in an anisotropic plasma. It contains also analysis of typical calculations and identified physical effects for the initial number density of hot electrons $n_0$ (and the entire plasma) within $10^{21}$\,--\,$1.7 \times 10^{22}$~cm$^{-3}$ and for a strong external field within the range of $13$\,--\,$2500$~T. (For the above densities, an external field weaker than $1$~T has almost no effect on the magnetic fields created by plasma.)
Subsection~\ref{sec:chWeibMagn:weak} outlines the features of similar physical effects and presents the results of calculations confirming them for a lower plasma number density equal to $n_0 = 10^{20}$~cm$^{-3}$, and for a weaker external field, in the range of $0.5$\,--\,$13$~T. (For the above density, an external field less than or of the order of $0.1$~T has little effect on the generation of quasi-magnetostatic structures.)

\subsection{Formulation and parameters of the initial-value problem}
\label{sec:chWeibMagn:InitCond}

\begin{figure}[!b]
	\includegraphics[width=0.65\textwidth]{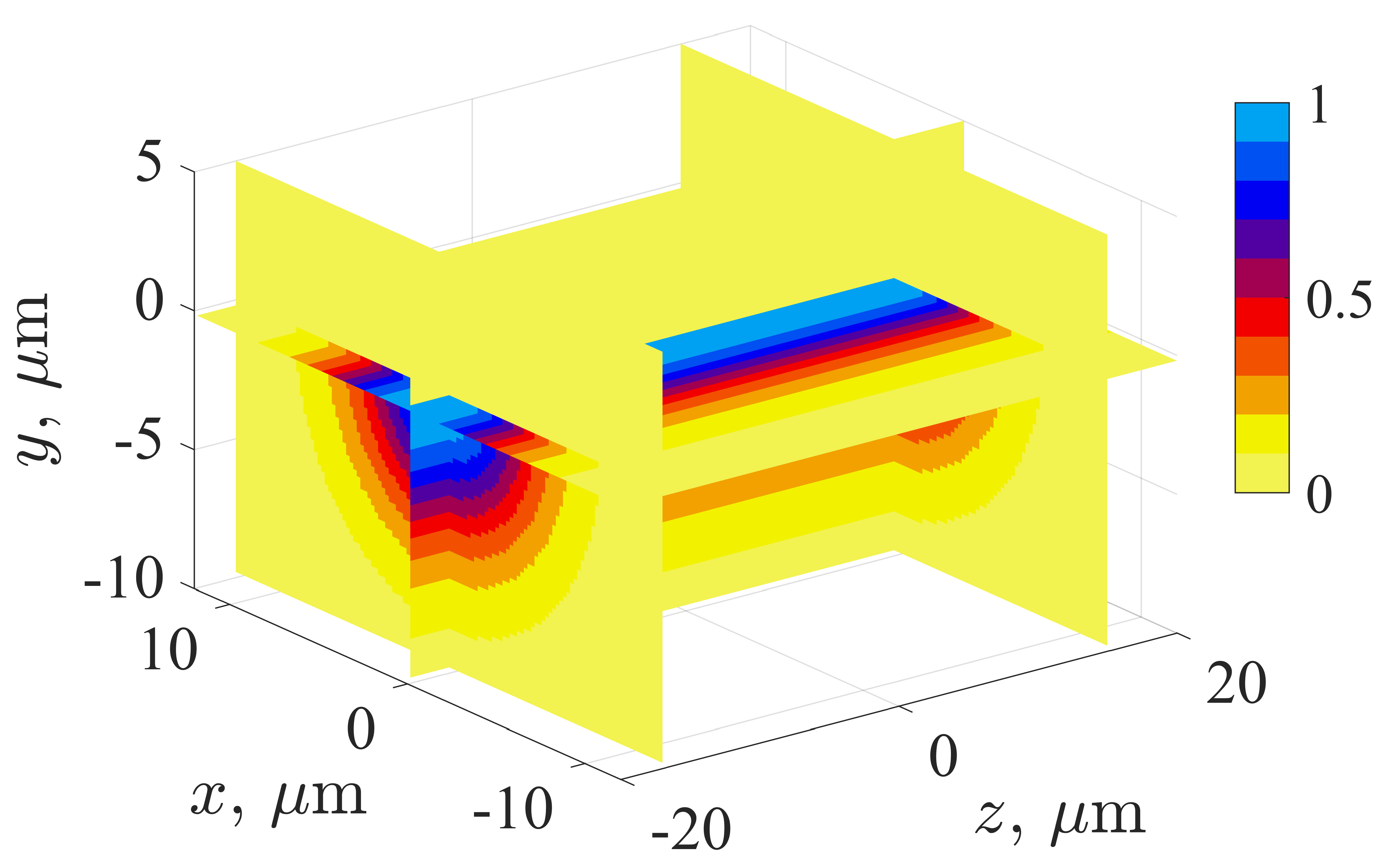}
	\centering
	\caption{
		The initial configuration of the discontinuity in a plasma. The quasi-1D heating region has the form of a half-cylinder with the axis ($z$-axis) lying on the planar plasma--vacuum boundary, $y = 0$. 
  Color shades show the distribution of the normalized initial electron temperature $T_{\mathrm{e} 0} / T_0$.
	}
	\label{fig9}
\end{figure}

Bearing in mind typical experiments on laser ablation of a flat target located in an external field~$\vec{B}_0$, we assume that at the zero moment of time the plasma number density~$n$ and ion temperature are homogeneous below the plane $y = 0$ of the target surface and are equal to $n(y \leq 0) = n_0$ and $T_{\mathrm{i} 0} \equiv 10 $~eV, correspondingly, with $n(y > 0) = 0$. The values considered are $n_0 = 1.7 \times 10^{22}$, $10^{21}$, $10^{20}$~cm$^{-3}$.
The initial temperature $T_\mathrm{e} (t=0) = T_{\mathrm{e} 0}$ of isotropically heated Maxwellian electrons (Fig.~\ref{fig9}) doesn't depend on the coordinate~$z$ along the laser-irradiated strip (of width~$2 r_0$) and varies in space, $T_{\mathrm{e} 0} = T_{\mathrm{i} 0} + \left( T_0 - T_{\mathrm{i} 0} \right) \exp\! \left( -\frac{x^2 + y^2}{r_0^2} \right)$. It has a maximum $T_0 = 1$~keV at the origin, $x=0, y=0$, and decreases with a scale $r_0 = 25$~$\mu$m in the case of plasma density $n_0 = 10^{20}$, $10^{21}$~cm$^{-3}$ or $r_0 = 5$~$\mu$m in the case of $n_0 = 1.7 \times 10^{22}$~cm$^{-3}$.
The external field $\vec{B}_0$ is uniform and directed along the $z$ or~$x$ axis.

Thus, we study a 2D initial-value problem (nothing depends on the coordinate $z$) on the expansion of hot electrons into vacuum from a region in the form of a half-cylinder, the axis of which coincides with the $z$ axis and the axial section lies on the surface of the plasma layer, $y = 0$. As we show below, the Weibel instability results in the development of current layers and/or filaments elongated along the~$z$ axis. 

In the fully 3D calculations, we use the periodic conditions for particles and fields at the borders of the simulation domain $z = \pm L_z / 2$ with~$L_z = 40$~$\mu$m. The particles are reflected, and the fields freely come out (are absorbed) on the lower boundary, $y = - L_y / 4$. Both fields and particles freely come out through the upper boundary, $y = 3 L_y / 4$. The periodic boundary conditions are also applied on the side boundaries, $x = \pm L_x / 2$.
The dimensions of the region are $L_{x,y} = 240$~$\mu$m for the densities $n_0 = 10^{20}$, $10^{21}$~cm$^{-3}$ and $L_{x,y} = 36$~$\mu$m for the density $n_0 = 1.7 \times 10^{22}$~cm$^{-3}$. A grid of $400 \times 400 \times 400$ ($1200 \times 1200$) cells is employed in the 3D~(2D) calculations. Initially the plasma layer occupies the lower quarter of the grid, $y < 0$.
Plasma modeling is carried out for a set of $2.5 \times 10^9$ ($2 \times 10^8$) particles of each fraction, electrons and ions.
Calculations are usually run until the time $\tau_\mathrm{R} = 6 \times 10^4 \, \omega_\mathrm{p}^{-1}$, when the transient processes under study have time to fully manifest themselves, but there are no yet qualitative differences between 3D3V and 2D3V calculations.
In dimensional units, one has $\tau_\mathrm{R} \approx 100$~ps and $8$~ps at $n_0 = 10^{20}$~cm$^{-3}$ and $1.7 \times 10^{22}$~cm$^{-3}$, respectively.

\begin{figure}[b]
    \includegraphics[width=1\textwidth]{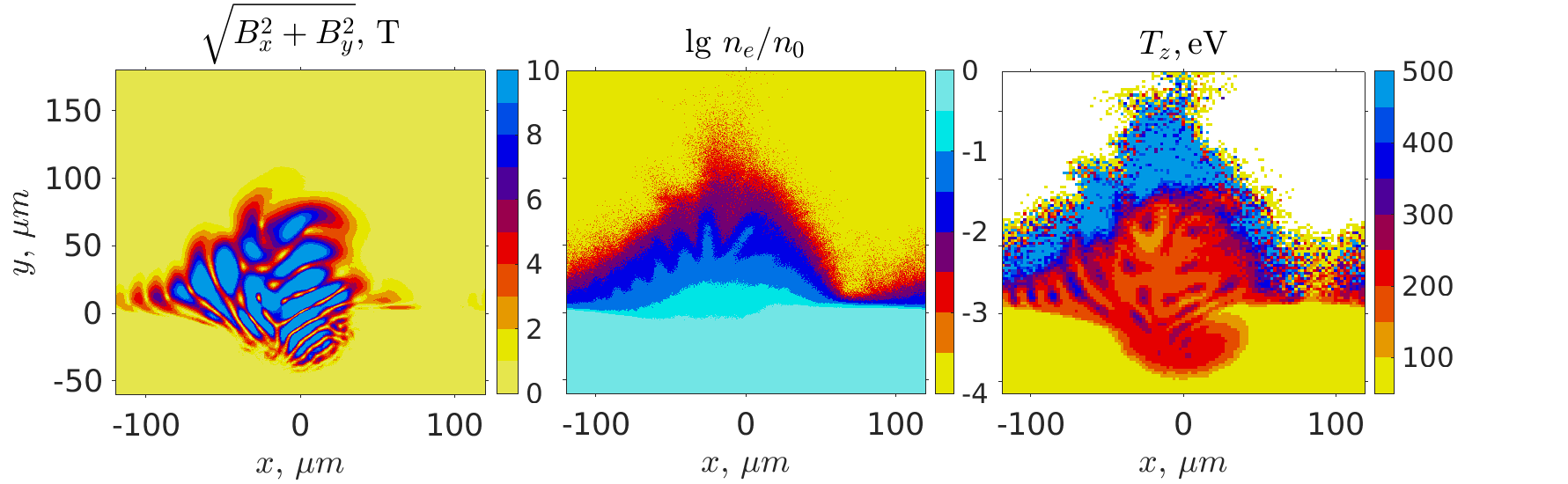}
	\centering
	\caption{
		Snapshots from the 2D3V simulation of the expansion of plasma containing hot electrons into vacuum subjected to an external field $B_{0z} = 13$~T at the moment $t = 25$~ps. 
        The plasma density at $t=0$ is $n_0 = 10^{20}$~cm$^{-3}$. 
        Left panel: the distribution of the transverse magnetic field $B_\bot$. 
        Middle panel: the logarithm of the normalized number density $n/n_0$.
        Right panel: the distribution of the effective temperature $T_z$ along the direction orthogonal to the simulation plane.
	}
	\label{fig10}
\end{figure}

For the initial number densities $n_0 = 1.7 \times 10^{22}$ and $10^{21}$~cm$^{-3}$, modeling involves either zero or strong external field with a single non-zero component $B_{0x}$ or $B_{0z}$ of values $13$, $250$, $2500$~T.
At $n_0 = 10^{20}$~cm$^{-3}$, a null or moderate external field $B_{0x} = 0.5$, $2$, $13$~T or $B_{0z} = 13$~T is used.
For the last case with $B_{0z} = 13$~T, Fig.~\ref{fig10} provides representative examples of computing the distributions of the magnitude of the field projection onto $xy$ plane, plasma number density~$n$ and effective temperature $T_z$. In the $z$ direction the electrons cool the least during their expansion.

Estimates confirm that, for the chosen parameters of plasma, the collisionless kinetics approximation is valid in the region above the surface of target, where the number density of plasma expanding into vacuum quickly decreases. Already at a density of about $0.1 \, n_0 = 10^{19}$\,--\,$1.7 \times 10^{21}$~cm$^{-3}$, the hot electrons have the mean free path (see, for example,~\citep{Trubnikov1965}) of about $5000$\,--\,$50$~$\mu$m, which is significantly larger than $r_0$, $L_{x,y}$ and the characteristic dimension of the region where Weibel magnetic fields are generated. For denser plasma inside the target, where the collision frequency is high, modeling by the particles-in-cell method is not correct.

\subsection{Dense plasma inhomogeneously heated in the presence of strong magnetic field}
\label{sec:chWeibMagn:strong}

At high initial plasma densities, for instance, $n_0 = 1.7 \times 10^{22}$~cm$^{-3}$, the explosive expansion of a cloud of hot electrons into vacuum is greatly suppressed by extremely strong external field of the order of $2500$~T, that is, of the value corresponding to $(8 \pi n_0 T_0)^{1/2}$. In less strong fields, $250$~T and $13$~T, the plasma kinetic pressure is by many times higher than the magnetic field pressure and expansion results in displacing the field at roughly ion-acoustic speed, $(T_0 / m_\mathrm{i})^{ 1/2}\! \sim 10^8$~cm/s, in the central part of the hot region and at a lower speed at the edges of hot region. Thus, the efficiency of flow deceleration is essentially the same for both selected external field values, which differ by almost $20$~times, as well as for the null external field assumed in section \ref{ch:raspad}.

\begin{figure}[b]
    \includegraphics[width=0.65\textwidth]{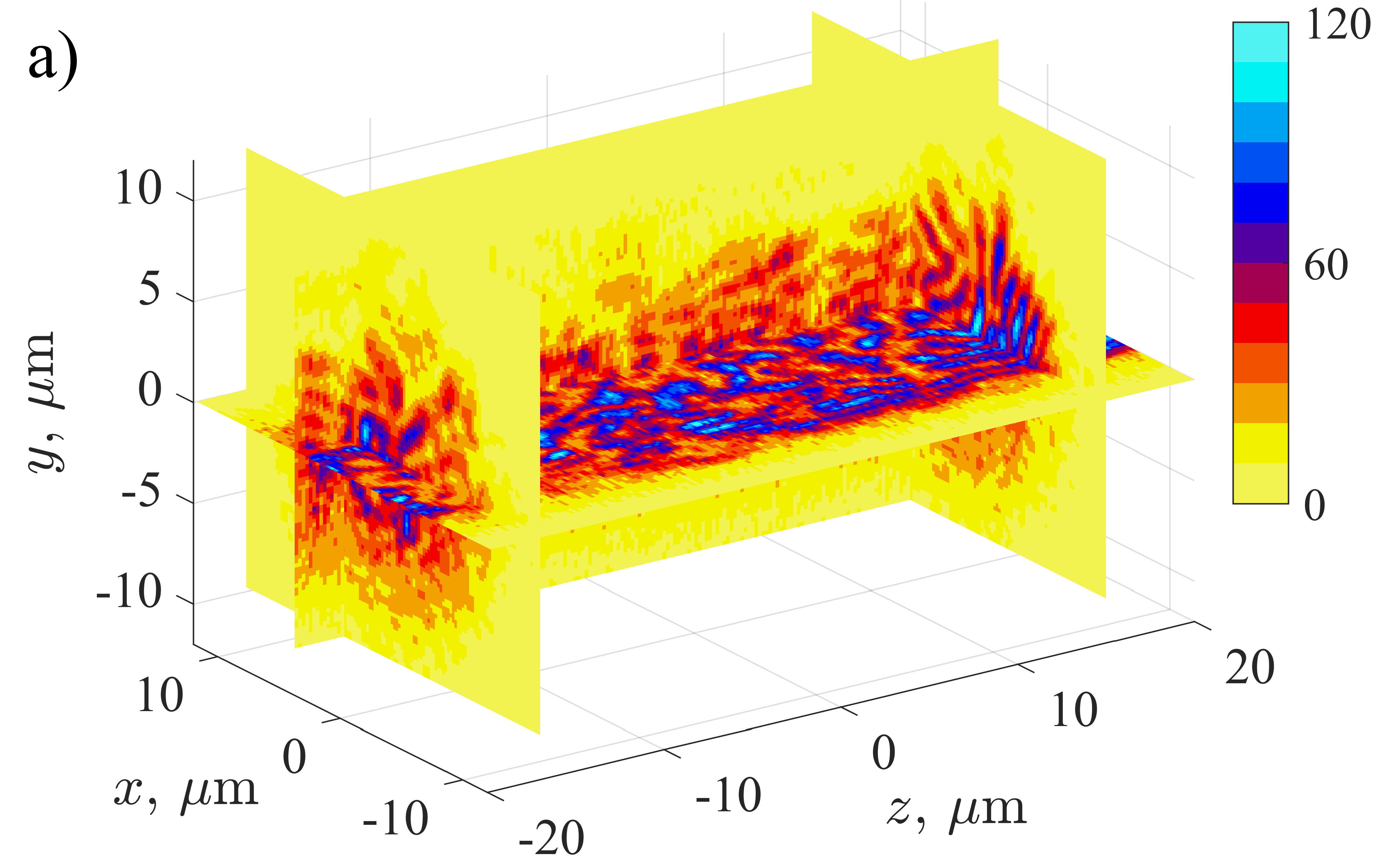}
    \includegraphics[width=0.65\textwidth]{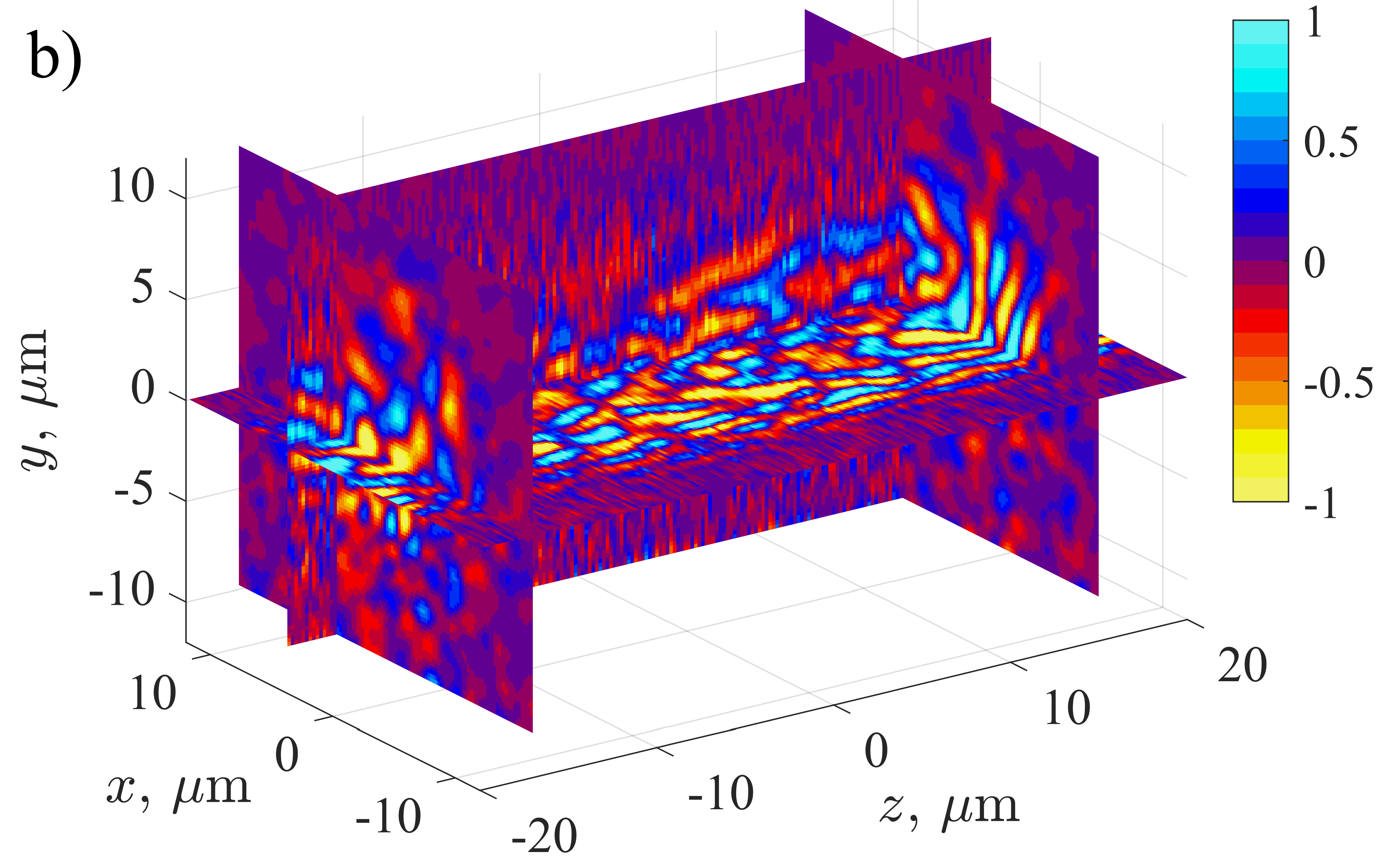}
	\centering
	\caption{
		a)~Structure of the transverse magnetic field $B_\perp$ (in Teslas) obtained in 3D3V modeling of the plasma expansion into a vacuum, $t = 3$~ps. 
		b)~Filamentary structure of the $z$-component of the electric current density $j_z$ (in units of~$10^{14}$A/m$^2$). 
		The initial number density of plasma is $n_0 = 1.7 \times 10^{22}$~cm$^{-3}$. 
		The external field is homogeneous and directed along $z$ axis, $B_{0z} = 250$~T.
	}
	\label{fig11}
\end{figure}

The results of 3D3V modeling of plasma expansion at the initial number density $n_0 = 1.7 \times 10^{22}$~cm$^{-3}$ and external field $B_{0z} = 250$~T oriented along the~$z$ axis are shown in Fig.~\ref{fig11} for the moment of time $t \approx 3 \times 10^4 \, \omega_\mathrm{p}^{-1} \approx 3$~ps.
By this moment, the expanding plasma displaces almost completely the external field from a small half-cylinder of a radius $\sim 3$~$\mu$m above the plane of initial discontinuity, $y = 0$, there is already a small half-cylinder with a radius of about $3$~$\mu$m. Electron currents produce the longitudinal magnetic field $B_z - B_{0z}$ opposing the external one and reaching a magnitude of the order of that of the external field. Distribution of the transverse magnetic field value, $B_{\perp} = \left( B_x^2 + B_y^2 \right)^{1/2}$, shown in Fig.~\ref{fig11}, is created by the small-scale current filaments. They are mainly parallel to the~$z$ axis, are formed predominantly by hot electrons and, during the expansion of plasma, are displaced along with its flow.

The mechanism of formation of such current structures resembling $z$-pinches, distorted by the plasma inhomogeneity, is completely similar to that discussed in the previous section. As hot electrons fly away, their effective temperatures $T_x$ and $T_y$ along the~$x$ and~$y$ axes quickly decrease, while the temperature $T_z$ along the~$z$ axis changes much more slowly due to the large extension of the half-cylinder of the initially heated plasma in this direction (in~2D3V calculations~--- due to the imposed uniformity of fields and currents along this axis).
As a result, the plasma becomes highly nonequilibrium, acquiring a significant value of the anisotropy parameter $A_\mathrm{e} = T_z / T_{x} - 1 \gtrsim 1$, which leads to Weibel instability of disturbances of the longitudinal current and transverse magnetic field.

For a qualitative analysis of the instability the electron distribution function may be described approximately by a bi-Maxwellian form.
In this case, the maximum growth rate is achieved for the perturbations with wave vectors orthogonal to the $z$ axis, that is, the direction of the highest electron temperature. The smaller the angle between the perturbation wave vector and the $z$ axis, the weaker the growth rate. According to 3D modeling at later times, the inclined filaments are getting more pronounced and the whole current structure becomes more and more irregular, acquiring dependence on the~$z$ coordinate. At the same time, the filaments directed along the $z$ axis get saturated by the self-consistent magnetic field. The central part of the expanding plasma clearly reveals the inclined filaments as per Fig.~\ref{fig11}.

The modeling shows that at the moment of saturation of transverse magnetic field, which reaches a value of $B_\perp \sim 80$~T, in the region of plasma expanding into vacuum, $0 < y < 5$~$\mu$m, the number density of electrons is $n_\mathrm{e} \sim 0.1 \, n_0 \approx 1.7 \times 10^{21}$~cm$^{-3}$, their longitudinal temperature is $T_z \sim 0.5$~keV and the degree of plasma anisotropy is $A_\mathrm{e} \sim 1 $.
Using the linear theory of Weibel instability (see~\citep{Weibel1959, Davidson1989, Vagin2014, Kocharovsky2016_UFN} and formulas in section~\ref{sec:chWeibExp:correl}), one can confirm that the hot-electron gyroradius, $r_\mathrm{L } = (2 {T_x} / m_\mathrm{e})^{1/2} \omega_B^{-1} \approx {0.6}$~$\mu$m, is approximately equal to the optimal half-period of Weibel perturbations, $\lambda_0/2 \sim 5 \, c / \omega_\mathrm{p} \, (n_0 / n_\mathrm{e})^{1/2} A_\mathrm{e}^{-{1/2}}\! \sim 0.7$~$\mu$m, and the electron gyrofrequency, $\omega_{B} = e B_\perp / (m_\mathrm{e} c)$, is about the maximum growth rate, $\omega_0 \sim 3 \, (T_x / m_\mathrm{e})^{1/2} \lambda_0^{-1} (1 + A_\mathrm{e}^{-1})^{-1} \approx 0.6 \, \omega_{B}$.
Thus, the magnitude of the magnetic field corresponds to the known criterion for saturation of Weibel instability~\citep{Kocharovsky2016_UFN}.
The growth time of the magnetic field energy $\tau_B$ that defines the duration of the linear stage of instability is also consistent with the Weibel mechanism: $\tau_B \sim 10 \, \omega_0^{-1}$.
The above confirms the assumption about the Weibel nature of the observed instability.
The 2D3V modeling, outlined below, leads to the similar results: $\lambda \sim r_\mathrm{L}$, $\omega_0 \sim \omega_{B}$ and $\tau_B \sim 10 \, \omega_0^{-1}$. 

\begin{figure}[b]
	\includegraphics[trim = 2cm 0cm 1.5cm 0, clip, width=0.6\textwidth]{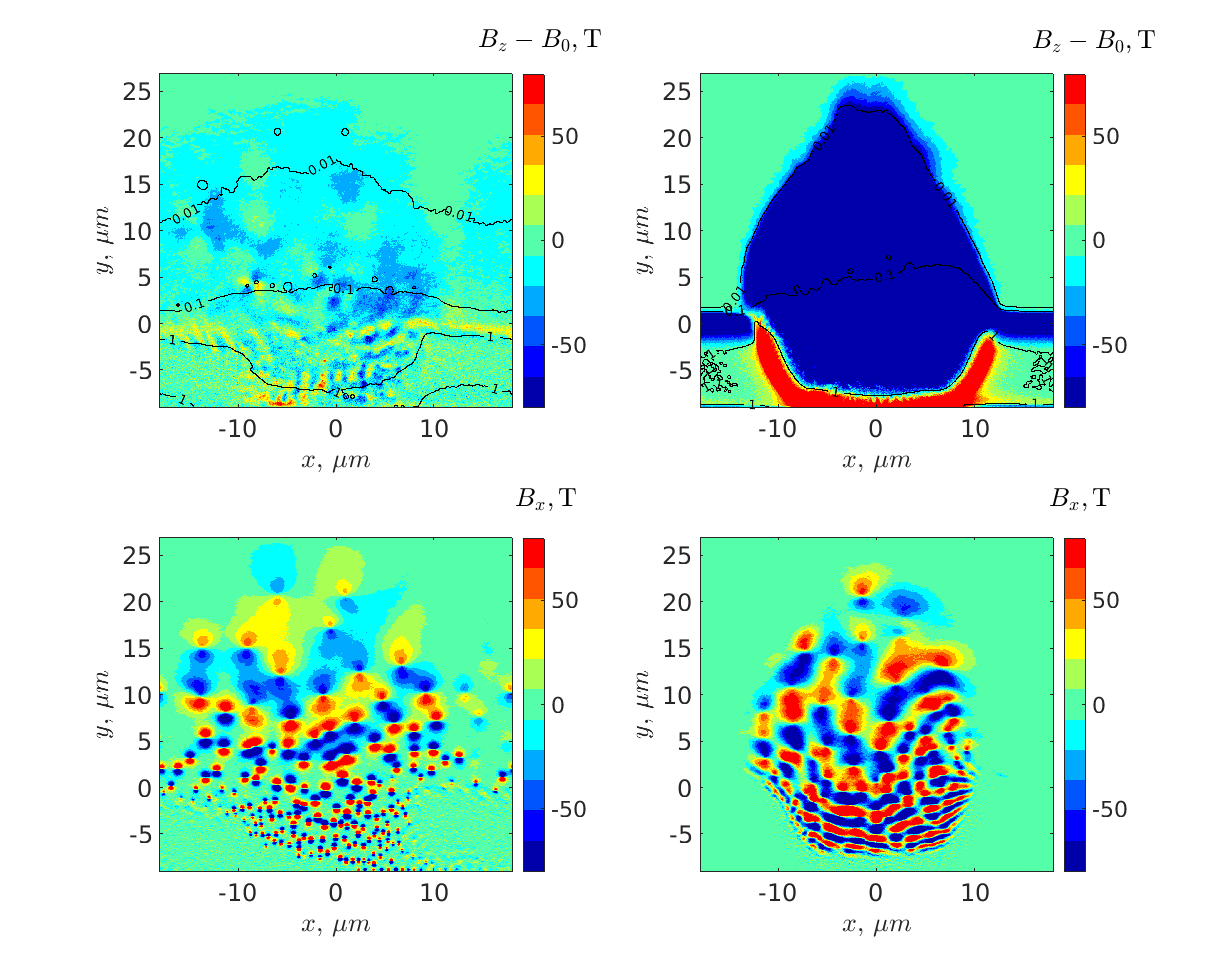}
	\centering
	\caption{
           Structure of magnetic field obtained in 2D3V modeling of plasma expansion into vacuum at $t = 7.5$~ps. The plasma density at $t=0$ is $n_0 = 1.7 \times 10^{22}$~cm$^{-3}$. The external magnetic field $B_{0z}$ is perpendicular to the simulation plane. For the left panels, it is $B_{0z} = 13$~T. Left top panel presents the $z$-component of the field minus the external one, $B_z - B_{0z}$. The isolines are the plasma density levels $n / n_0 = 0.01$, $0.1$,~$1$, correspondingly. Bottom left panel shows the magnetic field component $B_x$. 
           Right panels show the same for the simulation with $B_{0z} = 250$~T.
	}
	\label{fig12}
\end{figure}

\begin{figure}[b]
	\includegraphics[width=0.6\textwidth]{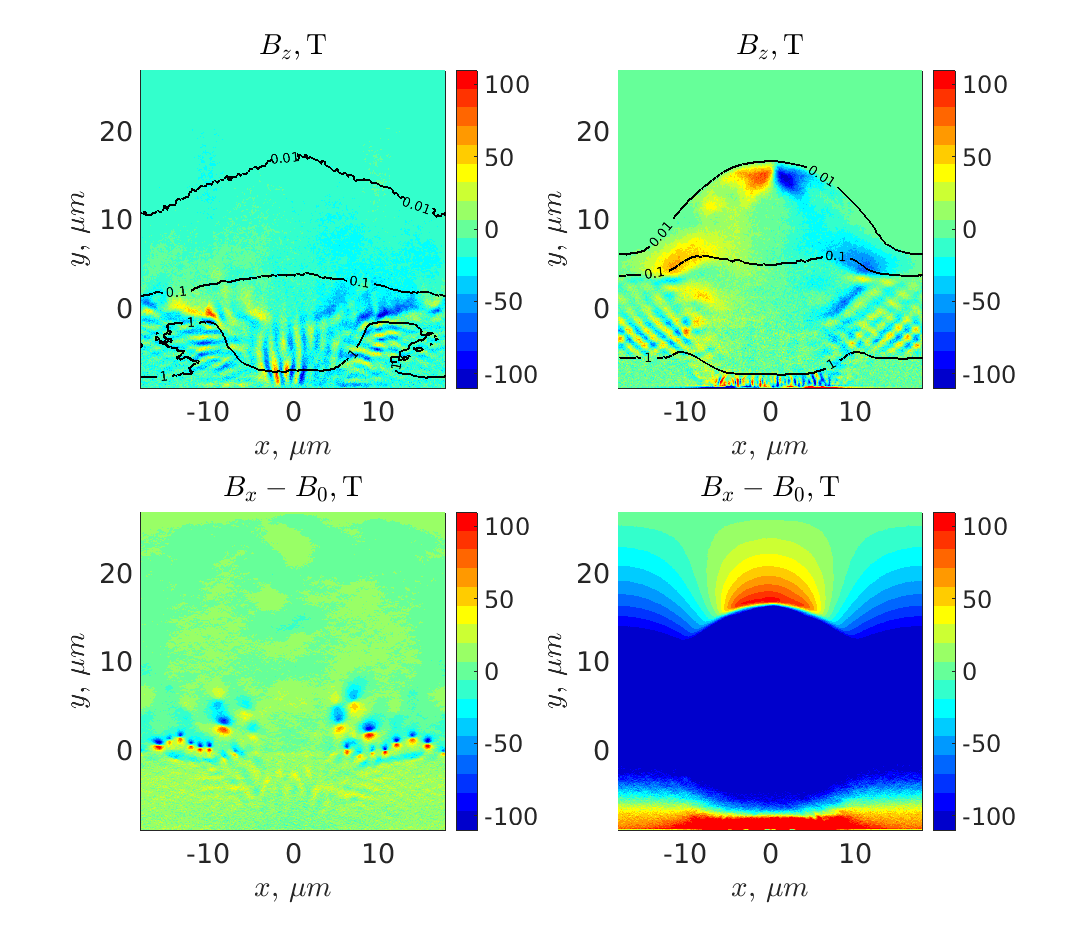}
	\centering
	\caption{Structure of magnetic field obtained in 2D3V modeling of plasma expansion into vacuum at $t = 7.5$~ps. The plasma density at $t=0$ is $n_0 = 1.7 \times 10^{22}$~cm$^{-3}$. The external magnetic field $B_{0x}$ lies in the simulation plane. For the left panels, it is $B_{0x} = 13$~T. Left top panel shows the magnetic field component $B_z$. The isolines are the plasma density levels $n / n_0 = 0.01$, $0.1$,~$1$, correspondingly. Bottom left panel presents the $x$-component of the field minus the external one, $B_x - B_{0x}$.
    Right panels show the same for the simulation with $B_{0x} = 250$~T.
	}
	\label{fig13}
\end{figure}

The case of a strong field oriented along the axis of a heated half-cylinder, $B_{0z} = 250$~T, i.e., across the simulation $xy$-plane in 2D modeling, stands out in that it leads to a cumulative (focusing) effect, as a result of which the rate of this field displacement is almost one and a half times greater, and the resulting plasma ejection in the form of a tongue is almost one and a half times narrower than that for the same magnitude field $B_{0x}$ in the simulation plane (see Figs.~\ref{fig12},~\ref{fig13}). The differences are associated with the structure of the emerging global currents, in the first case flowing predominantly in the $xy$ plane at the periphery of expanding plasma cloud and forming an inhomogeneous solenoid with the~$z$ axis, significantly weakening the field $B_{0z}$, and in the second case flowing mainly along the~$z$ axis, generally speaking, in different directions and forming an inhomogeneous configuration of magnetic field in $xy$ plane, weakening the external field $B_{0x}$ inside the plasma cloud.
A noticeable cumulative effect does not arise in the case of a weaker field, $13$~T or less, which for both orientations has almost no effect on the expanding-plasma density profile.

\begin{figure}[!b]
	\includegraphics[width=0.55\textwidth]{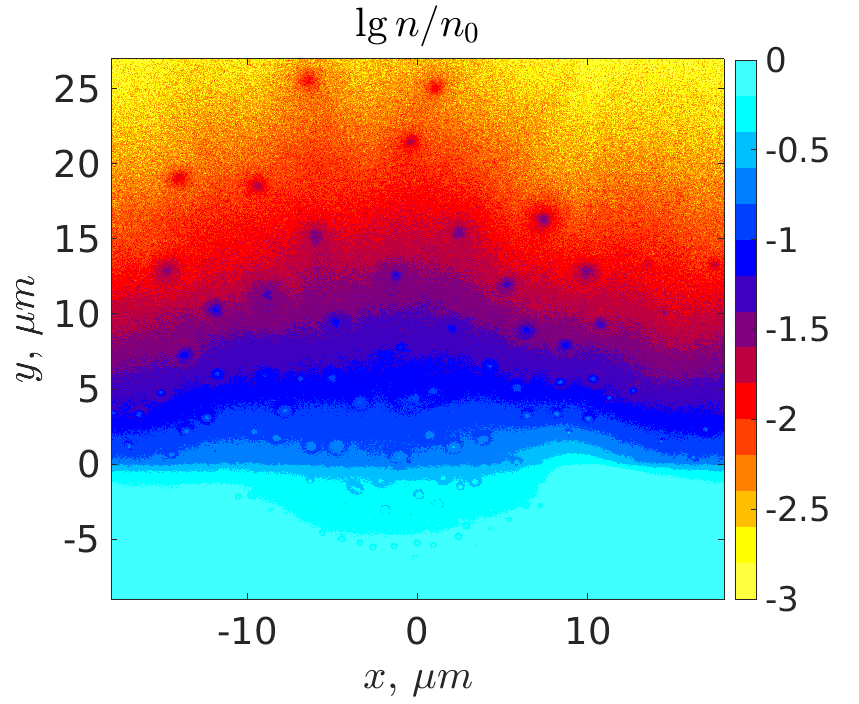}
	\centering
	\caption{
            The logarithm of the plasma density $n/n_0$ at $t = 9$~ps. The external magnetic field is $B_{0z} = 13$~T. The initial plasma density is $n_0 = 1.7 \times 10^{22}$~cm$^{-3}$. 
	}
	\label{fig14}
\end{figure}

At the same time, if the external field is oriented along the~$z$ axis, the symmetry of expansion becomes slightly violated because electrons are systematically displaced to the left by the Lorentz force. It is especially pronounced at the initial stage of discontinuity decay, when the external magnetic field is not yet displaced and the fraction of electrons flying upward along the~$y$ axis and interacting with the external field is still quite large (see also Fig.~\ref{fig10} for low plasma densities). A similar asymmetry in the plasma density distribution is clearly visible in Fig.~\ref{fig12} for a strong external field $B_{0z} = 250$~T. In the case of a weaker external field, $B_{0z} = 13$~T, the discussed symmetry violation remains noticeable in the distribution of the small-scale transverse magnetic field. The latter field, generated by the Weibel current filaments, turns out to be strong enough to compress not too dense plasma and form filaments of its density. They stretch along $z$ axis and shift due to the Lorentz force which tracks the plasma flow asymmetry (Fig.~\ref{fig14}).

In the case of an in-plane external field $B_{0x}$, there is no Lorentz force component along the $x$ axis. So, there is no asymmetry in geometry of the stream lines, plasma ejection, and the structure of magnetic fields in the simulation $xy$ plane (see~Fig.~\ref{fig13}). 

For both considered directions of the external field on the upper panels in Figs.~\ref{fig12} and~\ref{fig13}, to the left and right relative to the center of heated area near the origin, one can observe opposite in sign and approximately equal in magnitude $z$-components of the field generated by plasma.
Such a mirror-symmetrical distribution is induced by the vertical current component, which is parallel to the $y$ axis and composed of the hot electrons participating both in the ''fountain'' currents and the charge compensation currents penetrating the dense plasma.
A generally similar picture is observed in calculations without the external field, discussed in section~\ref{ch:raspad} (see~\citep{Kolodner1979, Sakagami1979, Albertazzi2015})

2D3V modeling clearly reveals a small-scale current structure in the form of $z$-pinches, which produces the ''dipole'' spots of the transverse field components $B_x$ and $B_y$ (minus the external field component $ B_{0x}$, if present; see~bottom panels in Fig.~\ref{fig12}, \ref{fig13}).
In this case, as modeling shows, the appearance of such $z$-pinches can be hindered by a strong external field $B_{0x}$ directed along the~$x$ axis. During the process of its displacement the large-scale multidirectional currents flowing along the~$z$ axis are created in the plasma cloud as well as the associated suppression of the growth of anisotropy of cooling electrons occurs, mainly due to a decrease in temperature~$T_z$.
The effect is similar to that discussed in section \ref{ch:raspad}, where a decrease in the degree of anisotropy in the regions of maximum current density at the nonlinear stage of Weibel instability of an expanding plasma is revealed. 
For the adopted parameters of plasma, as little as a weak external field $B_{0x} \lesssim 1$~T does not destroy the initial process of multiple production of $z$-pinches, despite the fact that their transverse field after the formation of a self-consistent nonlinear structure is almost two orders of magnitude higher (about $50$~T even for a significant external field $B_{0x} = 13$~T as per Fig.~\ref{fig13}).

At the same time, the the external field $B_{0z}$, which does not suppress the explosive decay of the discontinuity (and has a value less than $1000$~T for the parameters used), doesn't stop the anisotropic cooling of the plasma displacing the external field and allows multiple formation of pinch-like structures in the form of an irregular and inhomogeneous lattice, as demonstrated in the lower panels in Fig.~\ref{fig12} using the example of the $x$-component of field.
It turns out that in such an external field localized currents effectively arise as $z$-pinches with a thickness of the order of electron gyroradius and with transverse fields one to two orders of magnitude greater than the external field. 
According to the simulation, one can expect that their lifetime significantly exceeds the characteristic time of development of quasi-magnetostatic Weibel (weak) turbulence that arises in an initially homogeneous plasma with similar parameters and anisotropy in the absence of external field, when the formation of such strongly nonlinear $z$-pinches with a large increase in plasma density in their center usually does not occur (see, in particular, \citep{Borodachev2017_RadiophysEn, Nechaev23_JPP}) and the average value of the saturated magnetic field $B_\mathrm{s}$ does not reach allowable values bounded from above by the value of the order of $(n T_0)^{1/2}$.

The pattern of expansion and efficiency of formation of the current features resembling $z$-pinches as well as dynamics and the entire structure of magnetic fields during discontinuity decay are sensitive not only to the value of the magnetic field, but also to the total plasma density and fraction of heated electrons. If there is a considerable fraction of cold electrons, then the efficiency of displacement of external field, number density profile as well as the spatial distribution and number of $z$-pinches are changed significantly as per modeling with the same problem parameters, but with a reduced fraction of heated electrons (down to $20$\% instead of $100$\%).

\subsection{Rarefied plasma inhomogeneously heated in the presence of a moderate magnetic field}
\label{sec:chWeibMagn:weak}

As the initial number density $n_0$ of plasma with hot electrons decreases, the limiting values of the external field also decrease. They are the maximum value $B_\mathrm{max}$, which still allows the explosive expansion of electrons into a vacuum, and the minimum value $B_\mathrm{min}$ required for any pronounced influence on this spread. In accord with estimates (subsection~\ref{sec:chWeibMagn:intr}, \ref{sec:chWeibMagn:InitCond}) and modeling, the first depends radically on the initial number density as well as maximum temperature of hot electrons: $B_\mathrm{max} \sim ( 8 \pi n_0 T_0)^{1/2}$. The second, in addition, implicitly depends on the orientation of external field within the plane of initial plasma disruption as well as geometric factors of the region with heated electrons, affecting their anisotropic cooling and expansion. Hence, scaling (i) the value of $B_\mathrm{min}$ by varying the number density of hot electrons or even (ii) the entire dynamics of discontinuity decay requires detailed numerical modeling. Below we overview the modeling results inthe case of initial plasma density $n_0 = 10^{20}$~cm$^{-3}$, which is significantly less than that employed in the previous section, but quite representative for the relevant laser-plasma experiments. We focus on the effect of fairly weak external fields accessible in the current laboratory experiments on the decay of plasma discontinuity, considering the initial maximum electron temperature $T_0 = 1$~keV to be given.

\begin{figure}[b]
    \includegraphics[width=1\textwidth]{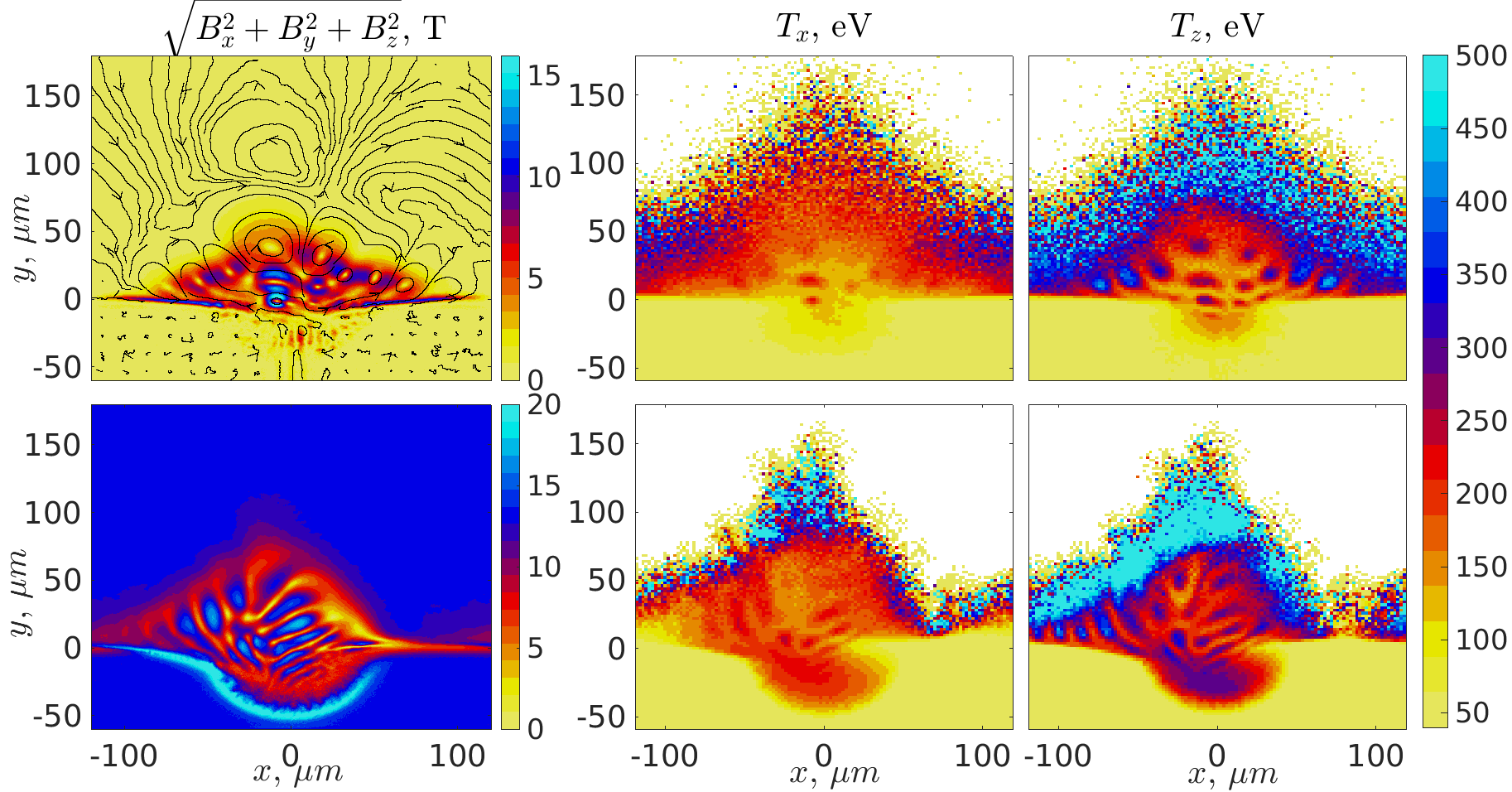}
	\centering
	\caption{
             Snapshot of an expansion of plasma containing hot electrons into vacuum at $t = 20$~ps. The initial plasma density is $n_0 = 10^{20}$~cm$^{-3}$.
             For the top panels, an external magnetic field is absent. For the bottom panels the external field is $B_{0z} = 13$~T. 
             Panels from left to right show the absolute value of total magnetic field, the effective temperatures along $x$ and $z$ axes, respectively (see also Fig.~\ref{fig10}).
	}
	\label{fig15}
\end{figure}

The starting point when comparing scenarios for the discontinuity decay in an external field of different strength is the zero-field variant shown in Fig.~\ref{fig15}. It reveals a far from trivial structure of plasma-generated magnetic fields. Contrary to the well-known fountain mechanism (which creates a large-scale field $B_z$, including that in the vicinity of the initial discontinuity plane), this small-scale field structure is a manifestation of the Weibel mechanism efficiently operating in the course of plasma expansion discussed here. As for the dense plasma, at the beginning of expansion the electron thermal velocities can lessen only in the~$xy$ plane, because their distribution is uniform along the~$z$ axis. Due to Weibel instability along this axis, the current filaments resembling $z$-pinches are formed, non-uniformly and rather randomly distributed over the simulation plane and gradually occupying an increasing area within it along with the expanding plasma. Quickly gathering into beams and producing currents reaching the level of nonlinear saturation of instability, corresponding to transverse magnetic field $B_\perp \sim 10$~T, electrons acquire a regular directed velocity along $z$ axis. At the same time, losing their longitudinal effective temperature $T_z$ and continuing to move in one way or another in the $xy$ plane, under the influence of the created magnetic fields and the electric fields induced by them, electrons gradually equalize their thermal velocities in all three orthogonal directions. In this process, as shown in Fig.~\ref{fig15}, a distinctive correlation arises (see section~\ref{ch:raspad}) between the effective electron temperatures (especially longitudinal one), spatial distributions of magnetic field strength, and plasma density.

If an external field in the plane of initial plasma-vacuum discontinuity is present, the expansion process described above experiences minimal distortions when this field is directed along the $z$ axis, if the currents flowing along it are not disturbed. Yet, according to the previous section, such a field asymmetrically deflects the plasma flow in $xy$ plane, unilaterally acting by the Lorentz force on fountain currents of electrons escaping from the gap (see~Fig.~\ref{fig10}).
In this case, contrary to the case of a null external field, the structure of currents flowing along the~$z$ axis turns out to be different. First, the current structure acquires the form of a system of curved ribbon-like current sheets. Later on they break up into isolated, deformed filaments resembling $z$-pinches. Finally, a further distortion of the self-consistent field $B_z$ appears and the average anisotropy, describing the electron velocity distribution, drops from a large magnitude, $A \sim 3$, to a small magnetude, $A \sim 0.1$--$0.3$. Thus, the longitudinal temperature, $T_z$, tends to~the transverse temperatures, $T_x$, $T_y$. The magnitude of the field in some regions reaches a value of the order of $20$~T, that is, about twice the magnitude of the external field displaced by the currents of plasma cloud. 

As a whole, a qualitative picture of the phenomenon remains similar to that in the case of a null external field. In particular, the correlations between the effective electron temperatures, inhomogeneities of the magnetic field, and plasma density persist. Yet, significant quantitative differences can be traced down to a very weak external field, $B_{\mathrm{min},z} \sim 10^{-2} B_\mathrm{max} \sim 1$~T, about two orders of magnitude weaker than the maximum value permitting plasma expansion, $B_\mathrm{max} \sim 100$~T.

\begin{figure}[t]
	\includegraphics[width=0.8\textwidth]{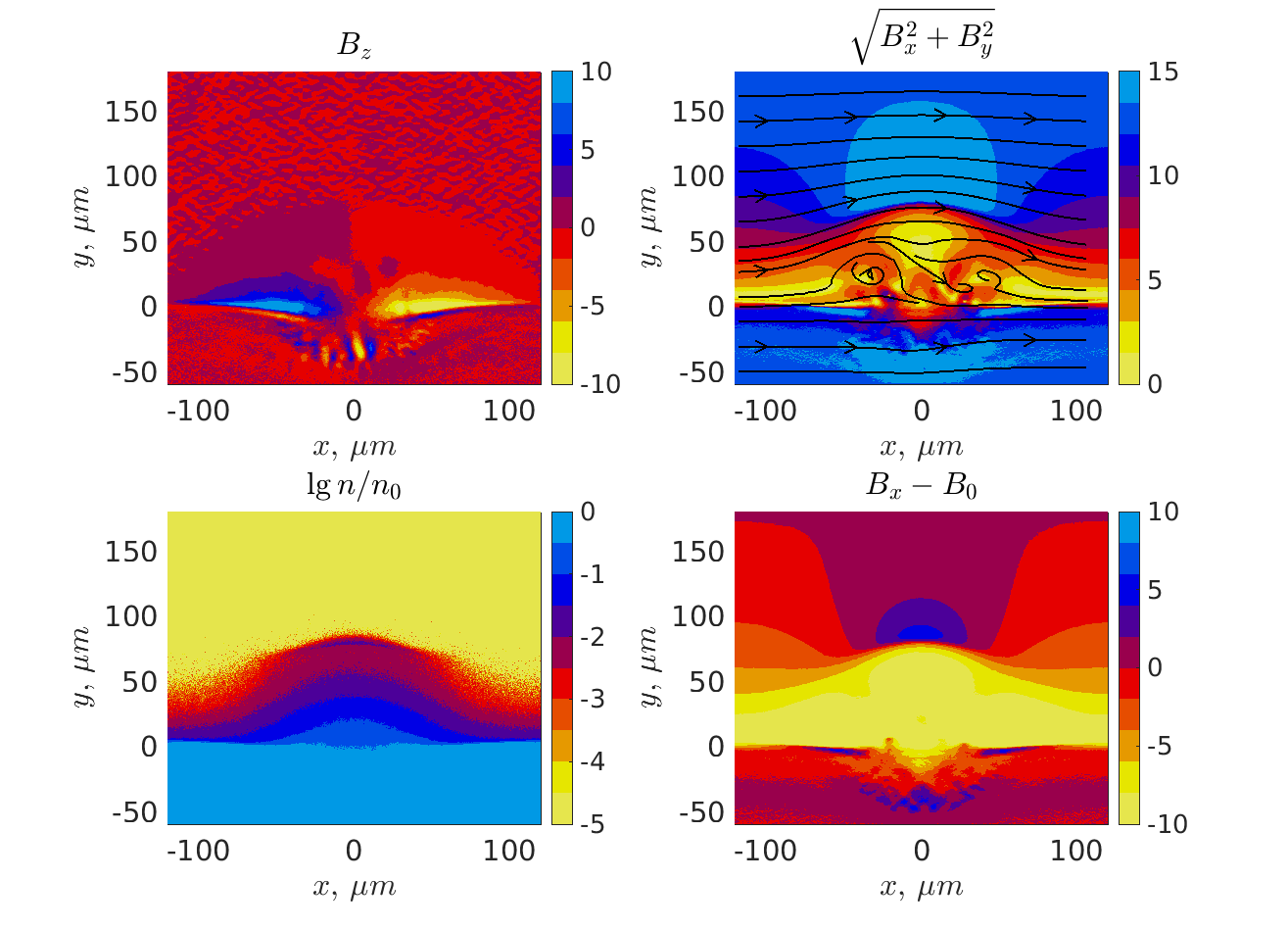}
	\centering
	\caption{
             Expansion of a plasma into vacuum with an external magnetic field $B_{0x} = 13$~T at $t = 20$~ps. The initial density is $n_0 = 10^{20}$~cm$^{-3}$. 
             Top left panel shows the field component $B_z$ (in Teslas).
             Top right panel shows the magnitude of the transverse field $B_\perp$ (in Teslas), its field lines are shown in black.
             Bottom left panel is the logarithm of the normalized plasma number density~$n/n_0$.
             Bottom right panel shows the field $B_x$ minus the external field, $B_x - B_{0x}$ (in Teslas).
	}
	\label{fig16}
\end{figure}

\begin{figure}[b]
	\includegraphics[trim = 1cm 0 2cm 0, clip, width=0.8\textwidth]{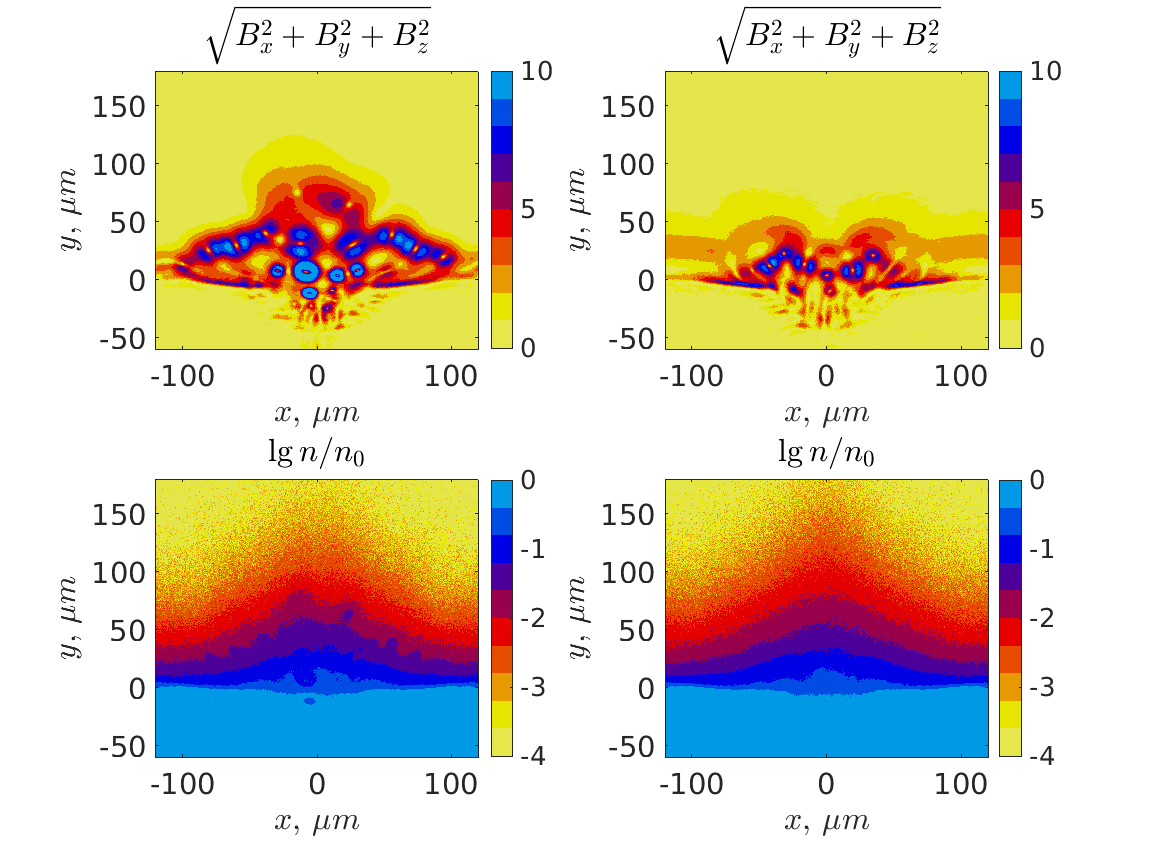}
	\centering
	\caption{
        Comparison of magnetic field and number density patterns in the course of expansion of a plasma at the time moment $t = 36$~ps for different values of the external field directed along the $x$ axis: $B_{0x} = 0$~T (left panels) and $0.5$~T (right panels). The initial number density is $n_0 = 10^{20}$~cm$^{-3}$. 
        Top panels present the absolute value of total field (in Teslas), bottom panels show the logarithm of the normalized plasma number density $n/n_0$ (cf. Fig.~\ref{fig10}).
	}
	\label{fig17}
\end{figure}

The greatest distortions in the decay process of the discontinuity in a plasma containing an electron heating region of a half-cylinder shape are established by an external field $B_{0x}$ oriented across the half-cylinder. Such a field prevents Weibel production of current filaments resembling $z$-pinches. As is clear from Fig.~\ref{fig16}, although the fountain mechanism of generation of global currents persists, this field does not create an appreciable cumulative effect and asymmetric ''fountain'' of hot electrons (see~subsection~\ref{sec:chWeibMagn:strong}). Moreover, the $B_{0x}$-field slows down its own deformation by inducing currents which support it and also makes it difficult for plasma cloud to escape. 
In the vicinity of the plasma cloud top, the long-existing plasma compaction and region of enhanced magnetic field are clearly visible in Fig.~\ref{fig16}. Such formations disappear just in a weak external field, $B_{0x} \lesssim 1$~T. Starting from a similarly weak field, the electron temperature along the $z$ axis remains above the temperatures in the $xy$ plane for a considerable interval of time after the initial phase of expansion, and an appreciable anisotropy of electron velocity distribution, $A \sim 0.1$, becomes possible. 
As a result, the current filaments along the $z$ axis appear and create an inhomogeneous small-scale transverse magnetic field in the inner regions of the expanding plasma cloud. For instance, such a characteristic field structure is clearly seen already in the case of a moderate external field, $B_{0x} = 2$~T. In a weaker field, $B_{0x} = 0.5$~T, the self-consistent current structures resemble those generated in the decay of a plasma discontinuity into vacuum in the absence of the external field (Fig.~\ref{fig17}), but create small-scale fields of considerably higher magnitude at the same moment of time.
Thus, for the orientation of external field in the simulation plane, quantitative features of the created currents and fields remain almost intact (as compared to the case of a null external field) only in the case of a very weak external field, $B_{\mathrm{min},x} \sim 10^{-3} B_\mathrm{max} \sim 0.1$~T and $B_{\mathrm{min},z} \sim 1$~T.

Comparison of the above simulations with the ones discussed in the previous subsection shows that the same electron Weibel mechanism operates in both scenarios of the dense plasma expansion into either background plasma in the absence of external field or a vacuum with external field.   
However, the structures of the small-scale magnetic field and current in those two cases are significantly different. Thus, for a wide range of parameters of dense and rarefied plasma, including those characteristic of astrophysical conditions, one should take into account such a difference in two-three orders of magnitude between the limiting values, $B_\mathrm{min}$ and $B_\mathrm{max}$, of external field. The external field in the range $B_\mathrm{min} < B_0 < B_\mathrm{max}$ has a nontrivial effect on the decay of discontinuity in the plasma near the surface of which an extended quasi-1D region with hot electrons is created. However, a discussion of the implementation of such scenarios in the space plasma physics and a general analysis of the results of modeling the corresponding process of small-scale filamentation of a current of hot electrons in the situation, when there is a background plasma of a finite density and an external field is present, are beyond the review's scope. For a discussion of the creation of a turbulent magnetic field by ion currents in the vicinity of front of a collisionless shock wave without taking into account any significant fraction of hot electrons, see references in subsection~\ref{sec:Laser}.

\section{Buildout of the orthogonal current structures in a plasma layer with injection of hot electrons}
\label{ch:chWeibInj}

\subsection{A boundary-value problem for a plasma expansion caused by particle injection}
\label{sec:chWeibInj:Intro}

Let us continue the overview of the processes of formation of current structures (sheets or filaments) in a nonequilibrium collisionless plasma within a different framework, in which the plasma with hot electrons is not initially prepared in the form of a cloud (as in sec.~\ref{ch:raspad} and~\ref{ch:chWeibMagn}), but is continuously injected from a surface into a relatively cold background plasma with a non-uniform number density.
In such a boundary-value problem, due to Weibel-type instability, the formation of mutually orthogonal current structures with sufficiently strong magnetic fields is possible in adjacent plasma layers.
We will describe the formation of these structures on the basis of 2D3V modeling for different injection durations as well as various spatial distributions and characteristic values of the number densities and electron and ion temperatures of the injected and background plasma, which initially has a monotonically decreasing number density profile. A 3D modeling performed for a typical set of parameters confirms the conclusions of 2D simulations, including the presence of Weibel-type instabilities.
As in the previous sections, the choice of parameters corresponds to a laser plasma created via ablation of a target by quasi-cylindrical focusing of a femtosecond laser beam.

In order to create the required initial profile of the background plasma density above the target surface, one can employ a sufficiently long low-energy prepulse generating plasma inside the target; see, for example, \citep{Ivanov2014}. The prepulse creates a thin inhomogeneous layer of a background plasma, which can noticeably expand and cool down by the time a powerful (tera- or petawatt) femtosecond pulse arrives, instantly heating electrons to keV temperatures in the surface layer. These hot electrons, escaping from the target, force the ions remaining there to draw in the cold electrons from the background plasma towards the target. At the same time, the hot electrons penetrate the background plasma layer and form a new inhomogeneous layer above it, acquiring an anisotropic velocity distribution. Such a new hot layer is subject to the Weibel instability, leading, in particular, to the formation of $z$-pinches (i.g., current sheets in 2D modeling) and associated magnetic field that differ in structure and orientation from the $z$-pinches (current sheets) and their field in the background plasma layer, which is also subject to the Weibel filamentation-type instability with the formation of counter currents of hot and cold electrons. A qualitative analysis and a number of experiments show that the transition region (i.e., both plasma layers described above) is filled with electron current filaments of different kinds. As a result, strong (mega-Gauss) small-scale magnetic fields emerge in the this region \citep{Stepanov22_TVTen, Borghesi1998, Chatterjee2017, Stepanov2018_LO2018, Stepanov20_LO20}. 

The transient process of the formation of current structures develops as follows (for details, see subsections~\ref{sec:chWeibInj:Model} and~\ref{sec:chWeibInj:Interpret}).
At the very early, subpicosecond, stage, the injection of hot electrons triggers the filamentation Weibel instability. It is associated with a multi-stream anisotropic velocity distribution of electrons and results in the small-scale current structures located within a certain layer of appropriately dense cold plasma. These structures are nothing else as the current sheets lying along the injection direction and produced by counter streams of cold and hot electrons. The streams are consistent with a gradually increasing drop of the electrostatic potential in the expanding cloud of nonequilibrium plasma~\citep{Nechaev20_FPen}. This self-consistent large-scale electric field slows down and then returns the runaway hot electrons back.

At a later, picosecond, stage (and a greater distance from the target), the hot electrons, anisotropically cooled during expansion, accumulate in significant numbers in a layer of more rarefied plasma (see sections~\ref{ch:raspad},~\ref{ch:chWeibMagn}).
Here the anisotropy of electron velocity distribution has another type, resembling a bi-Maxwellian, so the type of the Weibel instability changes to what we call thermal. The direction of the highest electron temperature is parallel, not orthogonal, to the target and oriented along the laser-heated strip. Hence, in the vicinity of previously emerged current structures, current filaments resembling $z$-pinches form in the orthogonal direction. They have a larger spatial scale and gradually occupy an increasingly larger part of the transition layer where the hot electrons dominate. If the injection of hot electrons from a rather large area on the target is sufficiently long, a chaotic growth of the ensemble of current filaments and their subsequent deformation are possible. So, an inhomogeneous quasi-magnetostatic turbulence is developed in the upper part of the expending cloud.

In the layer of denser plasma, after the hot-electron injection ends (in actual experiments it can last for up to few nanoseconds), the initially formed small-scale current sheets are destroyed quite rapidly, within a few picoseconds. (To speed things up, in the simulations the injection was limited to several picoseconds.) Later on, in a less dense plasma, Current filaments have a larger spatial scale and can exist longer, for times greater than tens of picoseconds.  
The magnetic fields created by the current structures are so strong that they lead to a pronounced stratification of the plasma density and anisotropy of electron velocity distribution. Thus, these fields change the collisionless kinetics of particles significantly, although a shock wave is not formed for the chosen geometry of inmhomogeneous plasma and a low injection rate of hot electrons. 
The kinetics of ions, which are cold, is taken into account in the simulations, but could have a significant effect only on longer, subnanosecond stage. Below we discuss only picosecond times, when the processes involving electrons dominate.

\subsection{Local injection of plasma containing hot electrons into an inhomogeneous layer of background cold plasma}
\label{sec:chWeibInj:Model}

As in sections~\ref{ch:raspad} and~\ref{ch:chWeibMagn}, we base the subsequent analysis on the 2D3V simulations of the simplest quasi-2D problem in which the simulation domain lies in the $xy$ plane of the Cartesian coordinate system and the spatial dependence of all quantities on the $z$ coordinate is absent (for details, see~\citep{Garasev22_GA_InjEn}).

A low-energy prepulse, preceding the main femtosecond laser pulse, produces a relatively cold background plasma that is modeled by about a billion macroparticles (singly charged ions and electrons with masses $m_\mathrm{i} = 100\,m_\mathrm{e}$ and $m_\mathrm{e}$, respectively) and fills the entire simulation domain.
At the initial moment, thei plasma temperature is $T_0 = 250$~eV, and the number density $N$ decreases with distance from the target surface (coinciding with the plane $y = 0$) in accord with the exponential law, $N (y) = N_0 \exp(-y / L)$. Let a maximum number density be $N_0 = 1.7 \times 10^{22}$~cm$^{-3}$ and a characteristic scale $L$ be in the range $16$--$32$~$\mu$m for different calculations.
Hot electrons with an initial temperature $T = 100$~keV and cold ions with a temperature $T_0$ are injected into the region, i.e., are generated on the target's surface $y = 0$, continuously starting from the moment the simulation begins and throughout the time $t_\mathrm{inj}$, which amounts to $2$--$5$~ps in different calculations. The injected particles have Maxwellian unbiased velocity distributions and a Gaussian number density distribution $n^*$ along the $x$ axis, $n^*(x, y=0) = n_0^* \exp\left( -x^2 / r_0^2 \right)$. The length $2 r_0 = 25$~$\mu$m characterizes the transverse size of a long strip on the target's surface, from each point (i.e., a simulation cell) of which hot electrons (and cold ions), created as a result of the cylindrical focusing of femtosecond laser, are isotropically emitted.
The plasma containing hot electrons is assumed to be highly rarefied compared to the background plasma at the surface of target. The ratio of their number densities $n_0^* / N_0$ is chosen within the range of $0.03$\,--\,$0.003$, so that the densities of hot and cold electrons turn out to be comparable only at a distance greater than or about $100$~$\mu$m from the target.

On all of the simulation domain's boundaries, open boundary conditions for the particles and absorbing conditions for fields are adopted, i.e., all particles and wave fields freely leave the region and no longer influence the processes inside it. The dimensions of the region along the $x$ and~$y$ axes are equal to $120$~$\mu$m and $200$~$\mu$m, respectively. The simulation grid consists of $2400 \times 4000$ cells, so the size of each of them is significantly smaller than the Debye radius, calculated from the temperature and density of hot plasma electrons near the target and amounting to $0.1$--$0.3$~$\mu$m depending on the ratio $n_0^* / N_0$.
In plasma regions with a density $\sim 10^{20}$~cm$^{-3}$, where the current structures arise, even conventionally  cold background electrons with an energy of about $250$~eV have a mean free path of about tens of microns (see, for example,~\citep{Trubnikov1965}) that is greater than the characteristic dimensions of current structures.
This makes the collisionless plasma approximation correct for describing the processes of Weibel instability under consideration, including its nonlinear stage.

\begin{figure}[b]
	\includegraphics[width=1.0\textwidth]{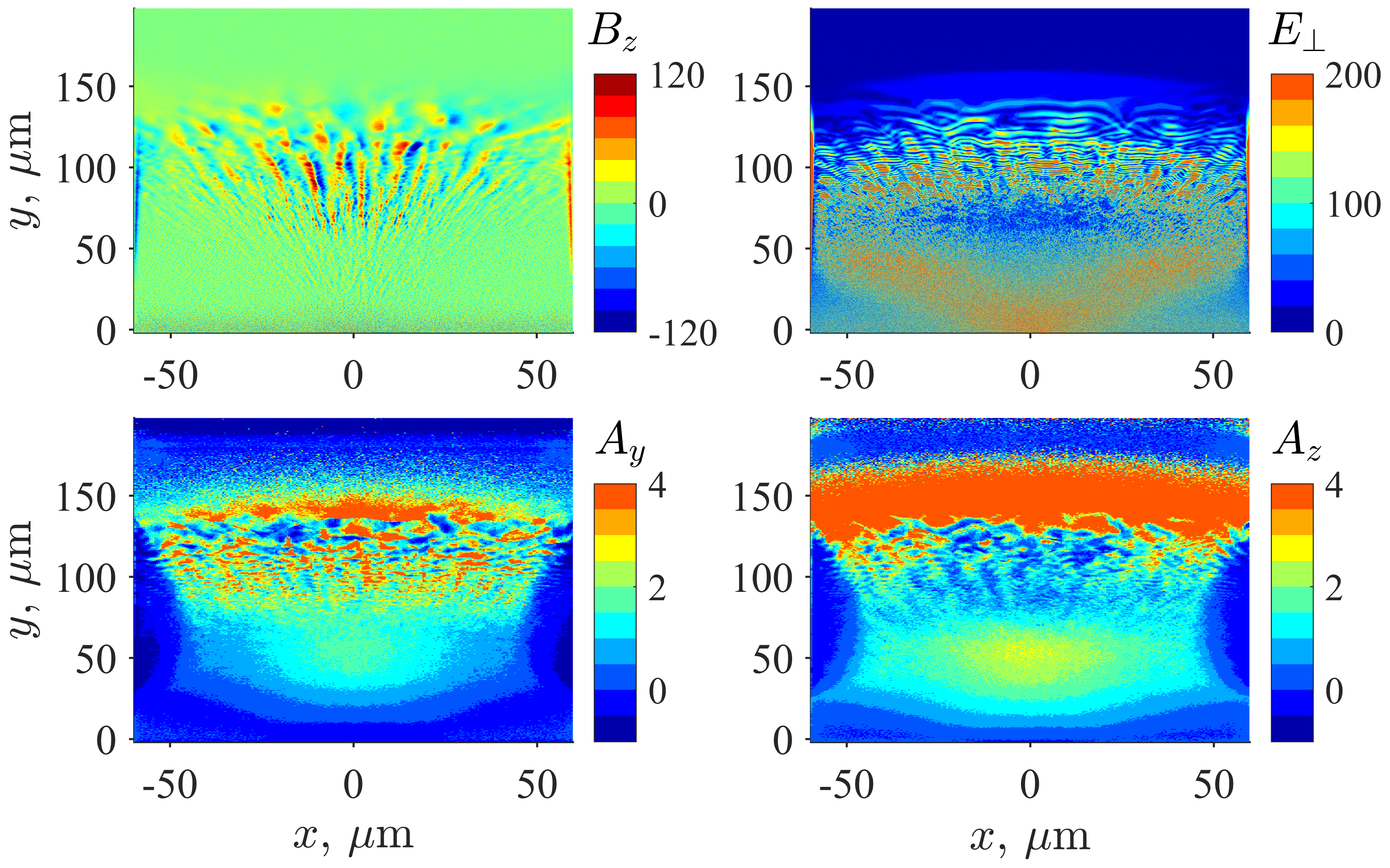}
	\centering
	\caption{
		Distributions of the field component $B_z$ (top left panel, in Teslas), the magnitude of the in-plane electric field $E_\perp$ (top right, in units of $3 \times 10^8$~V/m), and the anisotropy parameters $A_y = T_y / T_x - 1$ (bottom left) and~$A_z = T_z / T_x - 1$ (bottom right) at the moment $t = 0.7$~ps after the start of the injection of a plasma containing hot electrons into the background plasma with inhomogeneity scale of $L = 16$~$\mu$m at $n_0^* / N_0 = 0.03$.
	}
	\label{fig18}
\end{figure}

\begin{figure}[b]
	\includegraphics[width=0.9\textwidth]{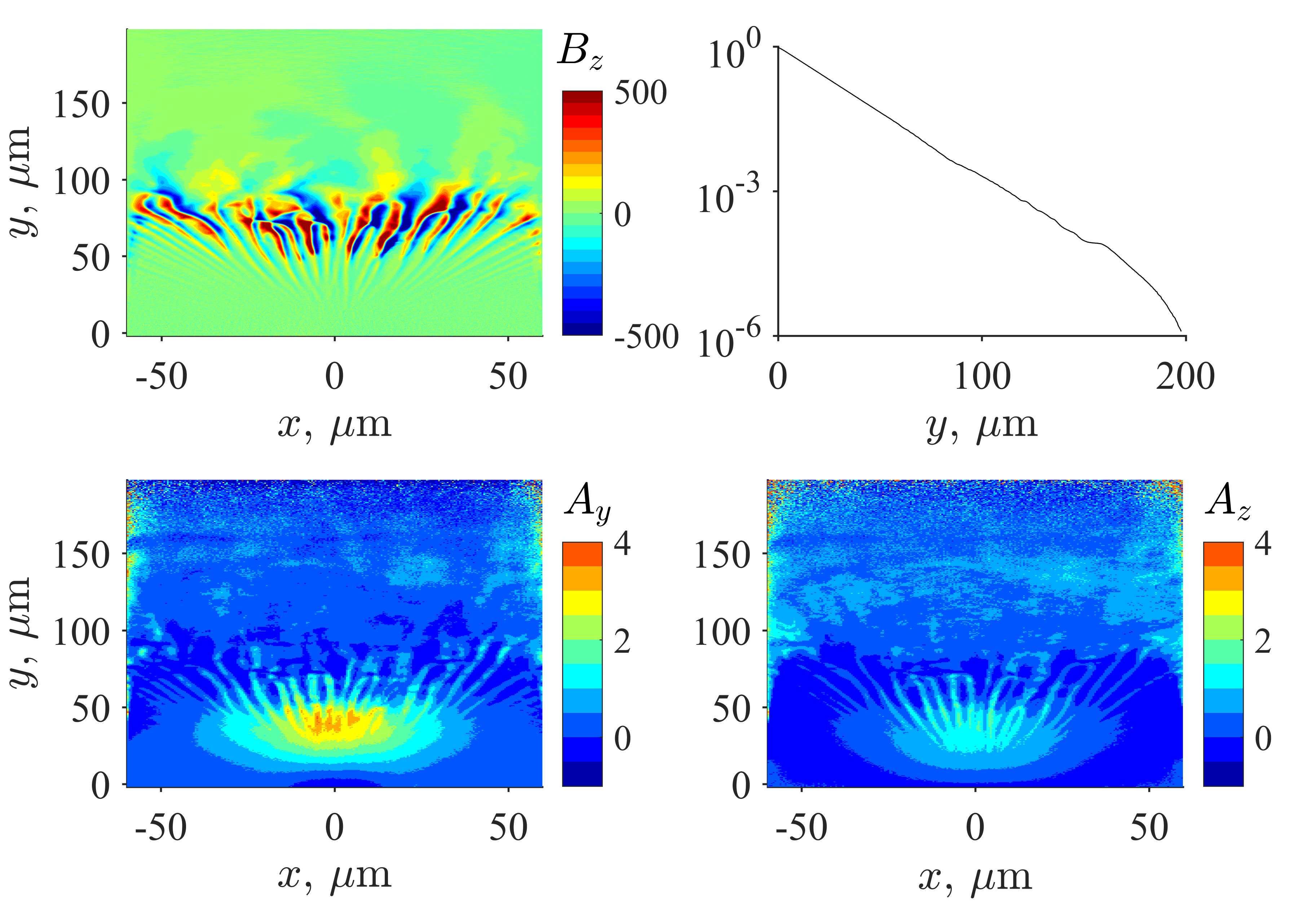}
	\includegraphics[width=0.9\textwidth]{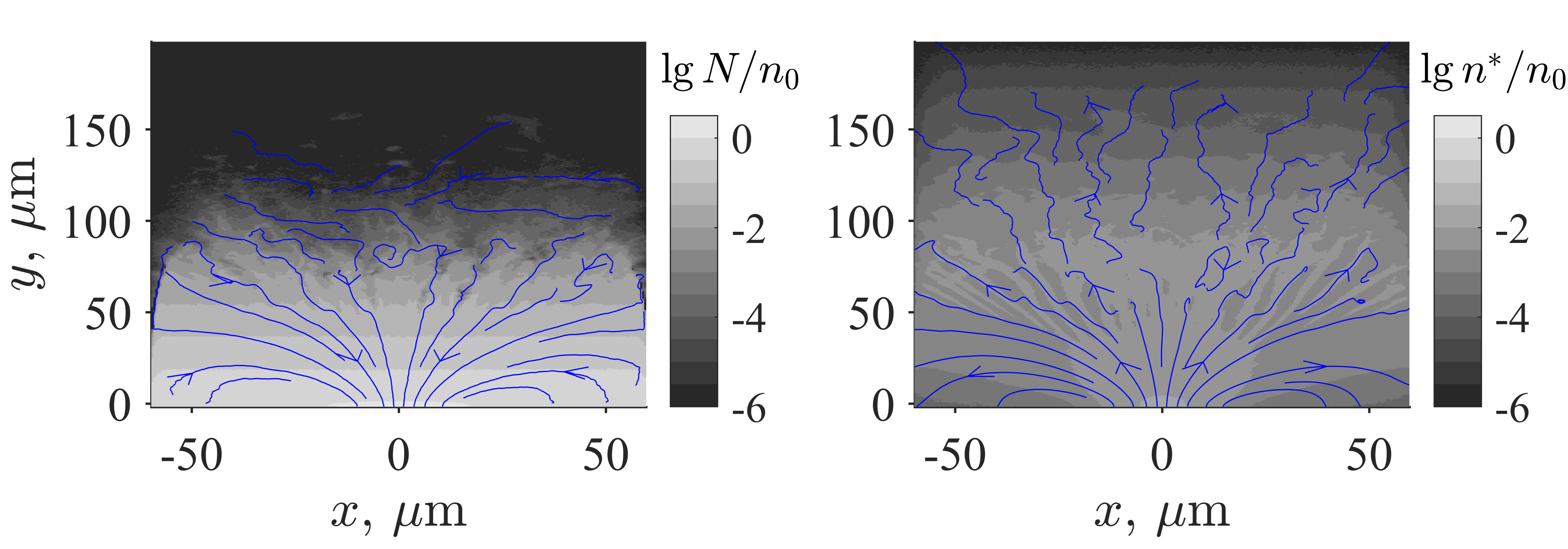}
	\centering
	\caption{
		The magnetic field's component $B_z$ (top left panel, in Teslas), the plasma number density profile $n(y) / n_0$ (top right), the anisotropy parameters $A_y = T_y / T_x - 1$ (middle left) and~$A_z = T_z / T_x - 1$ (middle right), and the number densities of cold electrons $N$ (bottom left) and injected hot electrons $n^*$ (bottom right) at the moment $t = 1.8$~ps.
		Bottom panels show the logarithms of the normalized number densities of cold and hot electrons, their streamlines are shown in blue. 
        Plasma parameters are the same as in Fig.~\ref{fig18}.
	}
	\label{fig19}
\end{figure}

\begin{figure}[t]
	\includegraphics[width=0.75\textwidth]{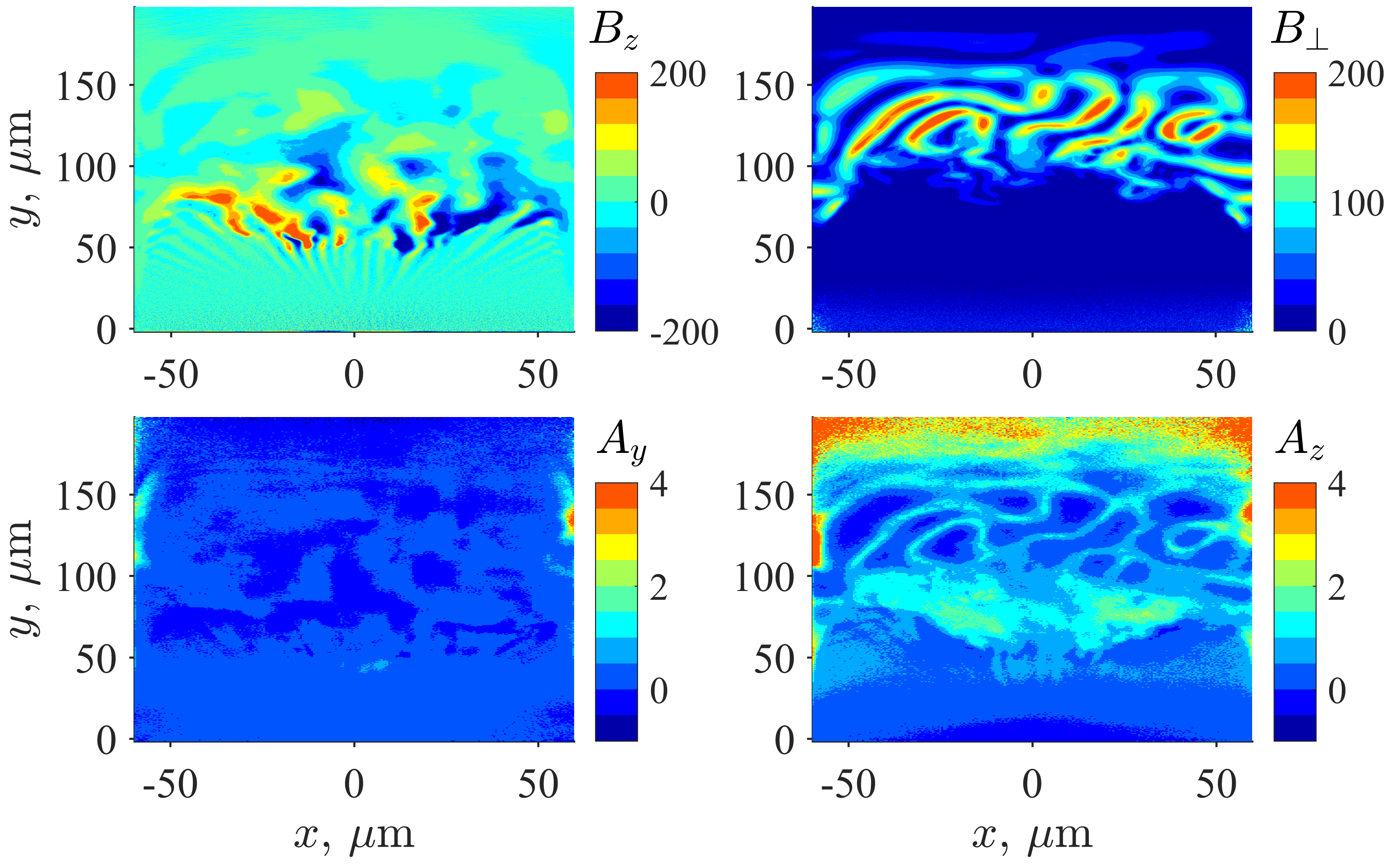}
	\centering
	\caption{
		The magnetic field's component $B_z$ (top left panel, in Teslas), the in-plane field $B_\perp$ (top right, in Teslas), the anisotropy parameters $A_y = T_y / T_x - 1$ (bottom left) and~$A_z = T_z / T_x - 1$ (bottom right) at the moment $t = 3.6$~ps after the start of the injection that lasted for $2$~ps. 
		Plasma parameters are the same as in Fig.~\ref{fig18}.
	}
	\label{fig20}
\end{figure}

\begin{figure}[t]
	\includegraphics[width=0.75\textwidth]{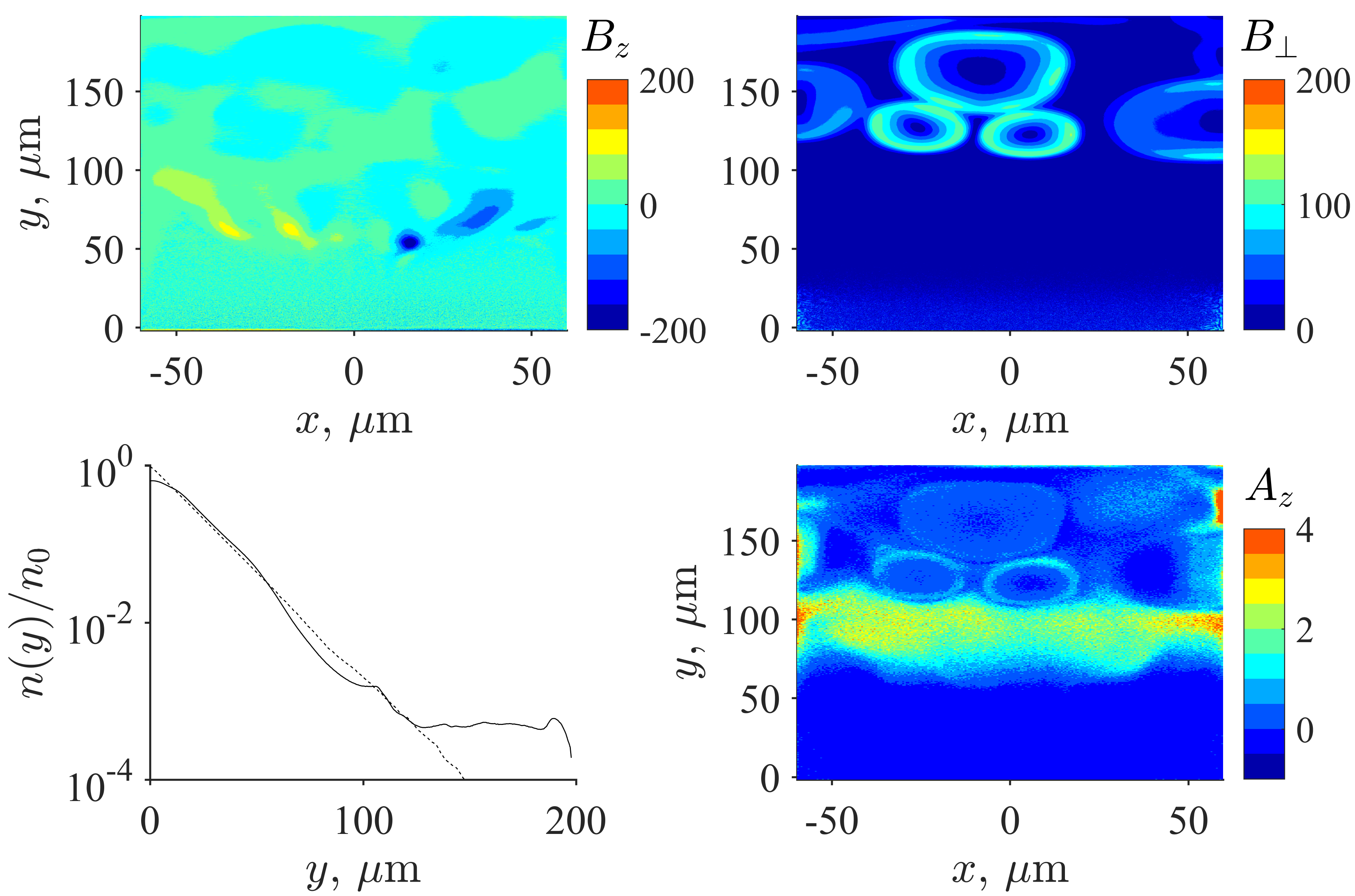}
	\centering
	\caption{
		The magnetic field's component $B_z$ (top left panel, in Teslas), the in-plane field $B_\perp$ (top right, in Teslas), the plasma number density profile $n(y) / n_0$ (bottom left, solid line for the moment $t = 10$~ps, dashed for $t = 1.8$~ps) and the anisotropy parameter $A_z = T_z / T_x - 1$ (bottom right) at the moment $t = 10$~ps after the start of the injection that lasted for $2$~ps. 
		Plasma parameters are the same as in Fig.~\ref{fig18}.
	}
	\label{fig21}
\end{figure}

For the following set of parameters, $L = 16$~$\mu$m, $t_\mathrm{inj} = 2$~ps, $n_0^* / N_0 = 0.03$, the results of calculations of the spatial distributions of the field components $B_z$ and $B_\perp = \left( B_x^2 + B_y^2 \right)^{1/2}\!$, anisotropy parameters $A_{y,z} = T_{y,z} / T_x - 1$, which determine the type and parameters of Weibel instability, as well as some other quantities are given in Figs.~\ref {fig18}--\ref{fig21} at four consecutive moments after the injection onset: $0.7$, $1.8$, $3.6$, $10$~ps. The local dispersion of the electron velocity components (cold and hot together) are characterized by the calculated effective temperatures $T_{x,y,z}$ (in energy units).
Based on this example, we briefly outline here the main stages of the transient process under study and physical phenomena which determine them. Important details of this process are discussed in the next subsection~\ref{sec:chWeibInj:Interpret} which involves calculations for other parameters of the problem.

From the very beginning of injection, a hot flow of radially expanding electrons penetrating a dense cold plasma causes a known beam-type instability and generation of a rapidly oscillating electric field $\vec{E}_\perp$. It is directed along the local flow velocity (in the $xy$-plane) and spatially modulated with a period of up to several microns (see~module $E_\perp = \left( E_x^2 + E_y^2 \right)^{1/2}$ in the upper right panel of Fig.~\ref{fig18}). Yet, these turbulent plasma waves are too short of a time to grow significantly and convert a significant portion of the energy of the directed motion of the scattering electrons into thermal energy. The point is that Weibel filamentation instability develops almost equally fast already at subpicosecond times in a wide layer of inhomogeneous dense plasma at a distance $\sim 30$--$100$~$\mu$m from the target. It is associated with the aperiodic generation of a magnetic field directed orthogonal to the $xy$ plane, which has azimuthal modulation (top left panel in Fig.~\ref{fig18}). Thus, counter flows of hot electrons moving radially from the region of injection and cold electrons moving towards it (as well as partially returning hot ones) are separated in space. They form alternating wedge-shaped current sheets located in the vicinity of neighboring magnetic field minima. For the chosen parameters, this process becomes essential at times greater than $\sim$$0.3$~ps.

In such a very regular ''fan'' structure, the total current is small, and the transverse dimensions of the sheets increase from approximately $1$ to $10$~$\mu$m with distance from the center of the injection region on the target's surface within the distance range from approximately $15$ up to $150$~$\mu$m. Due to the nonlinear effects of saturation of Weibel instability under the action of the created strong magnetic field (its value is of up to $120$~T), the indicated thicknesses of the current sheets are of the order of local gyroradii of hot electrons. The related stratification of the anisotropy parameter $A_y$ responsible for this instability is shown in Fig.~\ref{fig18}, bottom left panel.
The degree of anisotropy $A_z$ shown in the lower right panel in Fig.~\ref{fig18} is also nonzero. However, in the region of space under consideration it is less than $A_y$ and increases later, when the magnetic field $B_z$ there becomes already high. The latter, apparently, suppresses the Weibel instability corresponding to $A_z > 0$.
Note that in a fully 3D simulation, where inhomogeneity along the $z$ axis is allowed, the filamentation Weibel instability corresponding to $A_y > 0$ would lead to formation of vertical current filaments instead of sheets (see, e.g.,~\citep {Huntington2015, Ruyer2020, Peterson2021}).

In the further evolution of the injected plasma at times longer than $1$~ps, the aforementioned separation of the radial flows of hot and cold electrons (see Fig.~\ref{fig19}, lower panels) is preserved only in a gradually narrowing layer (located at~distances up to approximately $100$~$\mu$m at $t = 1.8$~ps according to Fig.~\ref{fig19}, upper left panel).
This process is accompanied by a gradual increase in current density and generated magnetic fields in radially inhomogeneous current sheets.
In this case, the hot electrons which have escaped upward, reach a region of fairly rarefied background plasma, where they are comparable in number density with cold electrons and displace some of them downward.
As a result, a neighboring layer is formed above the aforementioned layer with dominant hot electrons, which have experienced partial anisotropic cooling in the cause of expansion in the $xy$ plane. Thus, they reduce their effective temperatures $T_{x,y}$, keeping the initial temperature $T_z$ practically unchanged due to the spatial homogeneity of the injection of hot electrons and the number density of cold background ones in the direction of $z$ axis. 
The boundaries of this new layer are visible for a long time in the plasma density profile (see the inflections on it at $y \sim 90$~and $150$~$\mu$m on the upper right panel in Fig.~\ref{fig19}, and also Fig.~\ref{fig21} in~section~\ref{sec:chWeibInj:Interpret}).
Inside a significant part of the layer, at times longer than or on the order of $2$~ps, the anisotropy level $A_z$ begins to exceed the level $A_y$ (cf.~middle panels of Fig.~\ref{fig19}), even if the hot electrons are still injected continuesly (see subsection~\ref{sec:chWeibInj:Interpret}). After the injection stops (in the present case, at the same moment of time $2$~ps), its dominance establishes even more quickly.

Thus, along with the previously appeared structure of radial current sheets, a new set of filaments resembling $z$-pinches orthogonal to the sheets, oriented along the $z$ axis, is formed. These filaments arise as a result of the development of a novel stage of Weibel instability, dynamics of which is no longer associated with cold background electrons. It is caused by a slow increase in the anisotropy parameter $A_z$ above a certain threshold value.
This process is similar to that discussed in sections~\ref{ch:raspad}, \ref{ch:chWeibMagn}.
As shown in Fig.~\ref{fig20} for the moment $t = 3.6$~ps, the scales of such $z$-pinches, amounting to $10$--$50$~$\mu$m, strongly exceed the thickness of the underlying current sheets. These scales are comparable with the initial scale of the inhomogeneous background plasma and with the local gyroradius of a hot electron in the instability-saturating transverse field $B_\perp$, the average value of which is $150$~T, reaching $200$~T in certain places.
Note that in this region there is also a noticeable longitudinal field $B_z$ with a magnitude of up to $100$~T in absolute value, and, consequently, large-scale azimuthal and radial currents.

According to calculations at long times, after the injection stops, radial currents, especially underlying small-scale ones, decay faster than currents directed along the~$z$ axis. In this case, the anisotropy $A_y$ is relatively weak and insignificant, while the anisotropy $A_z$ is still large and correlates (cf.~section~\ref{ch:raspad}) with the distribution of the transverse field $B_\perp$, and is significant even in the underlying layer of sufficiently dense background plasma ($50$~$\mu$m~$\lesssim y \lesssim 100$~$\mu$m), where the densities of cold and hot electrons are comparable.

The upper panels in Fig.~\ref{fig21} represent a significantly weakened magnetic field which consists of two current structures, already far from being orthogonal, at an even later stage of their decay, $t = 10$~ps. (According to the left lower panel, at such times the plasma density profile changes quite strongly.)
The longer existence of an upper ensemble of slowly deforming $z$-pinches is due to the nonlinear effects of the capture of some hot electrons in the regions of strong self-consistent magnetic field. There are few cold electrons inside an individual pinch, and the current of hot ones there can even be directed opposite to the total current near and outside its boundaries, where the cold electrons can contribute to the current strongly and the magnetic field $B_\perp$ varies sharply.
Fig.~\ref{fig21} also shows a strong magnetic field $|B_z| \sim 100$~T (with opposite signs of $z$-projections) within several long-lived localized regions, formed by currents in the $xy$ plane. They remain after the collapse of the ensemble of radial current sheets and form the overall large-scale structure. The listed localized long-lived current formations, due to the pressure of the strong magnetic field they create, lead to a noticeable stratification of the total plasma density.
Discussion of such details of the long-term nonlinear evolution of the considered transient process caused by the injection of hot electrons is beyond the review's scope.

\subsection{Self-consistent current structures and magnetic fields}
\label{sec:chWeibInj:Interpret}

Suppose a target contains a laser-heated strip which plays a part of an injector of electrons and ions. Let the escaping hot electrons move inside the non-uniform cold plasma adjoining the target. Consider the current structures created in such a transient process. It turns out that, in a wide range of problem parameters, their simulation yields structures very similar to the ones described above.
The first to form is a completely regular structure of wedge-shaped sheets with alternating directions of currents consisting of hot and cold electrons. Next, a chaotic structure of currents of hot electrons flowing in opposite directions orthogonal to these sheets, along the heated strip, and similar to $z$-pinches grows on top of it. They slowly deform and fly apart together with the expansion of the whole plasma cloud containing hot electrons.

The characteristics of these mutually orthogonal current structures are consistent with the hypothesis about the origin of both as a result of the electron instability of the Weibel type, filamentation and thermal, respectively (see overview in the Introduction, section \ref{sec:intro}).
A well-known analytical theory of its linear phase~\citep{Weibel1959, Davidson1989, Vagin2014, Kocharovsky2016_UFN, Silva2021} for a uniaxial bi-Maxwellian or two-stream~\citep{Ruyer2015_NonlinWeibel} electron velocity distribution with a maximum effective temperature $T_{y,z}$ along the axis~$y$ or~$z$ shows that magnetic field perturbations with wave vectors orthogonal to this axis have the maximum growth rate $\omega_0$. For a moderate anisotropy, when the parameter $A_{y,z} = T_{y,z} / T_x - 1$ does not significantly exceed unity, one can use approximate expressions for the optimal wavelength of disturbances $\lambda_0 \sim 10 \, c / \omega_\mathrm{p} \, (n_0 / n)^{1/2} A_{y,z}^{-{1/2}}\!$ and growth rate $\omega_0 \sim 3 \, (T_x / m_\mathrm{e}) ^{1/2} \lambda_0^{-1} (1 + A_{y,z}^{-1})^{-1}$, which were already given in subsection~\ref{sec:chWeibExp:correl}.
Here, for simplicity's sake, the wave vectors are considered to be directed along $x$ axis (i.e., the direction corresponding to the lowest temperature), plasma frequency $\omega_\mathrm{p} = (4\pi e^2 n_0 / m_\mathrm{e})^{1/2}$ is defined by the density of all electrons $n_0 \approx 1.7 \times 10^{22}$~cm$^{-3}$ at the center of heated part of the target surface, $n$ is the local electron number density within the region where the instability develops. The plasma period is $2\pi / \omega_\mathrm{p} \approx 1$~fs.

For estimates, we take a local value of the total electron number density $n$ near the places of maximum values of the anisotropy parameters $A_y$ and $A_z$ during the nucleation of two related orthogonal current structures depicted in Figs.~\ref{fig18} and~\ref{fig20}, at about the same distance from target, such as $y \approx 100$~$\mu$m.
For the first structure, at $t = 0.7$~ps, the value of $A_z$ there is about $A_y/2 \approx 1$ and the hot electron density is still small compared to the cold electron density $N \approx 3 \times 10^{ 19}$~cm$^{-3}$. So, the temperature is $T_x \sim 10$~keV and $\lambda_0 \sim 7$~$\mu$m, $\omega_0^{-1} \sim 80$~fs.
For the second structure, at $t = 3.6$~ps, one has $A_y \approx 0$ and $A_z \approx 0.5$, the total density is still the same $n \approx 3 \times 10^{19}$~cm$^{-3}$ (its profile changes little here, see Fig.~\ref{fig21}), but hot electrons already noticably prevail over the cold ones. So, the temperature is $T_x \sim 35$~keV and $\lambda_0 \sim 15$~$\mu$m, $\omega_0^{-1} \sim 170$~fs.
In both cases, formation of current structures takes several characteristic times $\omega_0^{-1}$, and the estimate of $\lambda_0 / 2$ (see figures) is in agreement with a width of current sheets, which is about $3$~$\mu$m. The diameter of $z$-pinches equals to approximately $10$~$\mu$m. 
In addition, the local gyrofrequency $\omega_{B} = e B / (m_\mathrm{e} c)$ and gyroradius $r_\mathrm{L} = (2 T_x / m_\mathrm{e})^{1/2} \omega_{B}^ {-1}$ of a representative electron in a typical magnetic field $B$ of an emerged current structure, amounting to $120$~T and $150$~T in the first and second case, respectively, are of the order of $\omega_0$ and $\lambda_0 / 2$, respectively. So, the following relations hold: $\omega_{B} \approx 2 \omega_0$, $r_\mathrm{L} \approx 0.4 \, \lambda_0$ in the first case and $\omega_{B} \approx 4 \omega_0$, $r_\mathrm{L} \approx 0.3 \, \lambda_0$ in the second. These approximate equalities are within the known criteria for saturation of Weibel instability (\citep{Kocharovsky2016_UFN}) and are verified also in sections~\ref{ch:raspad} and~\ref{ch:chWeibMagn}.

Importantly, the parts played by the aforesaid physical phenomena can significantly change with varying the parameters of the injection problem under study. As a result, the features of the hot and cold electron currents, anisotropic velocity distribution functions as well as created magnetic fields can also considerably change. Let us illustrate this fact with examples.

A twofold larger scale of the background inhomogeneity. $L = 32$~$\mu$m, yields about two times lengthening of the ''fan'' structure of alternating current sheets of cold and hot electrons. Simultaneously, their transverse dimensions decrease and the small-scale magnetic field $B_z$ weakens to a value of the order of $20$--$50$~T (cf.~Fig.~\ref{fig19} for the same moment of time $t = 1.8$~ps). The related spatial modulation of the anisotropy parameters $A_{y,z}$ in the dense background plasma also reduces its scale. Modulation of the effective temperature $T_x$ persists only in the places where the background plasma is more rarefied, meaning that its density is of the order of or does not greatly exceed the density of hot electrons.

After the injection stops, a cloud of mixed cold and hot electrons along with ions quickly expands in all directions. Thus, some hot electrons move towards the target, and~most cold electrons with all ions, on the contrary, move away from the target. Hence,~the picture of electron flows shown in the lower panels in Fig.~\ref{fig19} changes a lot. Namely, hot electrons quickly reach the zone of rarefied background plasma and experience anisotropic cooling. They dominate there over the isotropic cold electrons. As a result, a multiple formation of the current filaments resembling $z$-pinches takes place as soon as a sufficiently high layer accumulates the number of hot electrons required for switching on the Weibel instability. Then these $z$-pinches begin gradually deform due to the inhomogeneity and expansion of the plasma and create a consistent, highly inhomogeneous transverse field~$B_\perp$, complementing the former inhomogeneous longitudinal field~$B_z$. Such a scenario takes place also in other simulation examples corresponding to different parameters of the background cold and injected hot plasmas. 

\begin{figure}[!b]
	\includegraphics[width=0.9\textwidth]{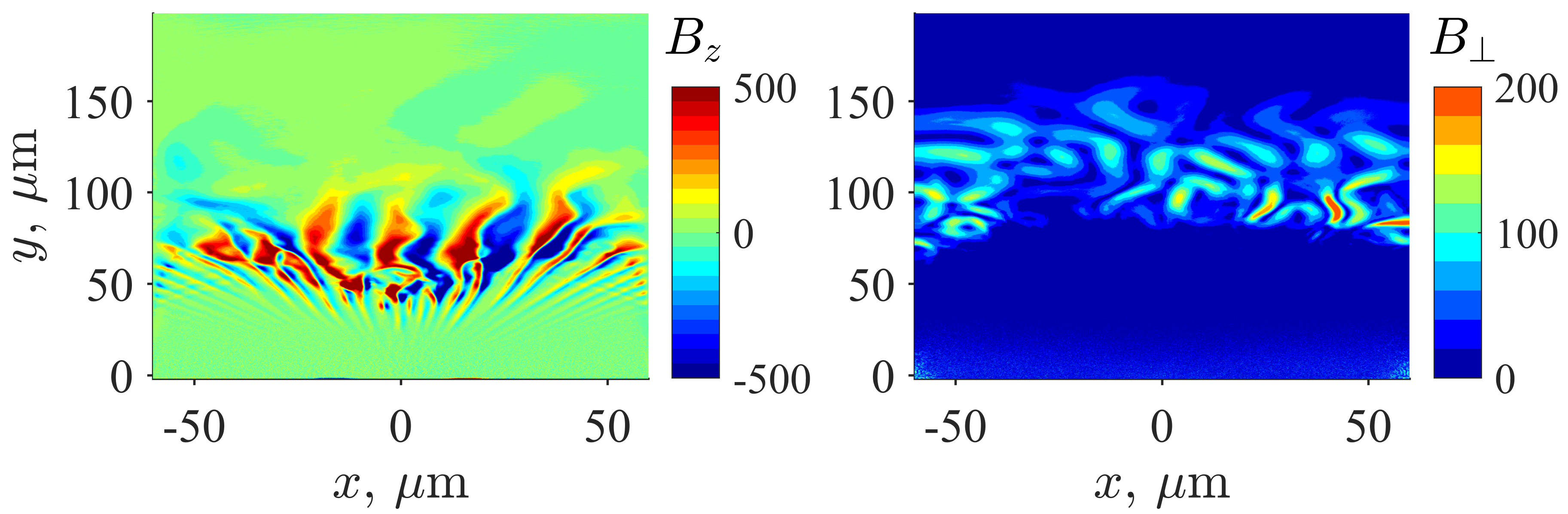}
	\centering
	\caption{
		The magnetic field component $B_z$ (left panel, in Teslas) and in-plane magnetic field $B_\perp$ (right, in Teslas) at the moment $t = 3.6$~ps after the start of injection that lasted $5$~ps. Other plasma parameters are the same as in Fig.~\ref{fig18}.
	}
	\label{fig22}
\end{figure}

The emerging nontrivial two-component picture of magnetic fields and currents, certainly, depends on the injection duration and, therefore, the number of hot electrons injected, as is exemplified by comparing Fig.~\ref{fig20} against Fig.~\ref{fig22}.
Both figures are plotted for the same parameters (see subsection~\ref{sec:chWeibInj:Model}) at the same moment $t = 3.6$~ps, but for different injection duration, $t_\mathrm{inj} = 2$~ps and $t_\mathrm{inj} = 5$~ps, respectively. With an increase in the number of injected electrons, the magnitude of transverse field $B_\perp$ and the scale of its inhomogeneity in the overlying layer decrease, while the field $B_z$ with a ''fan'' structure in the underlying layer survives in virtue of continued injection of the hot electrons and the oncoming flow of cold background ones, almost nullifying the total current.

If, for the same parameters as in subsection~\ref{sec:chWeibInj:Model}, the intensity of plasma injection is significantly reduced to $n_0^* / N_0 = 0.003$, maintaining the same injection time $t_\mathrm{ inj} = 2$~ps and, hence, reducing the number of particles injected from the target, then, say, at a time moment twice as long as the injection time, both orthogonal structures of currents modulated in space appear weaker and of a larger scale. The magnetic field they create behaves accordingly and reaches on average only a relatively small value $30$--$50$~T (Fig.~\ref{fig20}).
At the same time, the fountain current of hot electrons comes to the fore (see section~\ref{ch:raspad}), creating a global field oriented mainly along or against $z$ axis in the left and right halves of expanding plasma, respectively. It is maximum in the gap between the plasma layers under discussion, where the densities of cold and hot electrons are comparable and the horizontal components of their fountain currents are considerable. Such a fountain structure is clearly visible in Fig.~\ref{fig20}.

In general, according to simulations, the two Weibel instabilities considered above and the corresponding mutually orthogonal structures of current sheets and $z$-pinches do not appear if the fraction of hot electrons is too small, $n_0^* / N_0 < 10^{-3}$, or their injection is too short, $t_\mathrm{inj} < 0.3$~ps, or the background plasma is too homogeneous, $L > 100$~$\mu$m.
In the latter case, with a sufficiently long injection of a sufficiently dense plasma with hot electrons, instead of the phenomenon considered here, a collisionless shock wave is formed~\citep{Lyubarsky2006, Garasev2016, Fox2018, Moreno2020, Nishigai2021, Kropotina2023}.

Finally, note that the figures presented in this section demonstrate the global symmetry of magnetostatic structures relative to the~$y$ axis. Of course, in small details the right and left parts of each figure differ slightly due to the noise amplification inherent to random nature of Weibel instability.
According to the simulation discussed in section~\ref{ch:chWeibMagn}, a not too strong external field with a sufficiently uniform $B_z$ component, which does not completely preclude Weibel instability, would violate the stated symmetry and slow down to a certain extent the growth of current filaments, in particular, preventing formation of the filaments parallel to the direction of hot electron injection in $xy$ plane, and to a lesser extent the filaments parallel to the~$z$ axis.
An external field of not too great a magnitude, the field lines of which mainly lie within the $xy$ plane and are mostly directed radially from the injection spot into the rarefied surrounding plasma, in contrast, would strongly interfere with the formation of current filaments resembling $z$-pinches, but just slightly suppress or even promote formation of radial current structures.
An external field, the field lines of which are mainly orthogonal to the above radial direction in $xy$ plane, that is, cover the electron injection region like semicircles, would preclude formation of current structures of both kinds.
Such effects of an external field on the expansion of plasma subjected to particle injection are analyzed in the next section.

\section{Effect of an external field on the generation of strong small-scale magnetic fields in the course of plasma injection into a cold plasma layer}
\label{ch:DAN23}

\subsection{Particle injection into a magnetized background plasma}
\label{sec:DAN23:init}

Let us now consider the case of combined presence of both background plasma and an external magnetic field, which is realized, as in the previous section, for a laser plasma created by ablation of the target with a femtosecond pulse having a prepulse. 
This section is focused on the effect of an external homogeneous magnetic field $B_0$ on a similar process of filamentation of the current density $j(r,t)$ and generation of small-scale quasi-magnetostatic structures. The number density of cold background plasma with a temperature $T_0 < 1$ keV is considered to decrease with distance from the target surface (along the $y$ axis) according to the exponential law $n_0 \exp (-y/L)$ with a characteristic scale $L \sim 10$~$\mu$m and a near-surface value of the order of $n_0 = 10^{21}$~cm$^{-3}$. Again, we have in mind the cylindrical focusing of a femtosecond laser pulse onto a target. When heated, a narrow strip of a width $2 x_0$ (from several to tens of microns) for a long time $t_\mathrm{inj}$ (picoseconds or more) ejects a Maxwellian plasma which contains fairly cold ions (with the temperature $T_0$) and very hot electrons with a temperature $T$ up to hundreds of keV. On the considered scales of the order of tens of microns, for these electrons Coulomb collisions are rare and the plasma is actually collisionless. To be specific, we assume that its number density is distributed along the $x$ axis according to the law $0.3 n_0 \exp [-(x/x_0)^2]$. 

We choose the external field in the range 5--500 T and consider it being parallel to the target $xz$ plane and oriented either along the hot strip ($z$ axis) or orthogonal to it ($x$ axis), according to the end of subsection~\ref{sec:chWeibMagn:weak}. Above the specified maximum value of field, satisfying the condition $B_{0\mathrm{max}} \sim (n_0 T)^{1/2}$, i.e., lying near the magnetization boundary of hot electrons, the expansion of plasma is strongly suppressed and the generation of currents and magnetic fields practically does not occur. For the field values less than the specified minimum value, the hot electron gyroradius significantly exceeds the scale of plasma inhomogeneity and the influence of the magnetic field turns out to be insignificant.

The PIC-modeling is carried out with several billion macroparticles in a numerical box with the side lengths $L_x$, $L_y$, $L_z$ in the range of 60--200 $\mu$m for a duration of $t_\mathrm{sim} = 5$--$10$~ps. For uniformity in all calculations, the injection time $t_\mathrm{inj} = 2$~ps is taken equal by the order of magnitude to the time of flight of a hot electron through the numerical box, the walls of which across the $x$ and $y$ axes are open (along the $z$ axis the boundary conditions are periodic). In this case, the total number of injected electrons is of the order of the initial number of cold electrons in the background plasma layer. Below, to illustrate rich physics of the processes under consideration, both simplified 2D3V (with the spatial dependence on the $z$ coordinate being neglected) and fully 3D3V calculations are used; their results are qualitatively similar for each case examined.

Let's briefly list the mechanisms and main stages of the formation of the current configurations and their magnetic fields, indicating the reasons for their dependence on the external magnetic field directed along the target surface. Its orientation significantly affects the resulting structures and in the next two subsections two qualitatively different, orthogonal orientations are considered, similar to section~\ref{ch:chWeibMagn}.

The currents of the largest scale observed in this problem are the well-known ''fountain'' currents (see, for example, \citep{Albertazzi2015, Kolodner1979, Sakagami1979}) of hot electrons escaping from a strip of width $\sim 2 x_0$ from the target surface mainly upward along the $y$ axis  and at moderate angles to it. At a weak external magnetic field and a not very thick layer of the background plasma, they form a current sheet that expands along the $x$ axis with distance from the target and creates a magnetic field, which in the regions $x < 0$ and $x > 0$ is directed against and along the $z$ axis, respectively.

\subsection{The case of an external field parallel to the injection strip}
\label{sec:DAN23:parallel}

\begin{figure}[t]
	\includegraphics[trim = 2cm 0 1cm 0, clip, width=0.6\linewidth]{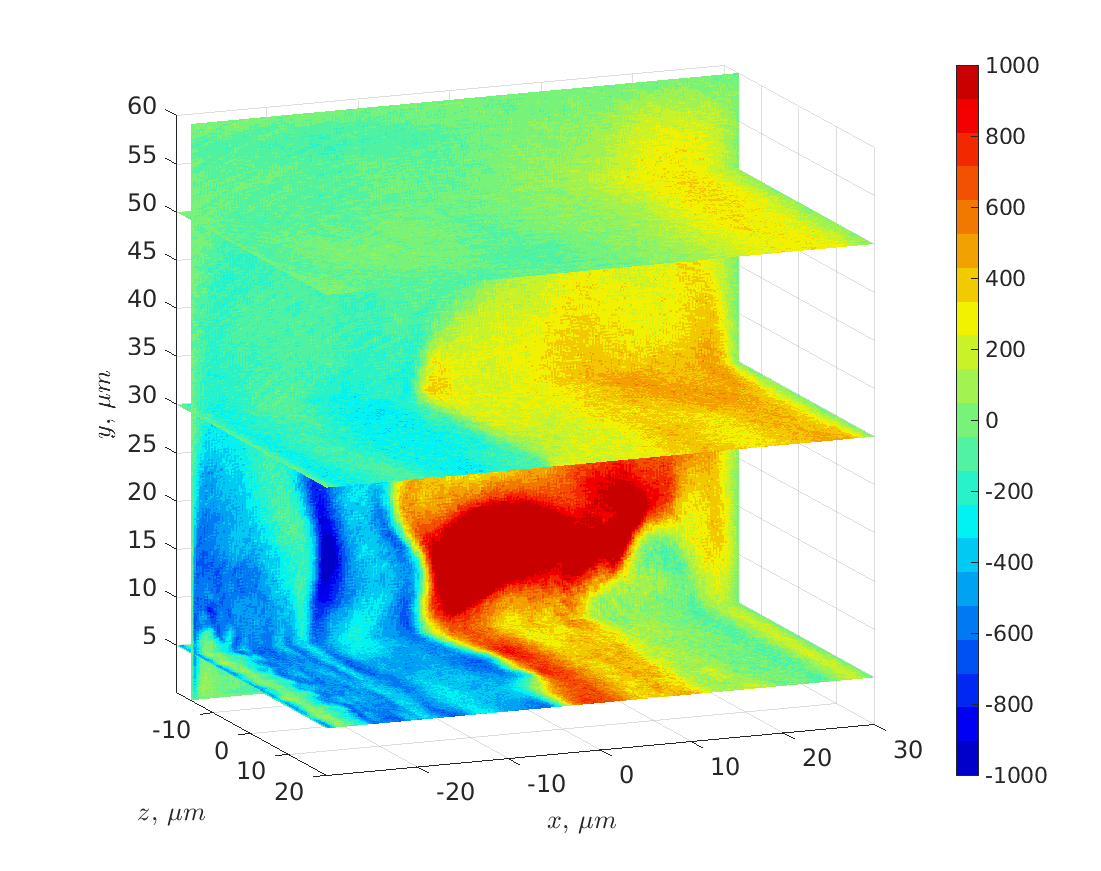}
	\centering
	\caption{
         Distribution of the field component $B_z - B_{0z}$ (in Teslas) in the 3D3V simulation at the time moment $t = 2$~ps with an external field $B_{0z} = 100$~T. For clarity, the color scale is limited by the value $\pm 1$~kT; in some regions the field value reaches $2$~kT.		
	}
	\label{fig23}
\end{figure}

\begin{figure}[t]
	\includegraphics[trim = 3.5cm 0 3cm 0, clip, width = 0.9\linewidth]{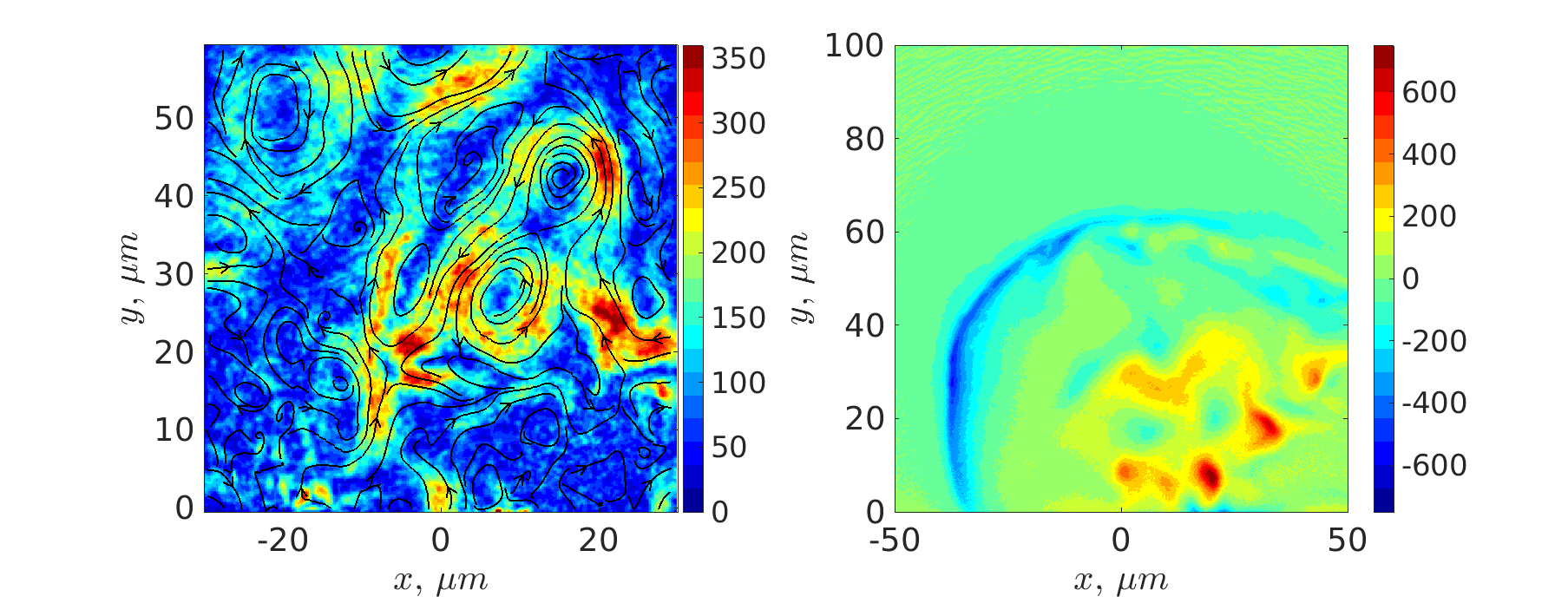}
	\centering
	\caption{
        Left panel: Distribution of the absolute value of transverse field, $\left(B_x^2 + B_y^2\right)^{1/2}$ (in Teslas), and its field lines as per 3D simulation at $t = 2.5$~ps in the plane $z = 30$~$\mu$m.
        Right panel: Distribution of the magnetic field component $B_z - B_{0z}$ in the 2D simulation at $t = 4$~ps.
        In both simulations, the external field is $B_{0z} = 100$~T.
	}
	\label{fig24}
\end{figure}

If the electrons are high-energy and the injection strip is narrow, then very quickly, as shown in Fig.~\ref{fig23} where $T = 300$ keV and $2 x_0 = 8$ $\mu$m, this kind of current sheet is formed in the presence of strong field oriented along the injection strip ($B_{0z} = 100$~T). This field to a certain extent even narrows the current sheet, of course, bending it by the Lorentz force, and creates in the regions $x < 0$ and $x > 0$ oppositely directed large-scale fields which exceed the external one markedly, even several times in magnitude in some localized regions (especially near the target, where current eddies can form). In the region with accumulating and anisotropically cooling electrons, due to Weibel instability, $z$-pinches and associated transverse magnetic field, also strong and oriented predominantly in the $xy$ plane, quickly form and exist for quite a long time (Fig.~\ref{fig24}).

At the same external field for less energetic electrons and a wider injection strip, according to calculations, the depth of penetration of the fountain current upward from the target turns out to be smaller, this current has a solenoidal shape and is concentrated closer to the boundary of the region out of which the external field is displaced. However, this region is much wider along the $x$ axis and inside it not only the above $z$-pinches emerge, but due to eddy currents, including currents of cold electrons of the background plasma, quite far from the target even stronger small-scale magnetic fields are now formed, directed along the $z$ axis and exceeding the external field by an order of magnitude.

\begin{figure}[!b]
	\includegraphics[width=0.6\linewidth]{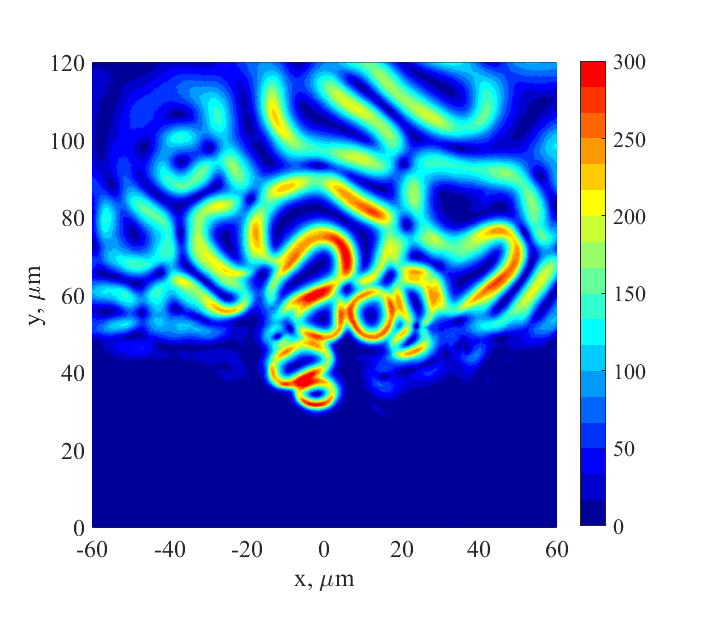}
	\centering
	\caption{
        The absolute value of transverse field $\left(B_x^2 + B_y^2\right)^{1/2}$ (in Teslas) as per the 2D simulation with an external magnetic field $B_{0z} = 20$~T at the time moment $t = 4$~ps.	
	}
	\label{fig25}
\end{figure}

If the external field directed along $z$ axis is many times weaker, say, $B_{0z} = 20$~T (Fig.~\ref{fig25}), then both the fountain currents of hot electrons and corresponding large-scale $B_z$ field and even more numerous ensemble of $z$-pinches and their small-scale transverse fields are no longer very different from those existing in the absence of external field. However, according to the simulations performed, the violation of symmetry and distortion of the overall structure of the fountain currents and the ensemble of $z$-pinches, for the selected parameters, are quite noticeable down to small values of external field $\sim B_{0z} = 5$ T. In all cases, the plasma region with hot electrons and the displaced external magnetic field expands from the injection strip at approximately the ion-acoustic speed.

\begin{figure}[t]
	\includegraphics[trim = 3cm 0 3cm 0, clip, width=0.6\linewidth]{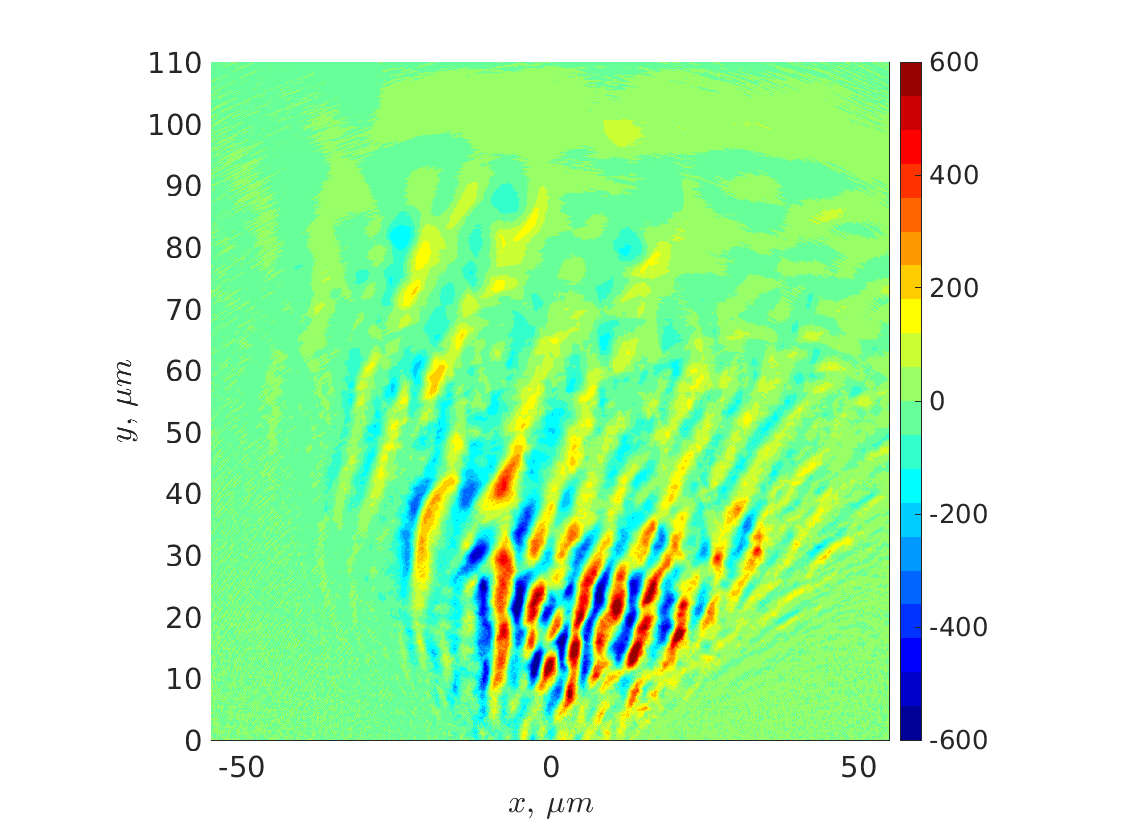}
	\centering
	\caption{
        The magnetic field's component $B_z - B_{0z}$ (in Teslas) in the 2D simulation with an external field $B_{0z} = 20$~T at the time moment $t = 0.6$~ps.
    }
	\label{fig26}
\end{figure}

Moreover, for all considered values of this field, $B_{0z} = 5$--$500$ T, the slope of the current filaments, formed due to the filamentation Weibel instability, is apparent from the very start of the injection of hot electrons into the cold background plasma (see Fig.~\ref{fig26}). The cold electrons, under the influence of the electric field of ions, are forced to move towards the target until the supply of electrons in the background plasma layer runs out or the injection of plasma with hot electrons ends.

\begin{figure}[t]
	\includegraphics[width=0.6\linewidth]{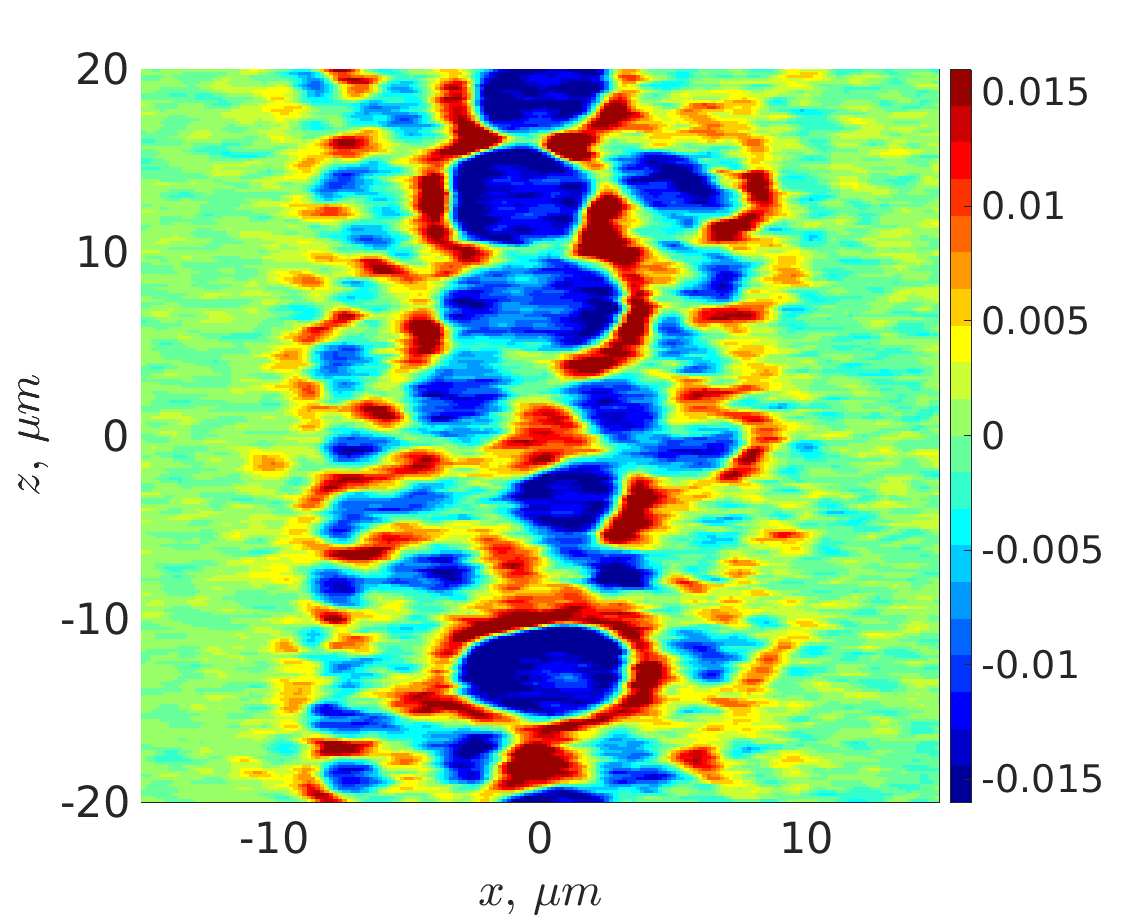}
	\centering
	\caption{
        Distribution of the current density $j_y / (e n_0 c)$ in the plane $y = 6$~$\mu$m  in the 3D simulation with an external field $B_{0z} = 100$~T at the time moment $t = 0.6$~ps.
	}
	\label{fig27}
\end{figure}

Note that in the simplified 2D3V modeling the pattern of these currents has the form of a ''fan'' of sheets emerging from the injection strip, while in the full 3D3V modeling, according to Fig.~\ref{fig27}, above this strip a dense system of filaments is observed, elongated along the $y$ axis, partially ordered into sheets along the $z$ axis and having different transverse scales -- larger for hot electrons and smaller for cold ones.

\subsection{The case of an external field orthogonal to injection strip}
\label{sec:DAN23:perpend}

\begin{figure}[t]
	\includegraphics[trim = 3cm 0 3cm 0, clip, width=0.6\linewidth]{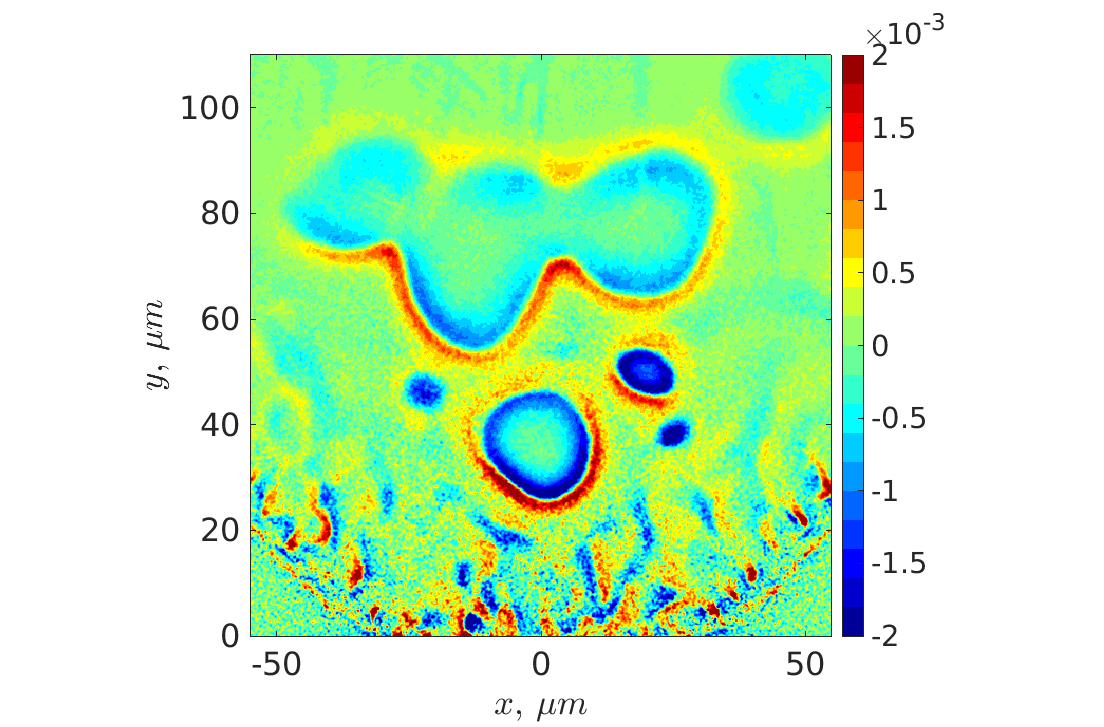}
	\centering
	\caption{
        Current density $j_z / (e n_0 c)$ in the 2D simulation with an external field $B_{0x} = 20$~T at the time moment $t = 4$~ps.
	}
	\label{fig28}
\end{figure}

Turning to the case of an external field directed along the $x$ axis, let us note that in all calculations performed the external field does limit significantly the injection of plasma with hot electrons, change fountain currents, and suppress the formation of small-scale current filaments and magnetic fields more strongly, other things being equal, than in the case discussed in subsection \ref{sec:DAN23:parallel}. This can be seen from comparing Figs.~\ref{fig25} and \ref{fig28} at a moderate field $B_{0x} = 20$ T and a high temperature $T = 100$ keV. Moreover, in Fig.~\ref{fig28} all small-scale current filaments of hot electrons have one sign of projection onto the $z$ axis (in contrast to filaments with alternating signs in Fig.~\ref{fig25}), since they are formed due to Weibel instability dragged by the background of a wider current which is directed along this axis and has arisen out of the fountain currents under the influence of external field in the process of its displacement. Large-scale current filaments near the boundary of the plasma region with the displaced magnetic field are not shown in Fig.~\ref{fig28}.

\begin{figure}[!b]
	\includegraphics[width=0.6\linewidth]{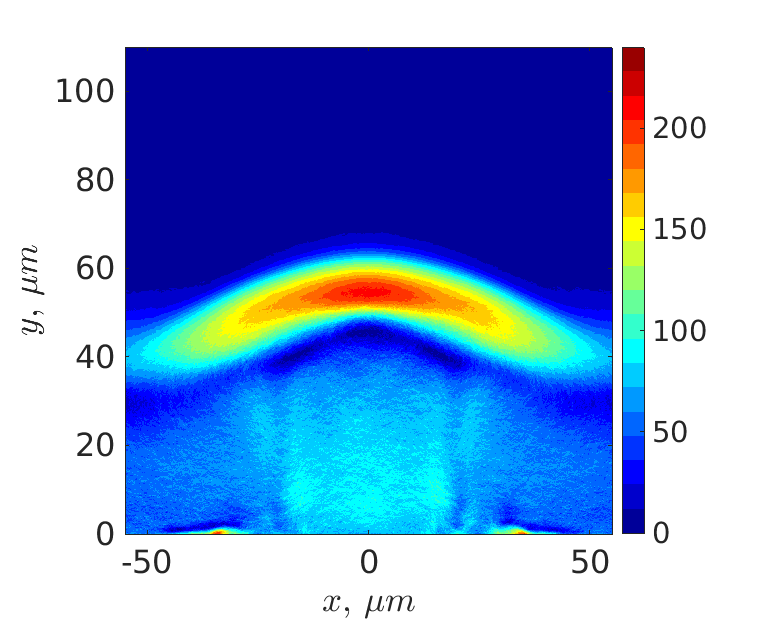}
	\centering
	\caption{
        Absolute value of transverse field minus the external one, $\left( [B_x - B_{0x}]^2 + B_y^2 \right)^{1/2}$ (in Teslas), as per 2D simulation at the moment $t = 9$~ps. The external magnetic field is $B_{0x} = 100$~T.	
	}
	\label{fig29}
\end{figure}

In a stronger field $B_{0x} = 100$ T with the same other parameters of plasma, the formation of small-scale current filaments is suppressed. However, even after a long time after the end of injection, according to Fig.~\ref{fig29}, hot electrons still form two pairs of large-scale stratified $z$-pinches, ensuring displacement of external field from the region pressed to the target and stretched along it.

\begin{figure}[t]
	\includegraphics[trim = 3cm 0 1cm 0, clip, width=0.7\linewidth]{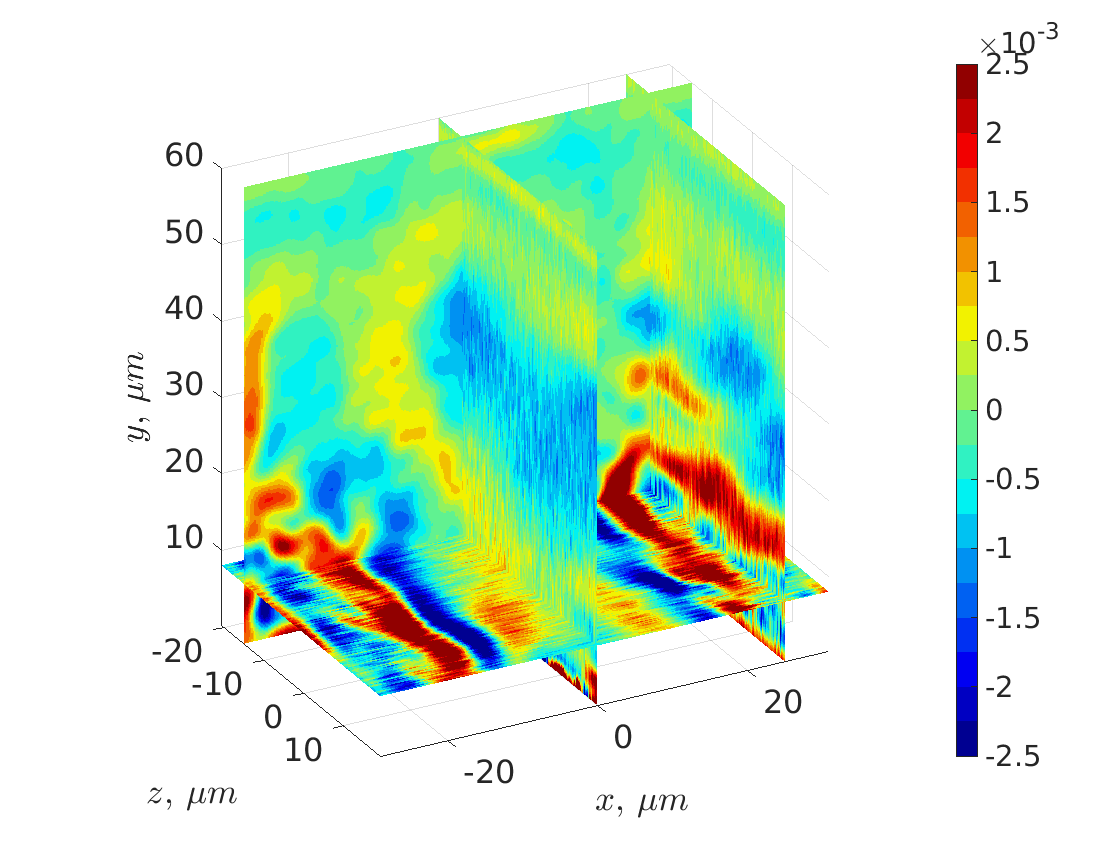}
	\centering
	\caption{
        Current density $j_z / (e n_0 c)$ in the 3D simulation with an external field $B_{0x} = 100$~T at the time moment $t = 2$~ps.	
	}
	\label{fig30}
\end{figure}

During the formation of such a region in the case of injection of plasma with more energetic electrons ($T = 300$ keV) out of a less wide band ($2 \, x_0 = 8$~$\mu$m) at the same external field $B_{0x} = 100$ T, as shown in Fig.~\ref{fig30}, small-scale current filaments exist not for long and only at the very start of the injection process, mainly at the edges of injection strip and not far from the target, where the roles of both hot and cold electrons are substantial.

It is necessary to emphasize that, as in the previous sections, a few examples of the electron current filamentation in the course of the injection of plasma containing hot electrons into an inhomogeneous layer of cold plasma in external magnetic field cannot address all important kinetic phenomena observed in numerical modeling. Their dependence on various parameters of the problem, even in the considered configurations, remain to be studied.

\section{Discussion and conclusions}
\label{sec:concl}

Thus, the electron Weibel-type instability turns out to be decisive for the creation of small-scale strong magnetic fields in all cases of expansion of a collisionless laser plasma analyzed above. This statement is valid for the boundary-value problem related to the injection of a plasma containing hot electrons into a cold plasma layer, including the magnetized one, as well as the initial-value problem related to the decay of a strong discontinuity between a plasma containing hot electrons and a cold plasma or vacuum subjected to a magnetic field.
Yet, it remains to study the effect of particle collisions, albeit rare but existing in an actual laser plasma, especially near the target where the plasma is dense, on the aforesaid mechanism of the magnetic field generation.

\subsection{Basic scenarios of plasma discontinuity decay and current filamentation}
\label{sec:Scenarios}

During the decay of a sharp discontinuity between a dense plasma containing hot electrons and a rarefied cold plasma (sec.~\ref{ch:raspad}), the Weibel instability takes place in an expanding plasma owing to a considerable, by a factor of two or more, cooling of hot electrons along the direction of their expansion, while in the transverse plane they cool much weaker. Consequently, in a large downstream region of the emerging quasi-electrostatic shock~\citep{Nechaev20_FPen} and in a fairly wide layer in the shock's upstream, a strong small-scale magnetic fields quickly form. They are oriented mostly along the front, have a spatial modulation in the orthogonal direction and bear an energy of up to an order of magnetude less than the energy of hot electrons. In laser plasma, the mega-Gauss magnitudes of the generated fields, their micrometer spatial scales, and picosecond growth times are feasible.

When the Weibel instability is saturated, a stage of gradual decay of the self-consistent quasi-static fields and currents starts. Their slow nonlinear deformation and increasing spatial scales are consistent with the expansion of both the shock wave and the region of pronounced anisotropy of the electron velocity distribution.
Using typical model examples, we establish a spatial correlation between the degree of this anisotropy and the magnitude of magnetic field, describe the dynamics of the spatial spectra of the field and current density, and point out the dependence of the developing electron anisotropy and the features of the emerging magnetic field structure on the parameters of the initial discontinuity between the cold rarefied plasma and dense plasma with hot electrons.

At the same time, the numerical modeling and qualitative analysis, described both in section~\ref{ch:raspad} and in the subsequent sections, represent only the beginning of a detailed study of the outlined class of processes occurring in the course of decay of strong discontinuity. In particular, the question remains open about (i) a possibility of effective generation of strong magnetic fields due to counter flows of cold electrons from the background plasma and (ii) plasma parameters on opposite sides of the discontinuity required for this generation to occur. Another important question concerns the optimal parameters of the discontinuity, including the ratio of the initial temperatures and number densities of cold and hot electrons, that are necessary for the utmost efficient generation of magnetic field. It is interesting also to what extent the created field can affect the dynamics of the nonequilibrium plasma's expansion, structure of the compacted layer at the shock front, temper of the anisotropic cooling of the hot electrons as well as heating of the cold ions. Also intriguing is the analysis of the field structure and the related spectra of electron and ion currents in the upstream region of the shock wave. Analysis of a nonlinear phase of Weibel instability faces the important problem of particle energy exchange and the problem of an accelerated diffusion, particularly if the decay of a discontinuity in a plasma is supplemented with the injection, i.e., the presence of a constant flow of plasma with hot electrons coming from the region of dense plasma (or from a heated target in the problem of laser ablation).

According to section~\ref{ch:chWeibMagn}, already a weak external field, with the pressure significantly less than that of a cloud containing hot electrons, expanding in the~$xy$ plane and homogeneous along the~$z$ axis significantly affects plasma spreading into vacuum and spatio-temporal pattern of the generated quasi-magnetostatic formations (turbulence). The most important point here is the anisotropic cooling of the ejected electrons. It significantly depends on the magnitude of external transverse magnetic field through a wide (several orders of magnitude) range. This dependence persists down to a fairly small field magnitudes not affecting the density dynamics of expanding plasma. 
On the basis of detailed numerical PIC-simulations, we reveal formation, long-term existence and pronounced motion of localized current structures ($z$-pinches and sheets) in the region where the external field becomes weakened. The associated small-scale fields can be stronger than the external field by many times. The density of plasma inside $z$-pinches can be notably increased.
These outcomes are the result of the nonlinear development of Weibel-type instability getting saturated when the energy of the quasistatic magnetic field approaches several percent of the energy of hot electrons. 

At the same time, there are fountains currents generated by the most energetic escaping electrons and the quasi-2D or volumetric currents generated by hot electrons inside the expanding plasma cloud. They both are associated with the large-scale magnetic fields displacing the external field inside the cloud. Symmetry of the expansion can be violated by an external field directed across the hot electron flow. In this case, the current structures of all scales will be inhomogeneously deformed and cumulation of the plasma flow could take place. 

In fact, the above analysis of a femtosecond laser ablation is based on the simplest model of an inhomogeneously heated plasma of a uniform density in a boundary layer of a target. The other factors could make the electron velocity distribution non-Maxwellian and anisotropic as well as the spatial distribution of the effective temperature and density of heated electrons non-quasi-1D
already from the very start of expansion. These factors include a complex polarization, cross section and temporal profile of the laser pulse, detuning of the optical frequency from the plasma frequency, non-planar geometry and finite thickness of the discontinuity layer, a significant influx of hot electrons continuously coming from deeper layers of the heated target, where particle collisions are important. The analysis of how these and other factors of the initial creation of a nonequilibrium plasma influence the generated currents and fields just begins and is beyond the scope of this review.

The above also applies to the features of the emergence, development and decay of the magnetic and current structures of various scales, described in sections 4 and 5 on the basis of a boundary-value problem. Namely, the problem when a rarefied plasma with hot electrons is injected into a layer of cold, denser plasma which has a monotonically decreasing number density and may be subjected to an external magnetic field. The considered model corresponds to the collisionless expansion of a high-energy-density laser plasma created as a result of femtosecond ablation of a flat target by a cylindrically focused laser beam, but can be characteristic of a number of other problems in the physics of space and laboratory plasma.

In a wide range of the problem parameters, there coexist two qualitatively distinct Weibel mechanisms of formation of mutually orthogonal structures of currents, creating strong small-scale magnetic fields in neighboring plasma layers. When crossing the boundary between these layers, the fraction of cold electrons in the total plasma number density jumps considerably. Correspondingly, the type of the electron velocity anisotropy, which is responsible for the Weibel-type instability, changes. In the denser layer the anisotropy axis is parallel to the direction of electron injection, while in the less dense layer it is orthogonal to that direction.

In the former layer the electron distribution contains a gradually relaxing hot flow and relatively cold thermal background. In the latter layer it reminds a bi-Maxwellian distribution of hot electrons. The currents in these layers are parallel to the direction of anisotropy axis, having the form of wedge-shaped inhomogeneous sheets in 2D simulations (filamentation type of Weibel instability) and filaments resembling deformed $z$-pinches (thermal type), respectively. When injection of hot electrons ceases, the filaments decay much slower than current sheets in virtue of the nonlinear trapping of electrons. A lifetime of the filaments considerably exceeds the injection duration.

As the simulation results reveal, the presence of external field in an inhomogeneous layer of cold plasma, into which the plasma with hot electrons is injected, in a wide range of system parameters significantly influences dynamics and spatial structure of the appearing small-scale currents and magnetic fields. These fields can be much stronger than the external field and, simultaneously, substantially depend on it. Such current structures and fields are generated not only by the hot electrons, but also by the cold electrons of background plasma, and, therefore, significantly depend on its number density and the scale of inhomogeneity. Thus, the characteristics of not only the femtosecond laser pulse, which determine the size of the heated spot on the target and the duration of the injection of hot electrons, are important, but also those of the prepulse, which implicitly sets the scale of the inhomogeneity and the initial number density of the background plasma. These characteristics affect the time of formation, number of filaments and current density in the emerging current systems as well as their subsequent restructuring, decay rate and total lifetime after injection of plasma containing hot electrons ceases.

In laser ablation experiments, it is possible to control the plasma discontinuity parameters, i.e., the properties of its hot and cold parts, by tuning the powerful femtosecond pulse that creates the hot part, and its prepulse that determines the cold one. Modern technologies also make it possible to create fairly uniform magnetic fields with different orientations and induction ranging from units to hundreds of Tesla (in pulsed mode). Finally, there are methods for diagnostics of the plasma number density and magnetic fields with high spatial and temporal resolution (see, e.g., \citep{Shaikh2017, Zhou2018, Borghesi1998, Chatterjee2017, Stepanov2018_LO2018, Stepanov20_LO20, Plechaty2009, ForestierColleoni2019}). All this gives reason to count on the possibility of experimentally studying the predicted phenomena in laser plasma. 
One can hope that the aforementioned patterns of transient phenomena of the formation of current structures and related magnetic fields of different scales and orientations in the course of the expansion of plasma with hot electrons take place for more complex geometries and laser ablation models and will prove useful for predicting and studying similar phenomena in astrophysical and magnetospheric plasmas.

\subsection{Some open problems in the theory of Weibel-type instabilities and magnetic turbulence}
\label{sec:Open problems}

Note that even in the approximation of a homogeneous plasma, without accounting for its expansion, a number of features of Weibel-type aperiodic instabilities of initialy anisotropic particle velocity distributions and associated spatio-temporal dynamics of emerging quasi-magnetostatic turbulence remain very poorly studied. Very fragmentary is the knowledge of even the linear dispersion properties of these instabilities, i.e., the dependencies of the growth rates of unstable modes of ordinary and extraordinary types on the direction and magnitude of their wave vectors for a somewhat representative class of particle velocity distribution functions (see, e.g., \citep{Kalman1968, Davidson1989, Vagin2014, Kocharovsky2016_UFN, Silva2021, Pierrard2010}).
Finding these dependencies turns out to be especially difficult in the presence of external field, when one has to use numerical methods~\citep{Camporeale2008, Hellinger2014, Xie2016, Bret2017, Verscharen2018, Lopez2019, Umeda2018, Moya2022, Lazar2023} or be limited to approximate analytical solutions for a very narrow range of system's parameters and wave vectors' directions or lengths~\citep{Landau1970, Meneses2018, Li2000, Gary2003, Pokhotelov2012, Grassi2017, Emelyanov2023, Stockem2006}.

Even less studied are the nonlinear properties of these instabilities, for example, the criteria for their saturation, which establish maximum of the root-mean-square value of the magnetic field in the emerging Weibel turbulence depending on the problem parameters. The known criteria are very approximate, quite partial and sensitive to the specific type of particle velocity distribution function (see, e.g.,~\citep{Davidson1972, Yang1994, Achterberg2007, Tautz2013, Kocharovsky2016_UFN, Kuznetsov22_FPen}).
The patterns of evolution of the Weibel turbulence after its growth has saturated are only beginning to be studied, in extremely simple cases and mainly by numerical methods. Examples of recently identified general patterns include partial self-similarity of the evolution of the field turbulence spectrum and the weakly nonlinear character of Weibel turbulence, largely governed by quasilinear dynamics (see~\citep{Davidson1972, Ruyer2015_NonlinWeibel, Kuznetsov22_FPen, Kuznetsov23_JETPen} and references therein).
In accordance with the latter, the direct 3- and 4-wave interaction of individual spatial modes is weak compared to their implicit, integral interaction due to the joint deformation of the spatially averaged particle velocity distribution function, which determines the instantaneous growth/decay rates of each mode. The effectiveness of the quasilinear approximation for a number of specific problems of the evolution of Weibel turbulence is shown using examples of bi-Maxwellian and Kappa distributions in~\citep{Kuznetsov22_FPen, Kuznetsov23_JETPen}.
These and some other works, say \citep{Garasev2021}, also present numerical simulations demonstrating a direct nonlinear interaction of Weibel modes, which in certain cases is resonant and goes beyond the quasilinear approximation at limited stages of the turbulence evolution, including the possibility of super-exponential or power-law growth of resonant harmonics and their power-law (non-exponential) decay.

As one of the few analytical achievements in the theory of nonlinear evolution of Weibel turbulence, formally valid for an arbitrary particle distribution function, let us point out the recently obtained~\citep{Nechaev23_JPP} universal expression for the plasma magnetization level. By definition, the latter is the ratio of the current value of the energy density of turbulent magnetic field $w_B$ to the initial thermal energy density of particles with an effective temperature $T_{z0}$ along a certain axis $z$ (determined, for example, by external field). It can be expressed as a function of the average wave number $k$, dominating in the spectrum of magnetic turbulence, instantaneous anisotropy parameter $A(t)$ and its initial value $A_0$:
\begin{equation}
    \frac{ 2 \, w_B (t)}{ n_0  T_{z 0} }  = \frac{ 2 k^2}{1 + k^2} \, \frac{\left( A_{0} + 1 \right)^{-1} \left( A_{0} - A \right)}{A + 2 k^2 + 3}
    \leq \frac{2}{2 \sqrt{6} + 5} \, \frac{A_{0}}{A_{0} + 1} .
\label{eq:mine}
\end{equation}
This approximate expression is obtained under the assumption of a weak inhomogeneity of the plasma number density ($n_0$) and weak turbulence along the specified spatial coordinate $z$, i.e., under the assumption of a two-dimensional spectrum of turbulence, including only wave vectors orthogonal to $z$ axis. It also follows from this equation that the magnetization value is limited by approximately $0.2$, which was verified by various 2D3V simulations presented in~\citep{Nechaev23_JPP, Kuznetsov23_JETPen}.

\subsection{Relevant problems in the cosmic plasma physics}
\label{sec:Cosmic}

The results stated above open up new possibilities for using the concept of the electron Weibel instability and nonlinear dynamics of currents and magnetic fields produced by it to solve problems of laboratory modeling and physical interpretation of a number of expected or observed phenomena in cosmic nonequilibrium plasma. The study of similar processes of current microstructuring in the course of the decay of a strong discontinuity in a partially magnetized cosmic plasma with hot electrons has not been carried out yet.
The studied transient phenomena may occur during explosive processes in cosmic plasma, for example, at the initial stage of a rapid local heating of density ducts in the atmospheres of stars or the magnetospheres of exoplanets. Particularly interesting in this regard are the problems of stellar (or solar) flares and winds, including the problem of injection of high-energy electrons and reconnection of magnetic field lines inside the coronas of active stars and inside magnetic clouds of stellar wind~\citep{Zaitsev2017, Dudik2017, Tautz2013, Priest2001, Zaitsev2008, Lazar2022}.

Thus, in a stellar (solar) wind~\citep{Dudik2017, DeForest2018, Yoon2017, Echim2011} one can expect the occurrence of a strong discontinuity of nonequilibrium plasma and the subsequent restructuring of magnetic fields as a result of contact interactions of individual plasma filaments containing hot electrons with extended magnetic clouds filled with much more rarefied and cold plasma. In the atmospheres of stars and the magnetospheres of planets (exoplanets), in particular, in the areas of high density~\citep{Baumjohann2010_ANGEO, Nakamura2018, Shuster2019, Voros2017, Dyal2006, Kelley2003, Zaitsev2015, Stockem2006}, as a result of various explosive processes, a significant local change in plasma number density and rapid heating of its electrons, for example by X-ray radiation, is possible. Such a heating, as well as an increase in plasma density, can lead to the displacement of the surrounding magnetic field, followed by small-scale structuring of the newly generated field throughout the entire region occupied by the plasma with anisotropically cooling, but still hot, electrons.

For the Sun and late-type stars~\citep{Zaitsev2017, Priest2014, Fletcher2008, Stepanov2022_Cowling, Fleishman2022, Malanushenko2022} it is not difficult to imagine a situation, where in an inhomogeneous coronal loop (a bundle of twisted magnetic tubes), filled with a sufficiently cold plasma, a region with hot electrons is formed, elongated along magnetic field and localized across it. This can be caused by various explosive processes, e.g., the injection of high-energy electrons by the convective fields of the photosphere or powerful Alfven waves rising from it, heating of the plasma by a longitudinal electric current when the loop moves in the chromosphere, or the reconnection of some of magnetic field lines of intersecting loops in the corona. Then, as our preliminary analytical estimates as well as numerical simulations reveal, a rapidly developing Weibel-type instability creates small-scale magnetic turbulence, which could not only change the balance of magnetic and kinetic pressures in the coronal loop, but also increase the effective (anomalous) conductivity of the plasma by many orders of magnitude. As a result, a restructuring of the original large-scale currents and an inhomogeneous magnetic field in the coronal loop may occur, multiple regions of magnetic reconnection may arise, and significant deformation of the loop structure or even its local ruptures may begin.

In other words, small-scale magnetic turbulence caused by the Weibel instability can initiate almost simultaneous (within tens of seconds) occurrence of a large number (up to billions) of so-called nanoflares associated with local processes of reconnection of magnetic field lines and heating of the coronal plasma in a separate loop. These solar nanoflares, as the attributes of statistical models of flares, have long been discussed and observed~\citep{Parker1998, Ulyanov2019, Purkhart2022, Vlahos1989, Klimchuk2006}, but their origin remained mysterious and was not associated with the Weibel-type instability or an anisotropic distribution of hot electrons elongated along the large-scale magnetic field of a loop (or a bundle of twisted flux tubes); see, e.g.,~\citep{Zaitsev2008, Zaitsev2015, Zaitsev2017, Zaitsev2019, Hudson2020, LiGan2006, Zhou2013_RAA}. 
It is the result of this small-scale instability that will be the rapid dissipation of both the current of injected energetic (keV) electrons and the existing large-scale current of background (rather cold) electrons of the loop, explaining the ''nanostructure'' of this type of solar flares.

The aforementioned and other problems of space physics and astrophysics related to (i)~the transient processes of plasma expansion in the presence of the hot electrons as well as (ii)~the modification of the kinetic and dynamic properties of plasma due to microstructuring of the self-consistent currents and magnetic fields seem relevant and promising for detailed studies.

\section*{Aknowledgements}
The work was supported partially by RSF grant n. 21-12-00416. The computer time for simulations were provided by the Joint Supercomputer Center of~the Russian Academy of~Sciences and by the Keldysh Institute of Applied Mathematics.

\section*{Declarations}

\bmhead{Conflict of interest}
The authors declare no conflict of interests.

\bibliography{biblio}

\begin{thebibliography}{149}
\providecommand{\natexlab}[1]{#1}
\providecommand{\url}[1]{{#1}}
\providecommand{\urlprefix}{URL }
\providecommand{\doi}[1]{\url{https://doi.org/#1}}
\providecommand{\eprint}[2][]{\url{#2}}
 \bibcommenthead

\bibitem[{Achterberg et~al(2007)Achterberg, Wiersma, and Norman}]{Achterberg2007}
Achterberg A, Wiersma J, Norman CA (2007) The {W}eibel instability in relativistic plasmas. {II}. nonlinear theory and stabilization mechanism. Astronomy {\&} Astrophysics 475(1):19--36. \doi{10.1051/0004-6361:20065366}

\bibitem[{Akhiezer et~al(1975)Akhiezer, Akhiezer, Polovin, Sitenko, and Stepanov}]{Akhiezer1975}
Akhiezer A, Akhiezer I, Polovin R, et~al (1975) Plasma Electrodynamics. Pergamon, \doi{10.1016/C2013-0-02586-8}

\bibitem[{Albertazzi et~al(2014)Albertazzi, Ciardi, Nakatsutsumi, Vinci, B{\'{e}}ard, Bonito, Billette, Borghesi, Burkley, Chen, Cowan, Herrmannsdörfer, Higginson, Kroll, Pikuz, Naughton, Romagnani, Riconda, Revet, Riquier, Schlenvoigt, Skobelev, Faenov, Soloviev, Huarte-Espinosa, Frank, Portugall, P{\'{e}}pin, and Fuchs}]{Albertazzi2014}
Albertazzi B, Ciardi A, Nakatsutsumi M, et~al (2014) Laboratory formation of a scaled protostellar jet by coaligned poloidal magnetic field. Science 346(6207):325--328. \doi{10.1126/science.1259694}

\bibitem[{Albertazzi et~al(2015)Albertazzi, Chen, Antici, B\"{o}ker, Borghesi, Breil, Dervieux, Feugeas, Lancia, Nakatsutsumi, Nicolaï, Romagnagni, Shepherd, Sentoku, Starodubtsev, Swantusch, Tikhonchuk, Willi, d{\textquotesingle}Humi{\`{e}}res, P{\'{e}}pin, and Fuchs}]{Albertazzi2015}
Albertazzi B, Chen SN, Antici P, et~al (2015) Dynamics and structure of self-generated magnetics fields on solids following high contrast, high intensity laser irradiation. Physics of Plasmas 22(12):123108. \doi{10.1063/1.4936095}

\bibitem[{Arber et~al(2015)Arber, Bennett, Brady, Lawrence-Douglas, Ramsay, Sircombe, Gillies, Evans, Schmitz, Bell, and Ridgers}]{Arber2015_EPOCH}
Arber TD, Bennett K, Brady CS, et~al (2015) Contemporary particle-in-cell approach to laser-plasma modelling. Plasma Physics and Controlled Fusion 57(11):113001. \doi{10.1088/0741-3335/57/11/113001}

\bibitem[{Balogh and Treumann(2013)}]{Balogh2013}
Balogh A, Treumann R (2013) Physics of Collisionless Shocks: Space Plasma Shock Waves. ISSI Scientific Report Series, Springer, New York

\bibitem[{Baumjohann and Treumann(2012)}]{Baumjohann2012}
Baumjohann W, Treumann R (2012) Basic Space Plasma Physics. Imperial College Press, London

\bibitem[{Baumjohann et~al(2010)Baumjohann, Nakamura, and Treumann}]{Baumjohann2010_ANGEO}
Baumjohann W, Nakamura R, Treumann RA (2010) Magnetic guide field generation in collisionless current sheets. Ann Geophys 28(3):789--793. \doi{10.5194/angeo-28-789-2010}

\bibitem[{Borghesi et~al(1998)Borghesi, MacKinnon, Bell, Gaillard, and Willi}]{Borghesi1998}
Borghesi M, MacKinnon AJ, Bell AR, et~al (1998) Megagauss magnetic field generation and plasma jet formation on solid targets irradiated by an ultraintense picosecond laser pulse. Physical Review Letters 81(1):112--115. \doi{10.1103/physrevlett.81.112}

\bibitem[{Borodachev et~al(2017)Borodachev, Garasev, Kolomiets, Kocharovsky, Martyanov, and Nechaev}]{Borodachev2017_RadiophysEn}
Borodachev LV, Garasev MA, Kolomiets DO, et~al (2017) Dynamics of a self-consistent magnetic field and diffusive scattering of ions in a plasma with strong thermal anisotropy. Radiophysics and Quantum Electronics 59(12):991--999. \doi{10.1007/s11141-017-9768-0}

\bibitem[{Bret(2009)}]{Bret2009}
Bret A (2009) Weibel, two-stream, filamentation, oblique, bell, {Buneman}...which one grows faster? ApJ 699(2):990--1003. \doi{10.1088/0004-637x/699/2/990}

\bibitem[{Bret and Dieckmann(2017)}]{Bret2017}
Bret A, Dieckmann ME (2017) Hierarchy of instabilities for two counter-streaming magnetized pair beams: {I}nfluence of field obliquity. Physics of Plasmas 24(6). \doi{10.1063/1.4985321}

\bibitem[{Camporeale and Burgess(2008)}]{Camporeale2008}
Camporeale E, Burgess D (2008) Electron firehose instability: {K}inetic linear theory and two-dimensional particle-in-cell simulations. J Geophys Res: Space Physics 113(A7):A07107. \doi{10.1029/2008ja013043}

\bibitem[{Chang et~al(2008)Chang, Spitkovsky, and Arons}]{Chang2008}
Chang P, Spitkovsky A, Arons J (2008) Long-term evolution of magnetic turbulence in relativistic collisionless shocks: Electron-positron plasmas. ApJ 674(1):378--387. \doi{10.1086/524764}

\bibitem[{Chatterjee et~al(2017)Chatterjee, Singh, Robinson, Blackman, Booth, Culfa, Dance, Gizzi, Gray, Green, Koester, Kumar, Labate, Lad, Lancaster, Pasley, Woolsey, and Rajeev}]{Chatterjee2017}
Chatterjee G, Singh PK, Robinson APL, et~al (2017) Micron-scale mapping of megagauss magnetic fields using optical polarimetry to probe hot electron transport in petawatt-class laser-solid interactions. Scientific Reports 7(1):8347. \doi{10.1038/s41598-017-08619-1}

\bibitem[{Chen and Fiuza(2023)}]{Chen2023}
Chen H, Fiuza F (2023) Perspectives on relativistic electron{\textendash}positron pair plasma experiments of astrophysical relevance using high-power lasers. Phys Plasmas 30(2):020601. \doi{10.1063/5.0134819}

\bibitem[{Davidson(1989)}]{Davidson1989}
Davidson RC (1989) Kinetic waves and instabilities in a uniform plasma. In: Galeev AA, Sudan RN (eds) Basic Plasma Physics: Selected chapters from the Handbook of Plasma Physics, vol. 1 and 2. North-Holland Publishing Company, Amsterdam, p 229

\bibitem[{Davidson et~al(1972)Davidson, Hammer, Haber, and Wagner}]{Davidson1972}
Davidson RC, Hammer DA, Haber I, et~al (1972) Nonlinear development of electromagnetic instabilities in anisotropic plasmas. Physics of Fluids 15(2):317. \doi{10.1063/1.1693910}

\bibitem[{DeForest et~al(2018)DeForest, Howard, Velli, Viall, and Vourlidas}]{DeForest2018}
DeForest CE, Howard RA, Velli M, et~al (2018) The highly structured outer solar corona. ApJ 862(1):18. \doi{10.3847/1538-4357/aac8e3}

\bibitem[{Dieckmann(2009)}]{Dieckmann2009}
Dieckmann ME (2009) The filamentation instability driven by warm electron beams: statistics and electric field generation. Plasma Physics and Controlled Fusion 51(12):124042. \doi{10.1088/0741-3335/51/12/124042}

\bibitem[{Dieckmann et~al(2018)Dieckmann, Moreno, Doria, Romagnani, Sarri, Folini, Walder, Bret, d{\textquotesingle}Humi{\`{e}}res, and Borghesi}]{Dieckmann2018}
Dieckmann ME, Moreno Q, Doria D, et~al (2018) Expansion of a radially symmetric blast shell into a uniformly magnetized plasma. Physics of Plasmas 25(5):052108. \doi{10.1063/1.5024851}

\bibitem[{Dud{\'{i}}k et~al(2017)Dud{\'{i}}k, Dzif{\v{c}}{\'{a}}kov{\'{a}}, Meyer-Vernet, Zanna, Young, Giunta, Sylwester, Sylwester, Oka, Mason, Vocks, Matteini, Krucker, Williams, and Mackovjak}]{Dudik2017}
Dud{\'{i}}k J, Dzif{\v{c}}{\'{a}}kov{\'{a}} E, Meyer-Vernet N, et~al (2017) Nonequilibrium processes in the solar corona, transition region, flares, and solar wind (invited review). Solar Physics 292(8). \doi{10.1007/s11207-017-1125-0}

\bibitem[{Dyal(2006)}]{Dyal2006}
Dyal P (2006) Particle and field measurements of the {Starfish} diamagnetic cavity. J Geophys Res Space Phys 111(A12):A12211. \doi{10.1029/2006ja011827}

\bibitem[{Echim et~al(2011)Echim, Lemaire, and Lie-Svendsen}]{Echim2011}
Echim MM, Lemaire J, Lie-Svendsen {\O} (2011) A review on solar wind modeling: Kinetic and fluid aspects. Surv Geophys 32(1):1--70. \doi{10.1007/s10712-010-9106-y}

\bibitem[{Emelyanov and Kocharovsky(2023)}]{Emelyanov2023}
Emelyanov NA, Kocharovsky VV (2023) Radiophysics and Quantum Electronics 66(10). (In press.)

\bibitem[{Fleishman et~al(2022)Fleishman, Nita, Chen, Yu, and Gary}]{Fleishman2022}
Fleishman GD, Nita GM, Chen B, et~al (2022) Solar flare accelerates nearly all electrons in a large coronal volume. Nature 606(7915):674--677. \doi{10.1038/s41586-022-04728-8}

\bibitem[{Fletcher and Hudson(2008)}]{Fletcher2008}
Fletcher L, Hudson HS (2008) Impulsive phase flare energy transport by large-scale {A}lfv{\'{e}}n waves and the electron acceleration problem. ApJ 675(2):1645--1655. \doi{10.1086/527044}

\bibitem[{Forestier-Colleoni et~al(2019)Forestier-Colleoni, Batani, Burgy, Sorbo, Froustey, Hulin, d{\textquotesingle}Humi{\`{e}}res, Jakubowska, Merzeau, Mishchik, Papp, and Santos}]{ForestierColleoni2019}
Forestier-Colleoni P, Batani D, Burgy F, et~al (2019) Space and time resolved measurement of surface magnetic field in high intensity short pulse laser matter interactions. Physics of Plasmas 26(7):072701. \doi{10.1063/1.5086725}

\bibitem[{Fox et~al(2018)Fox, Matteucci, Moissard, Schaeffer, Bhattacharjee, Germaschewski, and Hu}]{Fox2018}
Fox W, Matteucci J, Moissard C, et~al (2018) Kinetic simulation of magnetic field generation and collisionless shock formation in expanding laboratory plasmas. Physics of Plasmas 25(10):102106. \doi{10.1063/1.5050813}

\bibitem[{Garasev and Derishev(2016)}]{Garasev2016}
Garasev M, Derishev E (2016) Impact of continuous particle injection on generation and decay of the magnetic field in collisionless shocks. Monthly Notices of the Royal Astronomical Society 461(1):641--646. \doi{10.1093/mnras/stw1345}

\bibitem[{Garasev and Derishev(2021)}]{Garasev2021}
Garasev MA, Derishev EV (2021) Numerical simulation of nonlinear effects in the {W}eibel instability. Radiophys Quantum El 63(12):909--920. \doi{10.1007/s11141-021-10103-w}

\bibitem[{Garasev et~al(2017)Garasev, Korytin, Kocharovsky, Mal'kov, Murzanev, Nechaev, and Stepanov}]{Garasev2017_JETPLen}
Garasev MA, Korytin AI, Kocharovsky VV, et~al (2017) Features of the generation of a collisionless electrostatic shock wave in a laser-ablation plasma. {JETP} Letters 105(3):164--168. \doi{10.1134/s0021364017030067}

\bibitem[{Garasev et~al(2022{\natexlab{a}})Garasev, Kocharovsky, Nechaev, Stepanov, and Kocharovsky}]{Garasev22_GA_InjEn}
Garasev MA, Kocharovsky VV, Nechaev AA, et~al (2022{\natexlab{a}}) The coexistence of orthogonal current structures and the development of different-type {W}eibel instabilities in adjacent regions of a plasma transition layer with a hot electron flow. Geomagnetism and Aeronomy 62(S1):S10--S24. \doi{10.1134/S0016793222600436}

\bibitem[{Garasev et~al(2022{\natexlab{b}})Garasev, Nechaev, Stepanov, Kocharovsky, and Kocharovsky}]{Garasev22_JPP}
Garasev MA, Nechaev AA, Stepanov AN, et~al (2022{\natexlab{b}}) Multiscale magnetic field structures in an expanding elongated plasma cloud with hot electrons subject to an external magnetic field. J Plasma Phys 88:175880301. \doi{10.1017/S0022377822000423}

\bibitem[{Garasev et~al(2022{\natexlab{c}})Garasev, Nechaev, Stepanov, Kocharovsky, and Kocharovsky}]{Garasev22_GA_MagnEn}
Garasev MA, Nechaev AA, Stepanov AN, et~al (2022{\natexlab{c}}) Weibel instability and deformation of an external magnetic field in the region of decay of a strong discontinuity in a plasma with hot electrons. Geomagnetism and Aeronomy 62(3):182--198. \doi{10.1134/S0016793222030094}

\bibitem[{Gary(1993)}]{Gary1993}
Gary SP (1993) Theory of space plasma microinstabilities. Cambridge University Press, Cambridge, \doi{10.1017/cbo9780511551512}

\bibitem[{Gary and Nishimura(2003)}]{Gary2003}
Gary SP, Nishimura K (2003) Resonant electron firehose instability: Particle-in-cell simulations. Physics of Plasmas 10(9):3571--3576. \doi{10.1063/1.1590982}

\bibitem[{G\"{o}de et~al(2017)G\"{o}de, R\"{o}del, Zeil, Mishra, Gauthier, Brack, Kluge, MacDonald, Metzkes, Obst, Rehwald, Ruyer, Schlenvoigt, Schumaker, Sommer, Cowan, Schramm, Glenzer, and Fiuza}]{Gode2017}
G\"{o}de S, R\"{o}del C, Zeil K, et~al (2017) Relativistic electron streaming instabilities modulate proton beams accelerated in laser-plasma interactions. Physical Review Letters 118(19). \doi{10.1103/physrevlett.118.194801}

\bibitem[{Grassi et~al(2017)Grassi, Grech, Amiranoff, Pegoraro, Macchi, and Riconda}]{Grassi2017}
Grassi A, Grech M, Amiranoff F, et~al (2017) Electron {W}eibel instability in relativistic counterstreaming plasmas with flow-aligned external magnetic fields. Physical Review E 95(2):023203. \doi{10.1103/physreve.95.023203}

\bibitem[{Grismayer and Mora(2006)}]{Grismayer2006}
Grismayer T, Mora P (2006) Influence of a finite initial ion density gradient on plasma expansion into a vacuum. Physics of Plasmas 13(3):032103. \doi{10.1063/1.2178653}

\bibitem[{Gruzinov(2001)}]{Gruzinov2001}
Gruzinov A (2001) Gamma-ray burst phenomenology, shock dynamo, and the first magnetic fields. The Astrophysical Journal 563(1):L15--L18. \doi{10.1086/324223}

\bibitem[{Hellinger et~al(2014)Hellinger, Tr{\'{a}}vn{\'{\i}}{\v{c}}ek, Decyk, and Schriver}]{Hellinger2014}
Hellinger P, Tr{\'{a}}vn{\'{\i}}{\v{c}}ek PM, Decyk VK, et~al (2014) Oblique electron fire hose instability: {P}article-in-cell simulations. J Geophys Res: Space Physics 119(1):59--68. \doi{10.1002/2013ja019227}

\bibitem[{Hudson et~al(2020)Hudson, Sim{\~{o}}es, Fletcher, Hayes, and Hannah}]{Hudson2020}
Hudson HS, Sim{\~{o}}es PJA, Fletcher L, et~al (2020) Hot {X}-ray onsets of solar flares. MNRAS 501(1):1273--1281. \doi{10.1093/mnras/staa3664}

\bibitem[{Huntington et~al(2015)Huntington, Fiuza, Ross, Zylstra, Drake, Froula, Gregori, Kugland, Kuranz, Levy, Li, Meinecke, Morita, Petrasso, Plechaty, Remington, Ryutov, Sakawa, Spitkovsky, Takabe, and Park}]{Huntington2015}
Huntington CM, Fiuza F, Ross JS, et~al (2015) Observation of magnetic field generation via the weibel instability in interpenetrating plasma flows. Nature Physics 11(2):173--176. \doi{10.1038/nphys3178}

\bibitem[{Huntington et~al(2017)Huntington, Manuel, Ross, Wilks, Fiuza, Rinderknecht, Park, Gregori, Higginson, Park, Pollock, Remington, Ryutov, Ruyer, Sakawa, Sio, Spitkovsky, Swadling, Takabe, and Zylstra}]{Huntington2017}
Huntington CM, Manuel MJE, Ross JS, et~al (2017) Magnetic field production via the weibel instability in interpenetrating plasma flows. Physics of Plasmas 24(4):041410. \doi{10.1063/1.4982044}

\bibitem[{Ivanov et~al(2014)Ivanov, Shulyapov, Ksenofontov, Tsymbalov, Volkov, Savel'ev, Brantov, Bychenkov, Turinge, Lapik, Rusakov, Djilkibaev, and Nedorezov}]{Ivanov2014}
Ivanov KA, Shulyapov SA, Ksenofontov PA, et~al (2014) Comparative study of amplified spontaneous emission and short pre-pulse impacts onto fast electron generation at sub-relativistic femtosecond laser-plasma interaction. Physics of Plasmas 21(9):093110. \doi{10.1063/1.4896348}

\bibitem[{Kakad et~al(2016)Kakad, Lotekar, and Kakad}]{Kakad2016}
Kakad A, Lotekar A, Kakad B (2016) First-ever model simulation of the new subclass of solitons {\textquotedblleft}supersolitons{\textquotedblright} in plasma. Physics of Plasmas 23(11):110702. \doi{10.1063/1.4969078}

\bibitem[{Kalman et~al(1968)Kalman, Montes, and Quemada}]{Kalman1968}
Kalman G, Montes C, Quemada D (1968) Anisotropic temperature plasma instabilities. Phys Fluids 11(8):1797--1808. \doi{10.1063/1.1692198}

\bibitem[{{Kato} and {Takabe}(2008)}]{Kato2008}
{Kato} TN, {Takabe} H (2008) Nonrelativistic collisionless shocks in unmagnetized electron-ion plasmas. ApJL 681(2):L93. \doi{10.1086/590387}

\bibitem[{Kelley and Livingston(2003)}]{Kelley2003}
Kelley MC, Livingston R (2003) Barium cloud striations revisited.  Geophys Res Space Phys 108(A1):1044. \doi{10.1029/2002ja009412}

\bibitem[{King et~al(2019)King, Butler, Wilson, Capdessus, Gray, Powell, Dance, Padda, Gonzalez-Izquierdo, Rusby, Dover, Hicks, Ettlinger, Scullion, Carroll, Najmudin, Borghesi, Neely, and McKenna}]{King2019}
King M, Butler NMH, Wilson R, et~al (2019) Role of magnetic field evolution on filamentary structure formation in intense laser{\textendash}foil interactions. High Power Laser Science and Engineering 7:e14. \doi{10.1017/hpl.2018.75}

\bibitem[{Klimchuk(2006)}]{Klimchuk2006}
Klimchuk JA (2006) On solving the coronal heating problem. Solar Physics 234(1):41--77. \doi{10.1007/s11207-006-0055-z}

\bibitem[{Kocharovsky et~al(2016)Kocharovsky, Kocharovsky, Martyanov, and Tarasov}]{Kocharovsky2016_UFN}
Kocharovsky VV, Kocharovsky VV, Martyanov VY, et~al (2016) Analytical theory of self-consistent current structures in a collisionless plasma. Physics-Uspekhi 59(12):1165--1210. \doi{10.3367/ufne.2016.08.037893}

\bibitem[{Kocharovsky et~al(2023)Kocharovsky, Garasev, Derishev, Nechaev, and Stepanov}]{Kocharovsky23_DANen}
Kocharovsky VV, Garasev MA, Derishev EV, et~al (2023) Influence of a uniform magnetic field on the generation of strong small-scale magnetic fields during the injection of a plasma with hot electrons into an inhomogeneous cold plasma layer. Doklady Physics 68. (In press.)

\bibitem[{Kolodner and Yablonovitch(1979)}]{Kolodner1979}
Kolodner P, Yablonovitch E (1979) Two-dimensional distribution of self-generated magnetic fields near the laser-plasma resonant- interaction region. Physical Review Letters 43(19):1402--1403. \doi{10.1103/physrevlett.43.1402}

\bibitem[{{Kropotina} et~al(2023){Kropotina}, {Petrukovich}, {Chugunova}, and {Bykov}}]{Kropotina2023}
{Kropotina} JA, {Petrukovich} AA, {Chugunova} OM, et~al (2023) Weibel-dominated quasi-perpendicular shock: hybrid simulations and in situ observations. MNRAS 524(2):2934--2944. \doi{10.1093/mnras/stad2038}

\bibitem[{Kuznetsov et~al(2022)Kuznetsov, Kocharovsky, Kocharovsky, Nechaev, and Garasev}]{Kuznetsov22_FPen}
Kuznetsov AA, Kocharovsky VV, Kocharovsky VV, et~al (2022) Saturating magnetic field of {Weibel} instability in plasmas with bi-{Maxwellian} and bikappa particle distributions. Plasma Phys Rep 48(9):973--982. \doi{10.1134/S1063780X22600700}

\bibitem[{Kuznetsov et~al(2023)Kuznetsov, Nechaev, Garasev, and Kocharovsky}]{Kuznetsov23_JETPen}
Kuznetsov AA, Nechaev AA, Garasev MA, et~al (2023) Quasilinear modeling of the evolution of {W}eibel turbulence in anisotropic collisionless plasma. JETP 137(6). (In press.)

\bibitem[{Landau and Cuperman(1970)}]{Landau1970}
Landau RW, Cuperman S (1970) A temperature-anisotropy instability for electromagnetic waves propagating across a static magnetic field. J Plasma Phys 4(1):13--20. \doi{10.1017/s0022377800004785}

\bibitem[{Langdon(1980)}]{Langdon1980}
Langdon AB (1980) Nonlinear inverse bremsstrahlung and heated-electron distributions. Physical Review Letters 44(9):575--579. \doi{10.1103/physrevlett.44.575}

\bibitem[{Lazar et~al(2022)Lazar, L{\'{o}}pez, Shaaban, Poedts, Yoon, and Fichtner}]{Lazar2022}
Lazar M, L{\'{o}}pez RA, Shaaban SM, et~al (2022) Temperature anisotropy instabilities stimulated by the solar wind suprathermal populations. Frontiers in Astronomy and Space Sciences 8:777559. \doi{10.3389/fspas.2021.777559}

\bibitem[{Lazar et~al(2023)Lazar, L{\'{o}}pez, Moya, Poedts, and Shaaban}]{Lazar2023}
Lazar M, L{\'{o}}pez RA, Moya PS, et~al (2023) The aperiodic firehose instability of counter-beaming electrons in space plasmas. Astronomy \& Astrophysics 670:A85. \doi{10.1051/0004-6361/202245163}

\bibitem[{Li et~al(2019)Li, Tikhonchuk, Moreno, Sio, d{\textquotesingle}Humi{\`{e}}res, Ribeyre, Korneev, Atzeni, Betti, Birkel, Campbell, Follett, Frenje, Hu, Koenig, Sakawa, Sangster, Seguin, Takabe, Zhang, and Petrasso}]{Li2019}
Li C, Tikhonchuk V, Moreno Q, et~al (2019) Collisionless shocks driven by supersonic plasma flows with self-generated magnetic fields. Physical Review Letters 123(5):055002. \doi{10.1103/physrevlett.123.055002}

\bibitem[{Li and Habbal(2000)}]{Li2000}
Li X, Habbal SR (2000) Electron kinetic firehose instability. Journal of Geophysical Research: Space Physics 105(A12):27377--27385. \doi{10.1029/2000ja000063}

\bibitem[{Li and Gan(2006)}]{LiGan2006}
Li YP, Gan WQ (2006) {RHESSI} {X}-ray loop and coronal sources of an occulted flare. Proceedings of the International Astronomical Union 2(S233):393. \doi{10.1017/s1743921306002298}

\bibitem[{L{\'{o}}pez et~al(2019)L{\'{o}}pez, Lazar, Shaaban, Poedts, Yoon, Vi{\~{n}}as, and Moya}]{Lopez2019}
L{\'{o}}pez RA, Lazar M, Shaaban SM, et~al (2019) Particle-in-cell simulations of firehose instability driven by bi-{K}appa electrons. The Astrophysical Journal 873(2):L20. \doi{10.3847/2041-8213/ab0c95}

\bibitem[{Lyubarsky and Eichler(2006)}]{Lyubarsky2006}
Lyubarsky Y, Eichler D (2006) Are gamma-ray burst shocks mediated by the weibel instability? The Astrophysical Journal 647(2):1250--1254. \doi{10.1086/505523}

\bibitem[{Malanushenko et~al(2022)Malanushenko, Cheung, DeForest, Klimchuk, and Rempel}]{Malanushenko2022}
Malanushenko A, Cheung MCM, DeForest CE, et~al (2022) The coronal veil. ApJ 927(1):1. \doi{10.3847/1538-4357/ac3df9}

\bibitem[{{Malkov} et~al(2016){Malkov}, {Sagdeev}, {Dudnikova}, {Liseykina}, {Diamond}, {Papadopoulos}, {Liu}, and {Su}}]{Malkov2016}
{Malkov} MA, {Sagdeev} RZ, {Dudnikova} GI, et~al (2016) Ion-acoustic shocks with self-regulated ion reflection and acceleration. Physics of Plasmas 23(4):043105. \doi{10.1063/1.4945649}

\bibitem[{Marcowith et~al(2016)Marcowith, Bret, Bykov, Dieckman, Drury, Lemb{\`{e}}ge, Lemoine, Morlino, Murphy, Pelletier, Plotnikov, Reville, Riquelme, Sironi, and Novo}]{Marcowith2016}
Marcowith A, Bret A, Bykov A, et~al (2016) The microphysics of collisionless shock waves. Reports on Progress in Physics 79(4):046901. \doi{10.1088/0034-4885/79/4/046901}

\bibitem[{Marsch(2006)}]{Marsch2006}
Marsch E (2006) Kinetic physics of the solar corona and solar wind. Living Reviews in Solar Physics 3:1. \doi{10.12942/lrsp-2006-1}

\bibitem[{Medvedev and Loeb(1999)}]{Medvedev1999}
Medvedev MV, Loeb A (1999) Generation of magnetic fields in the relativistic shock of gamma-ray burst sources. The Astrophysical Journal 526(2):697--706. \doi{10.1086/308038}

\bibitem[{Medvedev(2014)}]{Medvedev2014}
Medvedev YV (2014) Evolution of a density disturbance in a collisionless plasma. Plasma Physics and Controlled Fusion 56(2):025005. \doi{10.1088/0741-3335/56/2/025005}

\bibitem[{Meneses et~al(2018)Meneses, Gaelzer, and Ziebell}]{Meneses2018}
Meneses AR, Gaelzer R, Ziebell LF (2018) The oblique firehose instability in a bi-kappa magnetized plasma. Phys Plasmas 25(11):112901. \doi{10.1063/1.5063537}

\bibitem[{Moiseev and Sagdeev(1963)}]{Moiseev1963}
Moiseev SS, Sagdeev RZ (1963) Collisionless shock waves in a plasma in a weak magnetic field. Journal of Nuclear Energy Part C, Plasma Physics, Accelerators, Thermonuclear Research 5(1):43--47. \doi{10.1088/0368-3281/5/1/309}

\bibitem[{Moreno et~al(2020)Moreno, Dieckmann, Folini, Walder, Ribeyre, Tikhonchuk, and d'Humi{\`{e}}res}]{Moreno2020}
Moreno Q, Dieckmann ME, Folini D, et~al (2020) Shocks and phase space vortices driven by a density jump between two clouds of electrons and protons. Plasma Physics and Controlled Fusion 62(2):025022. \doi{10.1088/1361-6587/ab5bfb}

\bibitem[{Moritaka et~al(2016)Moritaka, Kuramitsu, Liu, and Chen}]{Moritaka2016}
Moritaka T, Kuramitsu Y, Liu YL, et~al (2016) Spontaneous focusing of plasma flow in a weak perpendicular magnetic field. Physics of Plasmas 23(3):032110. \doi{10.1063/1.4942028}

\bibitem[{Moya et~al(2022)Moya, L{\'{o}}pez, Lazar, Poedts, and Shaaban}]{Moya2022}
Moya PS, L{\'{o}}pez RA, Lazar M, et~al (2022) Comparing the counter-beaming and temperature anisotropy driven aperiodic electron firehose instabilities in collisionless plasma environments. The Astrophysical Journal 937(2):49. \doi{10.3847/1538-4357/ac8cf8}

\bibitem[{Nakamura et~al(2018)Nakamura, Varsani, Genestreti, Contel, Nakamura, Baumjohann, Nagai, Artemyev, Birn, Sergeev, Apatenkov, Ergun, Fuselier, Gershman, Giles, Khotyaintsev, Lindqvist, Magnes, Mauk, Petrukovich, Russell, Stawarz, Strangeway, Anderson, Burch, Bromund, Cohen, Fischer, Jaynes, Kepko, Le, Plaschke, Reeves, Singer, Slavin, Torbert, and Turner}]{Nakamura2018}
Nakamura R, Varsani A, Genestreti KJ, et~al (2018) Multiscale currents observed by {MMS} in the flow braking region. J Geophys Res Space Phys 123(2):1260--1278. \doi{10.1002/2017ja024686}

\bibitem[{Nechaev et~al(2020{\natexlab{a}})Nechaev, Garasev, Kocharovsky, and Kocharovsky}]{Nechaev20_RadiophysEn}
Nechaev AA, Garasev MA, Kocharovsky VV, et~al (2020{\natexlab{a}}) Weibel mechanism of magnetic-field generation in the process of expansion of a collisionless-plasma bunch with hot electrons. Radiophysics and Quantum Electronics 62(12):830--848. \doi{10.1007/s11141-020-10028-w}

\bibitem[{Nechaev et~al(2020{\natexlab{b}})Nechaev, Garasev, Stepanov, and Kocharovsky}]{Nechaev20_FPen}
Nechaev AA, Garasev MA, Stepanov AN, et~al (2020{\natexlab{b}}) Formation of a density bump in a collisionless electrostatic shock wave during expansion of a hot dense plasma into a cold rarefied one. Plasma Physics Reports 46(8):765--783. \doi{10.1134/s1063780x2008005x}

\bibitem[{Nechaev et~al(2023)Nechaev, Kuznetsov, and Kocharovsky}]{Nechaev23_JPP}
Nechaev AA, Kuznetsov AA, Kocharovsky VV (2023) On the analytical description of the nonlinear stage of the weibel instability in collisionless anisotropic plasma. J Plasma Phys (Submitted.)

\bibitem[{Ngirmang et~al(2020)Ngirmang, Morrison, George, Smith, Frische, Orban, Chowdhury, and Roquemore}]{Ngirmang2020}
Ngirmang GK, Morrison JT, George KM, et~al (2020) Evidence of radial {W}eibel instability in relativistic intensity laser-plasma interactions inside a sub-micron thick liquid target. Scientific Reports 10(1):9872. \doi{10.1038/s41598-020-66615-4}

\bibitem[{{Nishigai} and {Amano}(2021)}]{Nishigai2021}
{Nishigai} T, {Amano} T (2021) Mach number dependence of ion-scale kinetic instability at collisionless perpendicular shock: {C}ondition for {W}eibel-dominated shock. Phys Plasmas 28(7):072903. \doi{10.1063/5.0051269}

\bibitem[{Palmer et~al(2019)Palmer, Campbell, Ma, Antonelli, Bott, Gregori, Halliday, Katzir, Kordell, Krushelnick, Lebedev, Montgomery, Notley, Carroll, Ridgers, Schekochihin, Streeter, Thomas, Tubman, Woolsey, and Willingale}]{Palmer2019}
Palmer CAJ, Campbell PT, Ma Y, et~al (2019) Field reconstruction from proton radiography of intense laser driven magnetic reconnection. Physics of Plasmas 26(8):083109. \doi{10.1063/1.5092733}

\bibitem[{Parker(1988)}]{Parker1998}
Parker EN (1988) Nanoflares and the solar x-ray corona. ApJ 330:474. \doi{10.1086/166485}

\bibitem[{Patel et~al(2021)Patel, Behera, Singh, Kumar, and Das}]{Patel2021}
Patel BG, Behera N, Singh RK, et~al (2021) A {3D} magnetohydrodynamic simulation of the propagation of a plasma plume transverse to applied magnetic field. Plasma Physics and Controlled Fusion 63(11):115020. \doi{10.1088/1361-6587/ac2617}

\bibitem[{Peterson et~al(2021)Peterson, Glenzer, and Fiuza}]{Peterson2021}
Peterson J, Glenzer S, Fiuza F (2021) Magnetic field amplification by a nonlinear electron streaming instability. Physical Review Letters 126(21). \doi{10.1103/physrevlett.126.215101}

\bibitem[{Pierrard and Lazar(2010)}]{Pierrard2010}
Pierrard V, Lazar M (2010) Kappa distributions: Theory and applications in space plasmas. Solar Physics 267(1):153--174. \doi{10.1007/s11207-010-9640-2}

\bibitem[{Plechaty et~al(2009)Plechaty, Presura, Wright, Neff, and Haboub}]{Plechaty2009}
Plechaty C, Presura R, Wright S, et~al (2009) Penetration of plasma across a magnetic field. Astrophysics and Space Science 322:195--199. \doi{10.1007/s10509-009-9997-6}

\bibitem[{Plechaty et~al(2013)Plechaty, Presura, and Esaulov}]{Plechaty2013}
Plechaty C, Presura R, Esaulov AA (2013) Focusing of an explosive plasma expansion in a transverse magnetic field. Phys Rev Lett 111(18):185002. \doi{10.1103/physrevlett.111.185002}

\bibitem[{Pokhotelov and Balikhin(2012)}]{Pokhotelov2012}
Pokhotelov OA, Balikhin MA (2012) Weibel instability in a plasma with nonzero external magnetic field. Annales Geophysicae 30(7):1051--1054. \doi{10.5194/angeo-30-1051-2012}, \urlprefix\url{https://doi.org/10.5194/angeo-30-1051-2012}

\bibitem[{Priest(2014)}]{Priest2014}
Priest E (2014) Magnetohydrodynamics of the {S}un. Cambridge University Press, Cambridge, \doi{10.1017/cbo9781139020732}

\bibitem[{Priest and Forbes(2001)}]{Priest2001}
Priest E, Forbes T (2001) The magnetic nature of solar flares. The Astronomy and Astrophysics Review 10(4):313--377. \doi{10.1007/s001590100013}

\bibitem[{Purkhart and Veronig(2022)}]{Purkhart2022}
Purkhart S, Veronig AM (2022) Nanoflare distributions over solar cycle 24 based on {SDO}/{AIA} differential emission measure observations. Astronomy {\&} Astrophysics 661:A149. \doi{10.1051/0004-6361/202243234}

\bibitem[{{Pusztai} et~al(2018){Pusztai}, {TenBarge}, {Csap{\'o}}, {Juno}, {Hakim}, {Yi}, and {F{\"u}l{\"o}p}}]{Pusztai2018}
{Pusztai} I, {TenBarge} JM, {Csap{\'o}} AN, et~al (2018) Low mach-number collisionless electrostatic shocks and associated ion acceleration. Plasma Physics and Controlled Fusion 60(3):035004. \doi{10.1088/1361-6587/aaa2cc}

\bibitem[{Quinn et~al(2012)Quinn, Romagnani, Ramakrishna, Sarri, Dieckmann, Wilson, Fuchs, Lancia, Pipahl, Toncian, Willi, Clarke, Notley, Macchi, and Borghesi}]{Quinn2012}
Quinn K, Romagnani L, Ramakrishna B, et~al (2012) Weibel-induced filamentation during an ultrafast laser-driven plasma expansion. Phys Rev Lett 108:135001. \doi{10.1103/PhysRevLett.108.135001}

\bibitem[{Romagnani et~al(2008)Romagnani, Bulanov, Borghesi, Audebert, Gauthier, Löwenbrück, Mackinnon, Patel, Pretzler, Toncian, and Willi}]{Romagnani2008}
Romagnani L, Bulanov SV, Borghesi M, et~al (2008) Observation of collisionless shocks in laser-plasma experiments. Phys Rev Lett 101:025004. \doi{10.1103/PhysRevLett.101.025004}

\bibitem[{Ruyer et~al(2015{\natexlab{a}})Ruyer, Gremillet, and Bonnaud}]{Ruyer2015}
Ruyer C, Gremillet L, Bonnaud G (2015{\natexlab{a}}) Weibel-mediated collisionless shocks in laser-irradiated dense plasmas: Prevailing role of the electrons in generating the field fluctuations. Physics of Plasmas 22(8):082107. \doi{10.1063/1.4928096}

\bibitem[{Ruyer et~al(2015{\natexlab{b}})Ruyer, Gremillet, Debayle, and Bonnaud}]{Ruyer2015_NonlinWeibel}
Ruyer C, Gremillet L, Debayle A, et~al (2015{\natexlab{b}}) Nonlinear dynamics of the ion weibel-filamentation instability: An analytical model for the evolution of the plasma and spectral properties. Physics of Plasmas 22(3):032102. \doi{10.1063/1.4913651}

\bibitem[{Ruyer et~al(2020)Ruyer, Bola{\~{n}}os, Albertazzi, Chen, Antici, B\"{o}ker, Dervieux, Lancia, Nakatsutsumi, Romagnani, Shepherd, Swantusch, Borghesi, Willi, P{\'{e}}pin, Starodubtsev, Grech, Riconda, Gremillet, and Fuchs}]{Ruyer2020}
Ruyer C, Bola{\~{n}}os S, Albertazzi B, et~al (2020) Growth of concomitant laser-driven collisionless and resistive electron filamentation instabilities over large spatiotemporal scales. Nature Physics 16(9):983--988. \doi{10.1038/s41567-020-0913-x}

\bibitem[{Sagdeev(1966)}]{Sagdeev1966}
Sagdeev RZ (1966) Cooperative phenomena and shock waves in collisionless plasmas. Rev Plasma Phys 4:23--91

\bibitem[{Sakagami et~al(1979)Sakagami, Kawakami, Nagao, and Yamanaka}]{Sakagami1979}
Sakagami Y, Kawakami H, Nagao S, et~al (1979) Two-dimensional distribution of self-generated magnetic fields near the laser-plasma resonant-interaction region. Physical Review Letters 42(13):839--842. \doi{10.1103/physrevlett.42.839}

\bibitem[{Sarri et~al(2011)Sarri, Murphy, Dieckmann, Bret, Quinn, Kourakis, Borghesi, Drury, and Ynnerman}]{Sarri2011}
Sarri G, Murphy GC, Dieckmann ME, et~al (2011) Two-dimensional particle-in-cell simulation of the expansion of a plasma into a rarefied medium. New Journal of Physics 13(7):073023. \doi{10.1088/1367-2630/13/7/073023}

\bibitem[{Sarri et~al(2012)Sarri, Macchi, Cecchetti, Kar, Liseykina, Yang, Dieckmann, Fuchs, Galimberti, Gizzi, Jung, Kourakis, Osterholz, Pegoraro, Robinson, Romagnani, Willi, and Borghesi}]{Sarri2012}
Sarri G, Macchi A, Cecchetti CA, et~al (2012) Dynamics of self-generated, large amplitude magnetic fields following high-intensity laser matter interaction. Physical Review Letters 109(20):205002. \doi{10.1103/physrevlett.109.205002}

\bibitem[{Schoeffler and Silva(2018)}]{Schoeffler2018}
Schoeffler KM, Silva LO (2018) General kinetic solution for the biermann battery with an associated pressure anisotropy generation. Plasma Physics and Controlled Fusion 60(1):014048. \doi{10.1088/1361-6587/aa883a}

\bibitem[{Schoeffler et~al(2016)Schoeffler, Loureiro, Fonseca, and Silva}]{Schoeffler2016}
Schoeffler KM, Loureiro NF, Fonseca RA, et~al (2016) The generation of magnetic fields by the biermann battery and the interplay with the weibel instability. Physics of Plasmas 23(5):056304. \doi{10.1063/1.4946017}

\bibitem[{Schou et~al(2007)Schou, Amoruso, and Lunney}]{Schou2007_LaserAblation}
Schou J, Amoruso S, Lunney JG (2007) Plume dynamics, Springer, pp 67--96. \doi{10.1007/978-0-387-30453-3}

\bibitem[{Scott et~al(2017)Scott, Brenner, Bagnoud, Clarke, Gonzalez-Izquierdo, Green, Heathcote, Powell, Rusby, Zielbauer, McKenna, and Neely}]{Scott2017}
Scott GG, Brenner CM, Bagnoud V, et~al (2017) Diagnosis of {W}eibel instability evolution in the rear surface density scale lengths of laser solid interactions via proton acceleration. New Journal of Physics 19(4):043010. \doi{10.1088/1367-2630/aa652c}

\bibitem[{Shaikh et~al(2017)Shaikh, Lad, Jana, Sarkar, Dey, and Kumar}]{Shaikh2017}
Shaikh M, Lad AD, Jana K, et~al (2017) Megagauss magnetic fields in ultra-intense laser generated dense plasmas. Plasma Physics and Controlled Fusion 59(1):014007. \doi{10.1088/0741-3335/59/1/014007}

\bibitem[{Shukla et~al(2020)Shukla, Schoeffler, Boella, Vieira, Fonseca, and Silva}]{Shukla2020}
Shukla N, Schoeffler K, Boella E, et~al (2020) Interplay between the {Weibel} instability and the {Biermann} battery in realistic laser-solid interactions. Physical Review Research 2(2):023129. \doi{10.1103/physrevresearch.2.023129}

\bibitem[{Shuster et~al(2019)Shuster, Gershman, Chen, Wang, Bessho, Dorelli, Silva, Giles, Paterson, Denton, Schwartz, Norgren, Wilder, Cassak, Swisdak, Uritsky, Schiff, Rager, Smith, Avanov, and Vi{\~{n}}as}]{Shuster2019}
Shuster JR, Gershman DJ, Chen LJ, et~al (2019) {MMS} measurements of the {Vlasov} equation: Probing the electron pressure divergence within thin current sheets. Geophys Res Lett 46(14):7862--7872. \doi{10.1029/2019gl083549}

\bibitem[{Silva(2006)}]{Silva2006}
Silva LO (2006) Physical problems (microphysics) in relativistic plasma flows. {AIP} Conference Proceedings 856:109. \doi{10.1063/1.2356387}

\bibitem[{Silva et~al(2021)Silva, Afeyan, and Silva}]{Silva2021}
Silva T, Afeyan B, Silva LO (2021) Weibel instability beyond bi-maxwellian anisotropy. Physical Review E 104(3):035201. \doi{10.1103/physreve.104.035201}

\bibitem[{Sironi and Spitkovsky(2009)}]{Sironi2009}
Sironi L, Spitkovsky A (2009) Particle acceleration in relativistic magnetized collisionless pair shocks: dependence of shock acceleration on magnetic obliquity. ApJ 698(2):1523--1549. \doi{10.1088/0004-637x/698/2/1523}

\bibitem[{{Sironi} and {Spitkovsky}(2011)}]{Sironi2011}
{Sironi} L, {Spitkovsky} A (2011) Particle acceleration in relativistic magnetized collisionless electron-ion shocks. ApJ 726(2):75. \doi{10.1088/0004-637X/726/2/75}

\bibitem[{Sironi et~al(2013)Sironi, Spitkovsky, and Arons}]{Sironi2013}
Sironi L, Spitkovsky A, Arons J (2013) The maximum energy of accelerated particles in relativistic collisionless shocks. ApJ 771(1):54. \doi{10.1088/0004-637x/771/1/54}

\bibitem[{Spitkovsky(2008)}]{Spitkovsky2008}
Spitkovsky A (2008) Particle acceleration in relativistic collisionless shocks: Fermi process at last? The Astrophysical Journal 682(1):L5--L8. \doi{10.1086/590248}

\bibitem[{Srivastava et~al(2019)Srivastava, Mishra, Jel{\'{\i}}nek, Samanta, Tian, Pant, Kayshap, Banerjee, Doyle, and Dwivedi}]{Srivastava2019}
Srivastava AK, Mishra SK, Jel{\'{\i}}nek P, et~al (2019) On the observations of rapid forced reconnection in the solar corona. The Astrophysical Journal 887(2):137. \doi{10.3847/1538-4357/ab4a0c}

\bibitem[{Stepanov et~al(2018)Stepanov, Garasev, Kocharovsky, Korytin, Malrkov, Murzanev, and Nechaev}]{Stepanov2018_LO2018}
Stepanov AN, Garasev MA, Kocharovsky VV, et~al (2018) Generation of magnetic fields behind the front of an electrostatic shock wave in a laser plasma. In: 2018 {International} {Conference} {Laser} {Optics} ({ICLO}). IEEE, Saint Petersburg, Russia, p 242, \doi{10.1109/LO.2018.8435840}

\bibitem[{Stepanov et~al(2020)Stepanov, Garasev, Kocharovsky, Korytin, Murzanev, Nechaev, Kartashov, and Samsonova}]{Stepanov20_LO20}
Stepanov AN, Garasev MA, Kocharovsky VV, et~al (2020) Investigation of the instabilities of an expanding plasma created during ablation of solid targets by intense femtosecond laser pulses. In: Proceedings of 2020 {International} {Conference} {Laser} {Optics} ({ICLO}). IEEE, Saint Petersburg, Russia, p 213, \doi{10.1109/ICLO48556.2020.9285395}

\bibitem[{Stepanov et~al(2022{\natexlab{a}})Stepanov, Garasev, Kocharovsky, Kocharovsky, and Nechaev}]{Stepanov22_TVTen}
Stepanov AN, Garasev MA, Kocharovsky VV, et~al (2022{\natexlab{a}}) Formation and expansion of current filaments during the decay of a cylindrical plasma region with hot electrons heated at the interface of a cold plasma and vacuum. High Temperature 60:287--291. \doi{10.1134/S0018151X22020080}

\bibitem[{Stepanov et~al(2022{\natexlab{b}})Stepanov, Zaitsev, and Kupriyanova}]{Stepanov2022_Cowling}
Stepanov AV, Zaitsev VV, Kupriyanova EG (2022{\natexlab{b}}) Cowling resistivity and {J}oule dissipation in solar atmosphere. In: Proceedings of The Multifaceted Universe: {T}heory and Observations - 2022. Sissa Medialab, \doi{10.22323/1.425.0052}

\bibitem[{Stockem et~al(2006)Stockem, Lerche, and Schlickeiser}]{Stockem2006}
Stockem A, Lerche I, Schlickeiser R (2006) On the physical realization of two-dimensional turbulence fields in magnetized interplanetary plasmas. The Astrophysical Journal 651(1):584--589. \doi{10.1086/507461}

\bibitem[{Stockem et~al(2014)Stockem, Grismayer, Fonseca, and Silva}]{Stockem2014}
Stockem A, Grismayer T, Fonseca RA, et~al (2014) Electromagnetic field generation in the downstream of electrostatic shocks due to electron trapping. Physical Review Letters 113(10):105002. \doi{10.1103/physrevlett.113.105002}

\bibitem[{Tautz and Triptow(2013)}]{Tautz2013}
Tautz RC, Triptow J (2013) Interstellar turbulent magnetic field generation by plasma instabilities. Astrophysics and Space Science 348(1):133--141. \doi{10.1007/s10509-013-1546-7}

\bibitem[{Thaury et~al(2010)Thaury, Mora, H{\'{e}}ron, and Adam}]{Thaury2010}
Thaury C, Mora P, H{\'{e}}ron A, et~al (2010) Self-generation of megagauss magnetic fields during the expansion of a plasma. Physical Review E 82(1). \doi{10.1103/physreve.82.016408}

\bibitem[{Treumann(2009)}]{Treumann2009}
Treumann RA (2009) Fundamentals of collisionless shocks for astrophysical application, 1. non-relativistic shocks. The Astronomy and Astrophysics Review 17(4):409--535. \doi{10.1007/s00159-009-0024-2}

\bibitem[{Trubnikov(1965)}]{Trubnikov1965}
Trubnikov BA (1965) Particle interactions in a fully ionized plasma. In: Leontovich MA (ed) Reviews of Plasma Physics, vol~1. Consultants Bureau, New York, p 105

\bibitem[{Ulyanov et~al(2019)Ulyanov, Bogachev, Loboda, Reva, and Kirichenko}]{Ulyanov2019}
Ulyanov AS, Bogachev SA, Loboda IP, et~al (2019) Direct evidence for magnetic reconnection in a solar {EUV} nanoflare. Solar Physics 294(9):128. \doi{10.1007/s11207-019-1472-0}

\bibitem[{Umeda and Nakamura(2018)}]{Umeda2018}
Umeda T, Nakamura TKM (2018) Electromagnetic linear dispersion relation for plasma with a drift across magnetic field revisited. Physics of Plasmas 25(10):102109. \doi{10.1063/1.5050542}

\bibitem[{Vagin and Uryupin(2014)}]{Vagin2014}
Vagin KY, Uryupin SA (2014) On the growth rate of aperiodic instability in plasma with an anisotropic bi-{M}axwellian electron velocity distribution. Plasma Physics Reports 40(5):393--403. \doi{10.1134/s1063780x14040096}

\bibitem[{Verscharen et~al(2018)Verscharen, Klein, Chandran, Stevens, Salem, and Bale}]{Verscharen2018}
Verscharen D, Klein KG, Chandran BDG, et~al (2018) {ALPS}: the arbitrary linear plasma solver. J Plasma Phys 84(4):905840403. \doi{10.1017/s0022377818000739}

\bibitem[{Vlahos(1989)}]{Vlahos1989}
Vlahos L (1989) Particle acceleration in solar flares. International Astronomical Union Colloquium 104(1):431--447. \doi{10.1017/s0252921100032048}

\bibitem[{V\"{o}r\"{o}s et~al(2017)V\"{o}r\"{o}s, Yordanova, Varsani, Genestreti, Khotyaintsev, Li, Graham, Norgren, Nakamura, Narita, Plaschke, Magnes, Baumjohann, Fischer, Vaivads, Eriksson, Lindqvist, Marklund, Ergun, Leitner, Leubner, Strangeway, Contel, Pollock, Giles, Torbert, Burch, Avanov, Dorelli, Gershman, Paterson, Lavraud, and Saito}]{Voros2017}
V\"{o}r\"{o}s Z, Yordanova E, Varsani A, et~al (2017) {MMS} observation of magnetic reconnection in the turbulent magnetosheath. J Geophys Res Space Phys 122(11):11442--11467. \doi{10.1002/2017ja024535}

\bibitem[{Weibel(1959)}]{Weibel1959}
Weibel ES (1959) Spontaneously growing transverse waves in a plasma due to an anisotropic velocity distribution. Physical Review Letters 2(3):83--84. \doi{10.1103/physrevlett.2.83}

\bibitem[{Xie and Xiao(2016)}]{Xie2016}
Xie H, Xiao Y (2016) {PDRK}: {A} general kinetic dispersion relation solver for magnetized plasma. Plasma Science and Technology 18(2):97--107. \doi{10.1088/1009-0630/18/2/01}

\bibitem[{Yang et~al(1994)Yang, Arons, and Langdon}]{Yang1994}
Yang TYB, Arons J, Langdon AB (1994) Evolution of the {W}eibel instability in relativistically hot electron{\textendash}positron plasmas. Physics of Plasmas 1(9):3059--3077. \doi{10.1063/1.870498}

\bibitem[{Yoon(2017)}]{Yoon2017}
Yoon PH (2017) Kinetic instabilities in the solar wind driven by temperature anisotropies. Reviews of Modern Plasma Physics 1(1):4. \doi{10.1007/s41614-017-0006-1}

\bibitem[{Zaitsev and Stepanov(2008)}]{Zaitsev2008}
Zaitsev VV, Stepanov AV (2008) Coronal magnetic loops. Physics-Uspekhi 51(11):1123--1160. \doi{10.1070/pu2008v051n11abeh006657}

\bibitem[{Zaitsev and Stepanov(2015)}]{Zaitsev2015}
Zaitsev VV, Stepanov AV (2015) Particle acceleration and plasma heating in the chromosphere. Solar Physics 290:3559--3572. \doi{10.1007/s11207-015-0731-y}

\bibitem[{Zaitsev and Stepanov(2017)}]{Zaitsev2017}
Zaitsev VV, Stepanov AV (2017) Acceleration and storage of energetic electrons in magnetic loops in the course of electric current oscillations. Solar Physics 292:141. \doi{10.1007/s11207-017-1168-2}

\bibitem[{Zaitsev et~al(2019)Zaitsev, Stepanov, and Melnikov}]{Zaitsev2019}
Zaitsev VV, Stepanov AV, Melnikov AV (2019) Dynamic model of magnetic flux ropes. Geomagnetism and Aeronomy 59(7):806--809. \doi{10.1134/s0016793219070272}

\bibitem[{Zhang et~al(2020)Zhang, Hua, Wu, Fang, Ma, Zhang, Liu, Peng, He, Huang, Marsh, Mori, Lu, and Joshi}]{Zhang2020}
Zhang C, Hua J, Wu Y, et~al (2020) Measurements of the growth and saturation of electron {W}eibel instability in optical-field ionized plasmas. Physical Review Letters 125(25):255001. \doi{10.1103/physrevlett.125.255001}

\bibitem[{Zhang et~al(2022{\natexlab{a}})Zhang, Wu, Sinclair, Farrell, Marsh, Hua, Petrushina, Vafaei-Najafabadi, Kupfer, Kusche, Fedurin, Pogorelsky, Polyanskiy, Huang, Lu, Mori, and Joshi}]{Zhang2022_PoP}
Zhang C, Wu Y, Sinclair M, et~al (2022{\natexlab{a}}) Electron {W}eibel instability induced magnetic fields in optical-field ionized plasmas. Physics of Plasmas 29(6):062102. \doi{10.1063/5.0089814}

\bibitem[{Zhang et~al(2022{\natexlab{b}})Zhang, Wu, Sinclair, Farrell, Marsh, Petrushina, Vafaei-Najafabadi, Gaikwad, Kupfer, Kusche, Fedurin, Pogorelsky, Polyanskiy, Huang, Hua, Lu, Mori, and Joshi}]{Zhang2022_PNAS}
Zhang C, Wu Y, Sinclair M, et~al (2022{\natexlab{b}}) Mapping the self-generated magnetic fields due to thermal weibel instability. Proceedings of the National Academy of Sciences 119(50):e2211713119. \doi{10.1073/pnas.2211713119}

\bibitem[{{Zhang} et~al(2018){Zhang}, {Cai}, and {Zhu}}]{Zhang2018}
{Zhang} Ws, {Cai} Hb, {Zhu} Sp (2018) The formation and dissipation of electrostatic shock waves: the role of ion-ion acoustic instabilities. Plasma Physics and Controlled Fusion 60(5):055001. \doi{10.1088/1361-6587/aab175}

\bibitem[{Zhou et~al(2018)Zhou, Bai, Tian, Sun, Cao, and Liu}]{Zhou2018}
Zhou S, Bai Y, Tian Y, et~al (2018) Self-organized kilotesla magnetic-tube array in an expanding spherical plasma irradiated by {kHz} femtosecond laser pulses. Physical Review Letters 121(25):255002. \doi{10.1103/physrevlett.121.255002}

\bibitem[{Zhou et~al(2013)Zhou, Wang, Li, Song, Melnikov, and Ji}]{Zhou2013_RAA}
Zhou TH, Wang JF, Li D, et~al (2013) The contracting and unshearing motion of flare loops in the {X 7.1} flare on 2005 {J}anuary 20 during its rising phase. Research in Astronomy and Astrophysics 13(5):526--536. \doi{10.1088/1674-4527/13/5/004}

\end{thebibliography}

\end{document}